\documentclass[12pt,journal,draftclsnofoot, onecolumn]{IEEEtran}

\usepackage{latexsym}
\usepackage[hidelinks]{hyperref}
\usepackage{tabularx}
\usepackage[T1]{fontenc}% optional T1 font encoding
\usepackage{algorithm}
\usepackage[noend]{algpseudocode}
\usepackage{graphicx}
\usepackage{amsmath,amssymb,amsfonts,stfloats}
\usepackage{array}
\usepackage{textcomp}
\usepackage[font=small]{caption}
\usepackage{paralist}
\usepackage{cite}
\usepackage{xcolor}
\usepackage{booktabs}
\usepackage[caption=false,font=footnotesize]{subfig}
\usepackage{url}
\usepackage{acronym}
\usepackage{latexsym}
\usepackage{hyperref}
\usepackage{soul}

\acrodef{COG}[CoG]{Center of Gravity}
\acrodef{DoF}[DoF]{Degrees of Freedom}
\acrodef{C-ITS}[C-ITS]{Cooperative Intelligent Transportation Systems}
\acrodef{BAT}[BAT]{Beam Alignment and Tracking}
\acrodef{NR}[NR]{New Radio}
\acrodef{FR}[FR]{Frequency Range}
\acrodef{BPA}[BPA]{Beamwidth and Power Adaptation}
\acrodef{BPC}[BPC]{Beamwidth \textit{and} Power Control}
\acrodef{BC}[BC]{Beam Control}
\acrodef{LOS}[LOS]{Line-Of-Sight}
\acrodef{NLOS}[NLOS]{Non-Line-Of-Sight}
\acrodef{EKF}[EKF]{Extended Kalman Filter}
\acrodef{V2X}[V2X]{Vehicle-to-Everything}
\acrodef{V2I}[V2I]{Vehicle-to-Infrastructure}
\acrodef{V2V}[V2V]{Vehicle-to-Vehicle}
\acrodef{MIMO}[mMIMO]{Multiple-Input Multiple-Output}
\acrodef{PDF}[PDF]{Probability Density Function}
\acrodef{CAD}[CAD]{Connected and Automated Driving}
\newcommand{\norm}[1]{\left\lVert#1\right\rVert}

\def\BibTeX{{\rm B\kern-.05em{\sc i\kern-.025em b}\kern-.08em T\kern-.1667em\lower.7ex\hbox{E}\kern-.125emX}}
\begin{document}

%\history{}
%\doi{}

\title{Sensor-Aided Beamwidth and Power Control for Next Generation Vehicular Communications}
\author{Dario~Tagliaferri, Mattia~Brambilla, Monica~Nicoli, Umberto~Spagnolini 	\thanks{Dario Tagliaferri is with Dipartimento di Elettronica, Informazione e Bioingegneria, Politecnico di Milano.}
\thanks{Mattia Brambilla and Monica Nicoli are with Dipartimento di Ingegneria Gestionale, Politecnico di Milano.}
\thanks{Umberto Spagnolini is with Dipartimento di Elettronica, Informazione e Bioingegneria, Politecnico di Milano, and he is Huawei Industry Chair.}}

%\title{Sensor-Aided Beamwidth and Power Control for Next Generation Vehicular Communications}

%\author{ \uppercase{Dario Tagliaferri}\authorrefmark{1} \IEEEmembership{Member, IEEE}, 
%		 \uppercase{Mattia Brambilla}\authorrefmark{1} \IEEEmembership{Member, IEEE},
%	  	 \uppercase{Monica Nicoli}\authorrefmark{2} \IEEEmembership{Member, IEEE}, AND 
%	  	 \uppercase{Umberto Spagnolini}\authorrefmark{1,3} \IEEEmembership{Senior Member, IEEE}}
%\address[1]{Dipartimento di Elettronica, Informazione e Bioingegneria (DEIB), Politecnico di Milano, Via Ponzio 34/5, 20133 Milano, Italy \\(\{dario.tagliaferri,mattia.brambilla,umberto.spagnolini\}@polimi.it)}
%\address[2]{Dipartimento di Ingegneria Gestionale (DIG), Politecnico di Milano, Via Lambruschini 4/b, 20156 Milano, Italy\\ (monica.nicoli@polimi.it)}
%\address[3]{Huawei Technologies Italia S.r.l., Via Andrea Verrocchio 2, 20129 Milano, Italy }

%\corresp{Corresponding author: D. Tagliaferri (e-mail: dario.tagliaferri@polimi.it).}

\maketitle

\begin{abstract}
\textbf{Ultra-reliable low-latency Vehicle-to-Everything (V2X) communications are needed to meet the extreme requirements of enhanced driving applications. Millimeter-Wave (24.25-52.6 GHz) or sub-THz (>100 GHz) V2X communications are a viable solution, provided that the highly collimated beams are kept aligned during vehicles' maneuverings. In this work, we propose a sensor-assisted dynamic Beamwidth and Power Control (BPC) system to counteract the detrimental effect of vehicle dynamics, exploiting data collected by on-board inertial and positioning sensors, mutually exchanged among vehicles over a parallel low-rate link, e.g., 5G New Radio (NR) Frequency Range 1 (FR1). The proposed BPC solution works on top of a sensor-aided Beam Alignment and Tracking (BAT) system, overcoming the limitations of fixed-beamwidth systems and optimizing the performance in challenging Vehicle-to-Vehicle (V2V) scenarios, even if extensions to Vehicle-to-Infrastructure (V2I) use-cases are feasible. We validate the sensor-assisted dynamic BPC on real trajectories and sensors' data collected by a dedicated experimental campaign. The goal is to show the advantages of the proposed BPC strategy in a high data-rate Line-Of-Sight (LOS) V2V context, and to outline the requirements in terms of sensors' sampling time and accuracy, along with the end-to-end latency on the control channel.}
\end{abstract}


\section{Introduction}
\label{sec:introduction}
%io qui ridurrei la lunghezza
The  development of \ac{C-ITS} in the framework of the fifth generation of cellular systems (5G) for \ac{CAD} will greatly improve the quality of mobility in terms of efficiency, safety and comfort, thus breaking the conventional paradigm of human-controlled driving \cite{andrews2014will,SAE}. In this context, \ac{V2X} communications are essential in enabling vehicular cloud network for fast sharing of massive mobility data. Differently from any other type of wireless systems, V2X and, in particular, \ac{V2V} communications for extended sensor functionalities in high Levels of Automation (LoA) are extremely critical due to the high mobility  (up to 250 km/h speed), end-to-end latency (< 10 ms), reliability (packet error rate < 10$^{-5}$) and data-rate (> 1 Gbps). These call for radio technologies in the boundary between ultra-reliable low-latency communications and enhanced mobile broadband 5G services \cite{ETSI:v2xreq_report,Heath2016,7244203,popovski20185g}.

\begin{figure}[]
	\centering
	\subfloat[Ideal \label{subfig:toyExampleA}]{	\includegraphics[width=0.8\columnwidth]{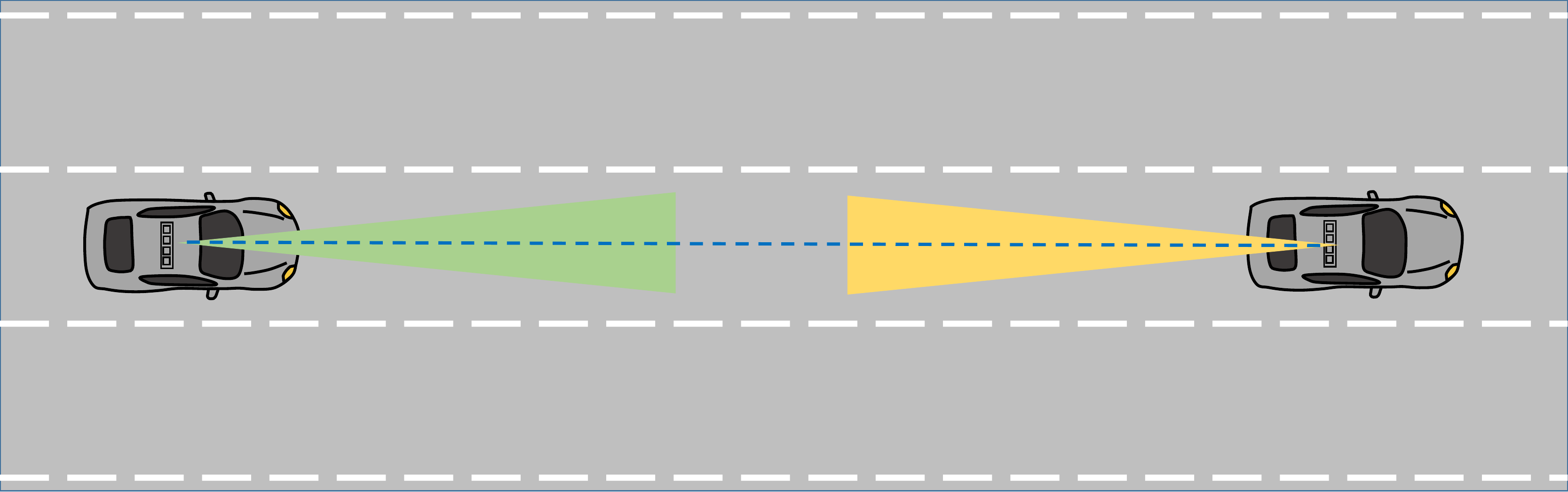}  } \\
	\subfloat[Impact of Rx position error \label{subfig:toyExampleB}]{	\includegraphics[width=0.8\columnwidth]{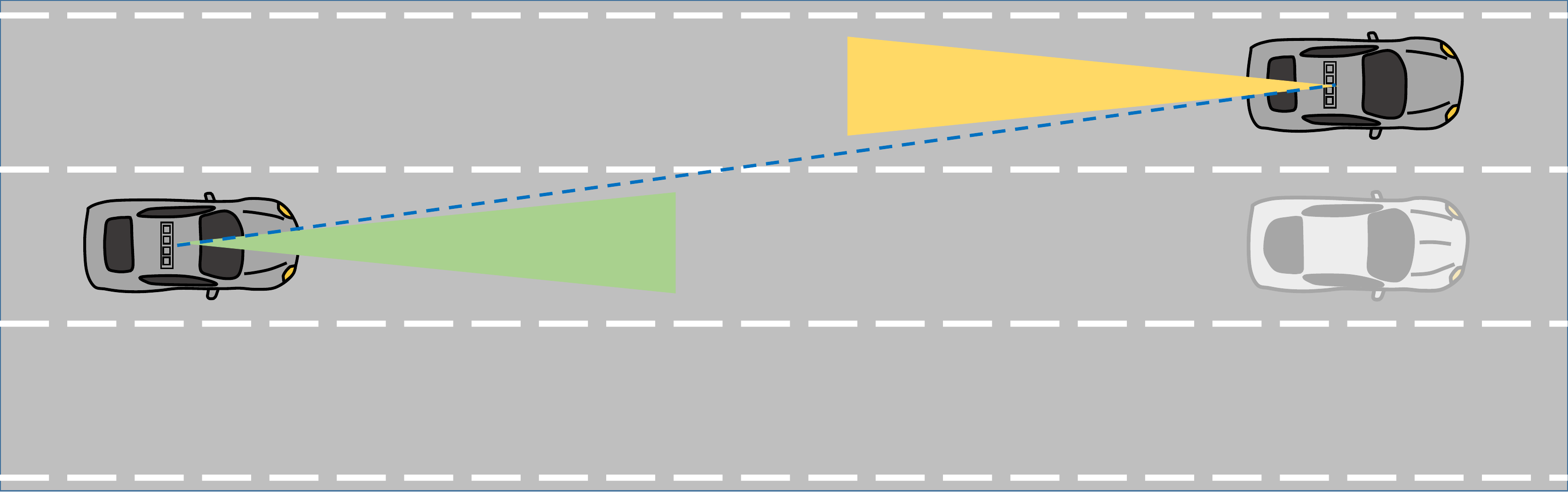}  } \\
	\subfloat[Impact of Tx/Rx position error and Tx orientation error\label{subfig:toyExampleC}]{	\includegraphics[width=0.8\columnwidth]{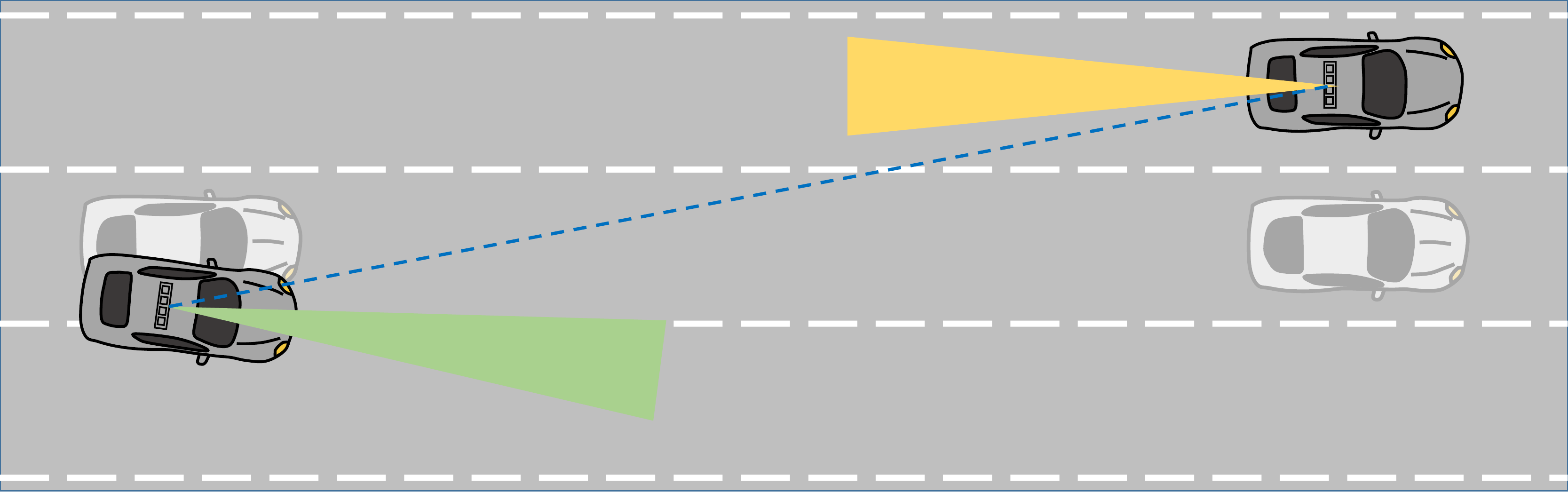}  } \\
	\subfloat[Adaptive BC \label{subfig:toyExampleD}]{	\includegraphics[width=0.8\columnwidth]{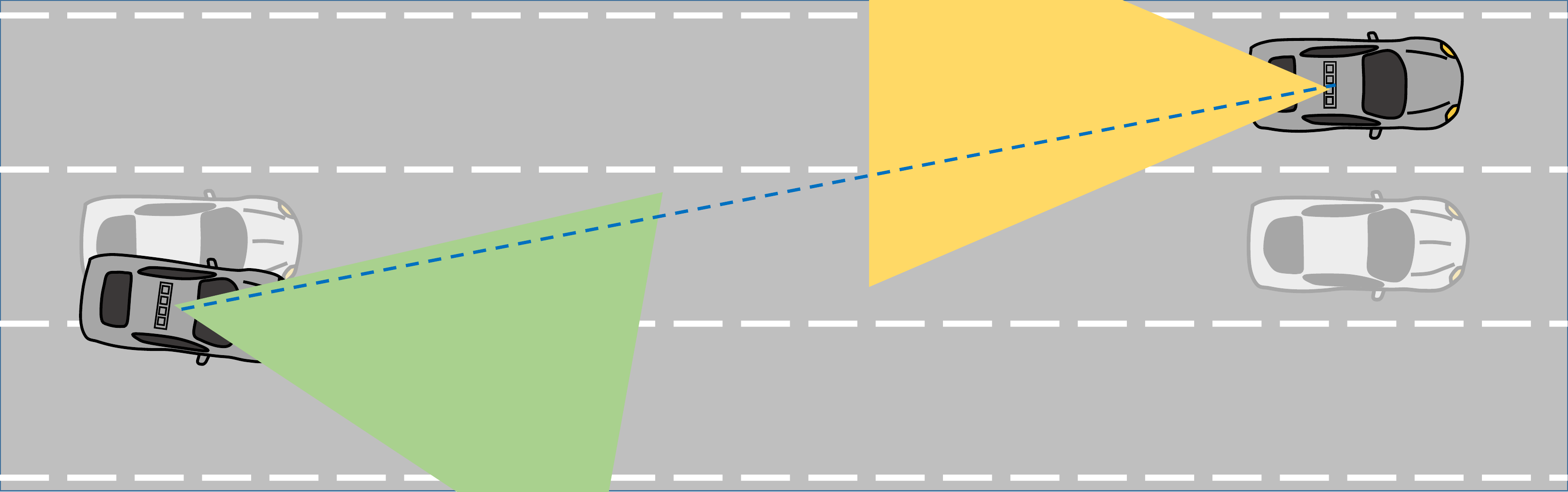}  } \\
	\caption{Graphical sketch of the importance of \ac{BC} in V2V systems.}
	\label{fig:toyexample}
\end{figure}

The most promising Radio Frequency (RF) technology capable of satisfying the extreme requirements of V2X communications for high-LoA is millimeter-Wave (mmW), thanks to the huge bandwidth (up to 3 GHz for 5G mmW) available in this spectrum portion. Nevertheless, the use of mmW presents several challenges, mainly related to the severe path-loss, blockage and mobility (Doppler effect) \cite{Zorzi2017}. In this regard, beam-based systems with high-directivity at both Transmitter (Tx) and Receiver (Rx) sides are able to counteract the strong path loss and minimize the inter-vehicle interference. Massive \ac{MIMO} systems allow shaping multiple highly directive radiation beams by packing hundreds of antennas in small arrays thanks to the reduced wavelength of mmW, e.g., 24.5-52.6 GHz systems of 5G New Radio (NR) Frequency Range 2 (FR2), or sub-THz (D-band). Alternatively, dielectric lenses could be employed \cite{dosSantos-2019}. Regardless of the specific array technology, a precise \ac{BC} is mandatory, especially when the scenario involves high mobility as in the V2X case. Indeed, the motion of vehicles is also affected by vibrations and tilting that easily induce misalignment for high collimated beams, hindering the continuity of communication. Conventional \ac{BAT} strategies relying on exhaustive or hierarchical search of the optimal Tx/Rx beam pair are too time demanding for vehicular applications \cite{IEEE802.11ad_Standard,Giordani2019}. For instance, if two vehicles point each other a beam of $10$ deg width, it is sufficient to mutually share their perfect position and orientation every $250$ ms to avoid mispointing, this timing ensures the communication even in case of a sharp turn (with a typical $20$ deg/s of angular velocity). However, when dealing with realistic position and orientation uncertainties, possibly augmented by vibrations, the signaling timing drops down to few-to-tens of ms, thus requiring a more sophisticated approach for \ac{BC}. Fig. \ref{fig:toyexample} illustrates an example of application of \ac{BC} in \ac{V2V} systems: when the two vehicles know and share their own position and orientations, \ac{BAT} is perfect and the beamwidths can be arbitrarily narrow (Fig. \ref{subfig:toyExampleA}). On the other hand, in real V2V systems the position and orientation estimates are affected by errors and the two vehicles can lose the connectivity (Fig. \ref{subfig:toyExampleB},\ref{subfig:toyExampleC}), driving the system to outage, unless a proper \ac{BC} is applied (Fig. \ref{subfig:toyExampleD}).

%State of the Art of BPC: describe briefly the BPC in communications, leaving abrief intro to BAt works 
The problem of \ac{BC} for V2X system is mostly related to \ac{BAT}, which has been largely addressed in the recent literature under many different perspectives, with the ultimate goal of improving the performance of conventional beam sweeping procedures, in terms of overhead reduction and increased spectral efficiency \cite{Wang2009,Heath2014,KadurFettweis2016}. Recent advances in dynamic \ac{BC} have shown the relevant role of the knowledge of vehicles' position and their instantaneous dynamics (orientation, vibrations) \cite{Heath2016b, Mohammadi2017-beamswitching,Mavromatis2017,Kang2018-BTparticlefilt,Bao2019,brambillaRF,brambillaFSO,brambillaVTC2020,Brambilla_2020}. 

Recent works focused on power control aspects in vehicular communications \cite{ElSawy2014,Lee2015,Ren2015,Zhang2018,Li2019} and, more general, in the context of Device-to-Device (D2D) communications. Beamwidth adaptation has been investigated in few works on the design of antenna arrays \cite{Debogovic2010,Ha2019}. Some recent works are related to Unmanned Aerial Vehicles (UAV) \cite{Yang2018-UAV,Yang2019-UAV,Tang2019-UAV,Krishna2020_drones}, with the effort to enable UAV-based communications. Contributions \cite{Kang-2020,Chung2020_beamwidthcontrol} are specifically focused on V2X scenarios, the former deriving the optimal Rx beamwidth in closed form in a Vehicle-to-Infrastructure (V2I) \ac{LOS} scenario, the latter proposing a beamwidth control in Infrastructure-to-Vehicle (I2V) systems using the output of a particle filter. The paper \cite{Feng2019-beamdesignV2V}, instead, proposes the Tx and Rx beamwidths adjustment in V2V communications by a constrained maximization of the average data-rate, leveraging inaccurate position information. The work in \cite{Feng2019-beamdesignV2V} is extended in \cite{Feng2020BC}, where V2V communications in a straight highway scenario with backward propagation of information (from front to rear vehicles) are considered. In particular, the authors in \cite{Feng2020BC} design a Monte Carlo-based beamwidth optimization problem to maximize the V2V throughput in the region of the receiver. However, they only consider the position uncertainty of the Tx vehicle and not of the joint Tx-Rx pair, and they do not account for orientation information (and related mispointing error).

Joint \ac{BPC} for vehicular scenarios has been recently raising some interest in \cite{Pradhan2018-JointBeamwidthEnergy,Zhu2019-JointRateBeamwidth,Saeed2019-JointBPCindoor,Gao2020-DeepRL}. In detail, a solution to the joint optimization of beamwidth and energy in multi-user mmW communications is in \cite{Pradhan2018-JointBeamwidthEnergy}, by solving a multi-objective optimization problem. A similar approach has been followed in \cite{Zhu2019-JointRateBeamwidth}, where the joint adaptation of rate and beamwidth is discussed with focus on I2V scenario. The specific \ac{BPC} for indoor mmW systems is targeted in \cite{Saeed2019-JointBPCindoor} and using a deep reinforcement learning approach in \cite{Gao2020-DeepRL}. 

To the best of our knowledge, no research has yet considered to leverage on-board vehicular sensors to perform a dynamic \ac{BPC}. Furthermore, validation on real vehicle dynamics data is of primary importance. The complexity of \ac{BPC} adaptation, in terms of required cooperation (control signaling) among vehicles, is also a key factor to be considered when dealing with V2X communication in dynamic scenario. Most of the aforementioned papers are focused on V2I applications, exploiting machine learning approaches or constrained maximization/minimization that are not yet proved to fit the rapidly time varying vehicles' dynamics of V2V communications.

\subsection*{Contribution}
This work aims to cover the above gap in the development and validation of a sensor-assisted \ac{BPC} method to enhance the V2V performance in highly dynamic scenarios. The proposed \ac{BPC} strategy extends the sensor-aided BAT method, detailed in our previous publications \cite{brambillaFSO,brambillaVTC2020,Brambilla_2020}. The proposed \ac{BPC} relies on cooperative processing of Global Positioning System (GPS) and Inertial Measurements Units (IMU) data from vehicles, mutually exchanged over a dedicated sub-6 GHz (e.g., 5G NR FR1) control link, so that the multi-gigabit link (e.g., 5G NR FR2) enables to meet the stringent requirements of eV2X applications. 

The contribution of the paper is as follows: \textit{(i)} we formulate the joint optimization of beamwidth and power, and propose a heuristic sensor-aided \ac{BPC}, in which both Tx/Rx beamwidths and Tx power are dynamically controlled on the basis of Tx and Rx positions/orientations and related uncertainties (covariances), obtained from a tracking filter; \textit{(ii)} we test the feasibility of the proposed \ac{BPC} solution via numerical simulations in a \ac{LOS} V2V scenario based on real GPS/IMU sensors data over a test road trajectory, collected during a dedicated experimental campaign in the urban area of Milano (Italy). Quaternion-based \ac{EKF}  fuses GPS and IMU data to extract position and orientation estimates and uncertainties to be used for both \ac{BAT} and \ac{BPC}; \textit{(iii)} framing the \ac{BPC} as an optimization problem, we show that the proposed \ac{BPC} allows to closely attain (up to 1-2 dB of excess power) the optimum performance of an exhaustive, multi-dimensional search for the optimal beamwidths/power; \textit{(iv)} we prove the advantages in employing a dynamic \ac{BPC} system in a V2V link compared to a fixed-beamwidth one, discussing the requirements in terms of Tx power, sensors' sampling time and performance, end-to-end latency on the control channel, by analyzing the Signal-to-Noise Ratio (SNR) and the corresponding outage probability. Notice that all links are assumed in LOS and without any blockage, as this is not accounted here.

\subsection*{Organization}
The paper is organized as follows: Section \ref{sec:SystemArchitecture} outlines the system layout and architecture considered in the paper, and presents the mathematical models describing vehicles' dynamics. Section \ref{sect:EKF} details the employed sensor fusion algorithm. The mmW V2V communication channel model is in Section \ref{sec:channel}, while the proposed \ac{BPC} strategy is in Section \ref{sec:BAT}. The experimental campaign is described in Section \ref{sect:campaign}. Section \ref{sec:results} reports  numerical results while Section \ref{sec:conclusion} draws the conclusions. Appendices \ref{app:appendix} and \ref{app:appendix2} provide details on the tracking filter implementation and on the beamwidth/power optimization, respectively.

\subsection*{Notation}
Bold upper- and lower-case letters describe matrices and column vectors. Any element of a matrix $\mathbf{A}$ is indicated with $\left[\mathbf{A}\right]_{ij}$, where $i$ is the row and $j$ is the column. $\mathbf{I}_N$ and $\mathbf{0}_{NM}$ denote, respectively, the identity matrix of size $N$ and a matrix of all zero entries of size $N\times M$. With  $\mathbf{a}\sim\mathcal{N}(\boldsymbol{\mu},\mathbf{C})$ we denote a multi-variate Gaussian random variable $\mathbf{a}$ with mean $\boldsymbol{\mu}$ and covariance matrix $\mathbf{C}$. Matrix transposition is indicated as $(\cdot)^{\mathrm{T}}$. Operator $\mathrm{tr}\left(\mathbf{A}\right)$ extracts the trace of matrix $\mathbf{A}$. $\mathbb{R}$ is the symbol for the set of real numbers. Operator $\norm{\cdot}_2$ represents the Euclidean norm. Quaternion multiplication and exponential are denoted by $\odot$ and $\exp_{\mathrm{q}}\{\cdot\}$ respectively while quaternion conjugation is $(\cdot)^{*}$.

\section{System Layout and V2X LOS Modeling}
\label{sec:SystemArchitecture}

\begin{figure}[]
	\centering
	\includegraphics[width=0.8\columnwidth]{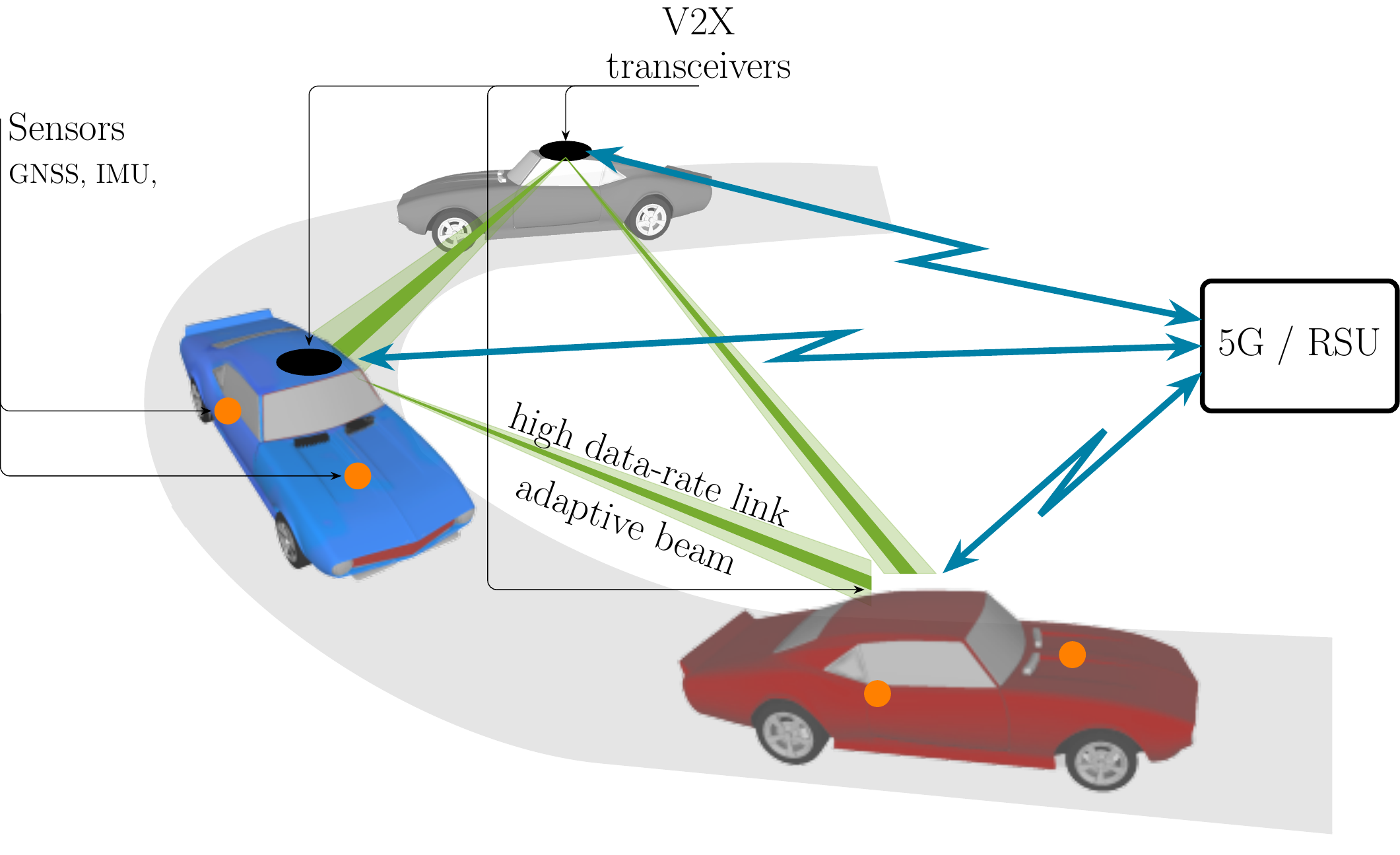}
	\caption{Overview of the proposed V2V system. Vehicles share their sensors' data with neighbors, by mutual pointing of high-frequency, directive links, possibly adaptive. \ac{BAT} and \ac{BPC} are enabled by the exchange of information over a reliable control link. }
	\label{fig:TwoVehiclesArch}
\end{figure}

The envisioned layout for V2V communications is represented in Fig. \ref{fig:TwoVehiclesArch} \cite{brambillaFSO,brambillaVTC2020,Brambilla_2020}. This comprises a number of inter-connected vehicles fully equipped with sensors such as IMU, cameras, GPS, radar, lidar, and others. The Road Side Units (RSUs), whenever existing, are expected to enhance the V2V network performance, e.g., by providing connectivity, forwarding control messages in case of \ac{NLOS} conditions or providing updated 3D road maps. The role of RSU can be covered by the radio access network such as 5G macro/micro cells. The main feature of the proposed system is the two-radio V2V communication in Fig. \ref{fig:TwoVehiclesArch}, which is based on \textit{(i)} a high data-rate, high-frequency, directive and adaptive beam-based link (mmW, sub-THz or even Free-Space Optics-FSO) for exchanging the raw sensor data required for high LoA cooperative perception (green beams in Fig. \ref{fig:TwoVehiclesArch}) and \textit{(ii)} a low data-rate link (either V2V or V2I) for exchanging locally processed information about vehicles' position and orientation (blue lines in Fig. \ref{fig:TwoVehiclesArch}).

Based on the sensors' signaling over the control link, the distributed processing unit of each vehicle predicts the geometrical arrangement of all the active V2V \ac{LOS} links to the surrounding road users (vehicles or RSU). Each link is described by position and orientation in a suitable reference system of the connected vehicle transceivers. In this way, all agents have the information needed to compute the time evolution of the beam pointing directions, thus providing superior mmW-based V2V communication performance, possibly by dynamically varying the system parameters, e.g., Tx power and beamwidth. The sharing of position information among vehicles does not require a high data-rate link so that the parallel control channel (\textit{ii}) can be provided by Cellular-V2X (C-V2X) networks (as detailed in \cite{Molina2017}) or 5G NR FR1. 
%Indeed, 3D position information can be encoded in $96$ bits, and the associated covariance in $12$ bits (diagonal components only) as indicated in \cite{SAE2}.

\begin{figure}[]
	\centering
	\includegraphics[width=0.8\columnwidth]{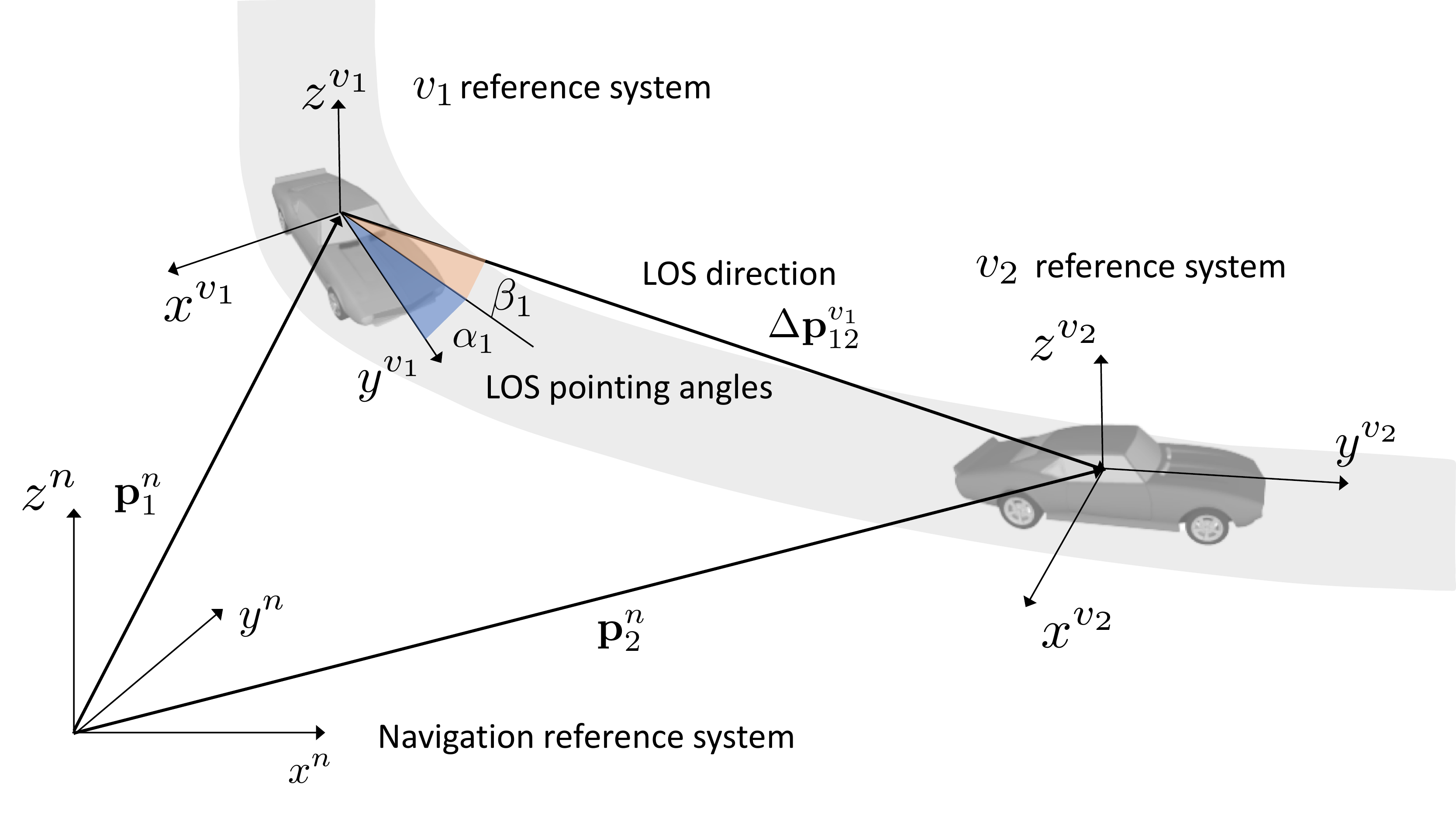}
	\caption{V2V geometry with navigation and vehicle reference systems, and LOS angles for Tx vehicle $v_1$.}
	\label{fig:TwoVehicles}
\end{figure}

\subsection{Spatial Reference Systems and V2X LOS Modeling}\label{subsec:ProblemFormalization}
For vehicle dynamics modeling we employ two Cartesian reference systems, as illustrated in Fig. \ref{fig:TwoVehicles}. The first one is the navigation reference system (superscript $n$), fixed with respect to the Earth and centered in a suitable point to describe the local vehicular network (e.g., a RSU), it is used to model the macro-scale variation of vehicle positions. The second one is the vehicle reference system (superscript $v$), dynamically evolving with the vehicle and centered at the mmW (or sub-THz) transceiver (or antennas) \cite{Grovesbook2015}.

The Tx/Rx equipment of a vehicle $v$ is described by the instantaneous 3D position, evaluated in the navigation reference system:
\begin{align}
\mathbf{p}_{v}^{n} = \begin{bmatrix} p_{x}^{n}\,\,\,p_{y}^{n}\,\,\,p_{z}^{n}\end{bmatrix}^{\mathrm{T}}\in\mathbb{R}^{3 \times 1 }\, ,
\end{align}
%and 3D velocity 
%\begin{align}
%	\mathbf{v}_{v}^{n} = \begin{bmatrix} v_{x}^{n}\,\,\,v_{y}^{n}\,\,\,v_{z}^{n}\end{bmatrix}^{\mathrm{T}}\in\mathbb{R}^{3 \times 1 }\, .
%\end{align}
and by its orientation, which can be represented by Euler angles or unit-norm quaternions \cite{kok2017using}. 
Euler angles are more intuitive and useful to define the beam pointing directions, representing rotation angles of the vehicle reference system ($v$-system) around $x^{n}$, $y^{n}$ and $z^{n}$ axes, which are called pitch, roll and yaw/heading, respectively:
\begin{align}
\boldsymbol{\gamma}_{v}^{n v} = \begin{bmatrix}\phi_{v}^{n v}\,\,\, \theta_{v}^{n v}\,\,\, \psi_{v}^{n v}\end{bmatrix}^{\mathrm{T}} \, .
\label{eq:EulerAngles}
\end{align}
Euler angles are associated to the corresponding rotation matrix $\mathbf{R}\left \{ \boldsymbol{\gamma}^{nv}_{v}\right \}$, allowing the transformation of a given quantity from the $v$-system to the $n$-system, and vice-versa \cite{Grovesbook2015}. By contrast, quaternions generalize the concept of complex exponential and are preferred in sensor fusion for complexity reduction \cite{kok2017using}. The quaternion describing the orientation of the $v$-system is:
\begin{equation}
\mathbf{q}_{v}^{nv}=\begin{bmatrix} q_{0,v}^{nv}\,\,\, q_{1,v}^{nv} \,\,\,  q_{2,v}^{nv}\,\,\,  q_{3,v}^{nv}\end{bmatrix}^{\mathrm{T}} \in \mathbb{R}^{4 \times 1},
\end{equation}
where $\norm{\mathbf{q}_{v}^{nv}}_2 = 1$. Quaternions and Euler angles are two equivalent parameterization of the orientation, and are used interchangeably in the paper \cite{kok2017using}. 
With quaternions, it is easy to switch from $v$-system to $n$-system by applying roto-translation: a translation by $\mathbf{p}_{v}^{n}$ and a rotation described $\mathbf{q}_{v}^{nv}$. More specifically, let $\mathbf{u}^{v}$ be a location (e.g., of a nearby vehicle) in the $v$-system, the same location expressed in the $n$-system is:
\begin{equation}\label{eq:quatrot}
\bar{\mathbf{u}}_{}^{n} = \mathbf{q}_{v}^{nv} \odot \bar{\mathbf{u}}_{}^v \odot (\mathbf{q}_{v}^{nv})^{*} \, ,
\end{equation}
in which $\bar{\mathbf{u}}_{}^n = \left[0 \,\,\, (\mathbf{u}_{}^n)^{\mathrm{T}}\right]^{\mathrm{T}} \in \mathbb{R}^{4 \times 1} $ is the quaternion representation of $\mathbf{u}^{n}_{}$. Similarly, $\bar{\mathbf{u}}_{}^v$ is the quaternion representation of $\mathbf{u}^v$.
The roto-translation is essential to define the relative position of any other vehicle in the network with respect to the vehicle at hand, and hence the \ac{LOS} angles for beam pointing. These are the input of the \ac{BAT} for Tx and Rx beamforming for any high-frequency communication (as for mmW and sub-THz), or by other technologies (as, for instance, a Micro Electro-Mechanical System (MEMS) mirror at the Tx side for a FSO-based communication). 

Let us consider a V2V \ac{LOS} communication between a Tx vehicle $v_{1}$ and a Rx vehicle $v_{2}$, as in Fig. \ref{fig:TwoVehicles}. The Tx-Rx pointing angles can be obtained as follows (a similar derivation is for Rx-Tx pointing). Let 
$\mathbf{p}^n_{1}$ and $\mathbf{p}^n_{2}$ be the Tx and Rx positions in the $n$-system, while quaternion $\mathbf{q}_{1}^{n v_1}$ and $\mathbf{q}_{2}^{n v_2}$ the Tx and Rx orientations, respectively. 
The relative position of $v_2$ with respect to $v_1$ is  computed as:
\begin{equation}\label{eq:estimdist}
\begin{split}
\Delta\bar{\mathbf{p}}_{12}^{v_{1}} & =\mathbf{q}_{1}^{v_1 n} \odot \underbrace{\left(\bar{\mathbf{p}}^{n}_{2}-\bar{\mathbf{p}}^{n}_{1}\right)}_{\Delta\bar{\mathbf{p}}_{12}^{n}} \odot (\mathbf{q}^{v_1 n}_{1})^{*} = \begin{bmatrix}
0\,\,\,\Delta \mathbf{p}_{12}^{v_{1},\mathrm{T}}\end{bmatrix}^{\mathrm{T}} ,
\end{split}
\end{equation}
where $\mathbf{q}_{1}^{v_1 n} = (\mathbf{q}_{1}^{n v_1})^*$ is the quaternion denoting the rotation from the $n$-system to the $v_1$-system, and the pointing \ac{LOS} direction for $v_1$ is identified by the azimuth ($\alpha_{1}$) and the elevation ($\beta_{1}$) angles, respectively defined as:
\begin{equation} \label{eq:AzimuthandElevation}
\alpha_{1}=\mathrm{atan}\left(\frac{\Delta {p}_{12,y}^{v_{1}}}{\Delta {p}_{12,x}^{v_{1}}}\right), \quad
\beta_{1}=\mathrm{asin}\left(\frac{\Delta {p}_{12,z}^{v_{1}}}{ \norm{\Delta\mathbf{p}_{12}^{v_{1}}}_2}\right).
\end{equation}

\section{Sensor Fusion for Tracking Vehicle Dynamics}
\label{sect:EKF}
Tracking the vehicle dynamics over time consists in evaluating the dynamics of the $v$-system (instantaneous position and orientation) with respect to the $n$-system. Since required accuracy is in the order of few centimeters, this represents a highly challenging task. As a matter of fact, GPS typically provides a position estimate but it is not accurate enough for eV2X applications. Among the augmentation possibilities, refinement using 5G cellular data has been proposed \cite{wymeersch20175g,Cui2016}, as well as leveraging localization with respect to reference features (e.g., streetlamps, traffic lights, tollbooths, etc.) \cite{Soatti2018,Brambilla2019} or map matching \cite{Mattern2011}. Relevant for the scope of this paper is the integration of GPS with IMU, e.g., GPS with gyroscope and velocity sensors \cite{Zhang2012}, Real-Time Kinematic (RTK), IMU and lidar \cite{Wan2018}. We refer to \cite{Kuutti2018} and references therein for an overview of other possible augmentation solutions. 

A Bayesian approach is used here to track the 3D position $\mathbf{p}_{v,t}^{n}$, 3D velocity $\mathbf{v}_{v,t}^{n} = [v_{x,t}\,\,v_{y,t}\,\,v_{z,t}]^{\mathrm{T}}$ and orientation $\mathbf{q}_{v,t}^{nv}$ at vehicle $v$, based on GPS and IMU data. All sensors are assumed to be synchronized; if not, an alignment procedure is required.
To ease the notation, in this section we drop the subscripts and superscripts of the state variable referring only on time index by subscript $t$ as the reference system it refers to can be easily inferred from the context (e.g., $\mathbf{p}_{t} \leftarrow \mathbf{p}_{v,t}^{n}$, $\mathbf{q}_{t} \leftarrow \mathbf{q}^{n v}_{v,t}$).  The ego vehicle state to be tracked is therefore:
\begin{align}\label{eq:state}
\boldsymbol{\theta}_{t} = \left[	\mathbf{p}_{t}^\mathrm{T}\,\,\,
\mathbf{v}_{t}^\mathrm{T}\,\,\,
\mathbf{q}_{t}^\mathrm{T}\right]^\mathrm{T} \in\mathbb{R}^{10 \times 1 }\,.
\end{align}
The state transition from time $t-1$ to $t$ is modeled as a non-linear function $\mathbf{f}(\cdot)$ that depends on the state itself $\boldsymbol{\theta}_{t-1}$, on a control input from the IMU $\mathbf{u}_{t-1}$ and some degree of randomness \cite{kok2017using}. Thus, the vehicle state transition is:
\begin{align}
\boldsymbol{\theta}_{t|t-1}& = \mathbf{f}\left(\boldsymbol{\theta}_{t-1},\mathbf{u}_{t-1},\mathbf{w}_{t} \right),
\label{eq:state_transition}
\end{align}
in which $\mathbf{w}_{t}$ is a zero mean Gaussian processes with time-varying covariance matrix $\mathbf{C}_{w,t}$. The input $\mathbf{u}_{t}$ consists of the 3D accelerations $\mathbf{z}_{a,t}$ and 3D angular velocities $\mathbf{z}_{\omega,t}$ that are measured by the IMU in its local reference system, here considered to be aligned with the $v$-system.  

The state $\boldsymbol{\theta}_{t}$ is hidden into a set of observations $\mathbf{z}_{t}$:
\begin{align}
 \mathbf{z}_{t}& = \mathbf{h}\left(\boldsymbol{\theta}_{t},\mathbf{n}_{t} \right) \, , \label{eq:measurement_equation}
\end{align}
where $\mathbf{h} (\cdot)$ is a non-linear function and $\mathbf{n}_{t}$ denotes a measurement error with time-varying covariance $\mathbf{C}_{n,t}$.
Different combinations can be used to get the observation $\mathbf{z}_{t}$, we consider the case where vehicle position, velocity and yaw/heading are retrieved from GPS data, while roll and pitch data come from the longitudinal and lateral accelerations measured by the on-board vehicle IMU after some specific mechanic-type calibration (see Section \ref{sect:campaign} for details). The observed data is thereby:
\begin{align}\label{eq:measurement_equation_split}
\begin{split}
 \mathbf{z}_{p,t} &=  \mathbf{p}_{t} + \mathbf{n}_{p,t} \, ,  \\
{z}_{v,t} &= \norm{\mathbf{v}_{t}}_2 + n_{v,t} \, , \\
\mathbf{z}_{q,t} & =\mathbf{q}_{t} + \mathbf{n}_{q,t} \, .
\end{split}
\end{align}
where $\mathbf{n}_t = \left[\mathbf{n}_{p,t},\,\,n_{v,t},\,\,\mathbf{n}_{q,t}\right]^{\mathrm{T}}\in\mathbb{R}^{7\times 1}$. It is important to mention that, usually, a direct measurement of the quaternion $\mathbf{q}_{t}$ (or, equivalently, of the Euler angles $\boldsymbol{\gamma}_{t}$) is achieved by means of magnetometers, not adopted here. Further details on pitch/roll estimation are given in Section \ref{sect:campaign}. The covariance matrices $\mathbf{C}_{w,t}$ - accounting for IMU noise and model uncertainty - and $\mathbf{C}_{n,t}$ - accounting for sensor measurement errors - have to be calibrated according to the vehicle dynamics and sensor performance. More details on the Bayesian tracking by using the \ac{EKF} are provided in Appendix \ref{app:appendix}, and can be also found in \cite{kok2017using}.

\section{Communication System Model}
\label{sec:channel}

The goal of this section is to set up the model for a beam-based \ac{LOS} communication link between two vehicles $v_1$ and $v_2$. We do not focus on a specific technological implementation for two reasons: \textit{(i)} practical mmW beam-based systems require the development of advanced beamforming algorithms (e.g., hybrid structures to cater with the huge number of antennas in a mMIMO setting \cite{Combi-Spagnolini_2019}), and their specific discussion is beyond the scope of this paper; \textit{(ii)} we are mainly interested in evaluating the benefits of the V2V \ac{BPC} system in a general framework, considering realistic Tx and Rx parameters without constraining the method to a specific communication technology. 

In this regard, we consider the two vehicles equipped with the same multi-antenna transceiver whose maximum gain is inversely proportional to the time-varying beamwidth $\mathbf{\Omega}_{v,t}= \left[\Omega_{v,t}^{\mathrm{az}}\,\,\, \Omega_{v,t}^{\mathrm{el}}\right]^\mathrm{T}$, for azimuth $\Omega_{v,t}^{\mathrm{az}}$ and elevation $\Omega_{v,t}^{\mathrm{el}}$ beamwidths. Given the \ac{LOS} direction $(\alpha_{v,t}\, ,\,\beta_{v,t})$ \eqref{eq:AzimuthandElevation} and the estimated one $(\hat{\alpha}_{v,t} \,,\, \hat{\beta}_{v,t})$, the array gain in case of mispointing is:
\begin{equation}\label{eq:G}
	G_{v,t} = G^{\mathrm{max}}_{t}\left(\mathbf{\Omega}_{v,t}\right) \,g(\delta \alpha_{v,t}\,,\,\delta \beta_{v,t} \,;\, \mathbf{\Omega}_{v,t}) \,,
\end{equation}
where $G^{\mathrm{max}}_{t}\left(\mathbf{\Omega}_{v,t}\right) \propto 1/\mathbf{\Omega}_{v,t}$ is the maximum gain and $g(\cdot)$ is the normalized array pattern ($\mathrm{max}\{g(\cdot)\}=1$), representing the gain loss in case of pointing errors $\delta \alpha_{v,t}  = \hat{\alpha}_{v,t} - \alpha_{v,t}$ and $\delta \beta_{v,t} = \hat{\beta}_{v,t} - \beta_{v,t}$. Without loss of generality, we consider Gaussian beam patterns \cite{Balanis2005}.
%to simulate a non-uniform gain distribution over the beamwidth while neglecting secondary lobes and out-of-beam radiation. The latter two complicate the analytical treatment in Section \ref{subsec:optimization} without adding significant value to the present work.
Notice that the beamwidth variation (enlargement/reduction) can be achieved, for instance, by activating a different number of antennas at a time, or by more sophisticated approaches as the Dolph-Chebyshev one \cite{WeiNOMABC2019}.

%It is worth underlining that the beamwidth variation (enlargement/reduction) in multi-antenna systems depends on how many antennas are activated at a time, and, in practice, any beamwidth adaptation is eased by the employment of hybrid beamforming architectures.

We choose as a performance metric the instantaneous Signal-to-Noise power Ratio (SNR) after beamforming, defined as:
\begin{equation} \label{eq:SNR}
	\mathrm{SNR}_{t}=\frac{P_{\mathrm{rx},t}}{P_{\mathrm{noise},t}}\,,
\end{equation}
where $P_{\mathrm{rx},t}$ denotes the received power and $P_{\mathrm{noise},t}$ the noise one at time $t$. The received power (in $\mathrm{dBm}$ scale, superscript) is:
\begin{equation}\label{eq:Prx}
P_{\mathrm{rx},t}^{\mathrm{dBm}}=P_{\mathrm{tx},t}^{\mathrm{dBm}}+G_{1,t}^{\mathrm{dB}}+G_{2,t}^{\mathrm{dB}}-\eta^{\mathrm{dB}}_{t}\,,
\end{equation}
where $\eta^{\mathrm{dB}}_{t}$ is the path-loss \cite{Akdeniz2014} and the Equivalent Isotropic Radiated Power (EIRP) $P_{\mathrm{tx},t}^{\mathrm{dBm}}+G_{1,t}^{\mathrm{dB}}$ is subject to regulatory limitations \cite{FCC}.
The gain in \eqref{eq:G} is maximum when the pointing and \ac{LOS} directions coincide. For instance, considering a $28$ GHz free-space mmW link with $400$ MHz bandwidth for two vehicles 100 m apart,  perfectly pointing one to the other with a beamwidth $\mathbf{\Omega}_{1,2}=\left[20 \,\,20\right]^\mathrm{T}$ deg, the requested Tx power to have a SNR of $10$ dB turns out to be $\approx\!0$ dBm, while for $\mathbf{\Omega}_{1,2}=\left[10\,\,10\right]^\mathrm{T}$ deg the value drops to $\approx\!-12.2$ dBm. However, deliberately narrowing the beams lead to uncontrolled outage when position/orientation are estimated with errors. This problem needs to be handled in the proposed architecture where vehicles take advantage of the locally estimated positions/orientations and related uncertainties. 
%It is worth stressing that, among all parameters impacting a V2V link budget, $G_{v,t}$ is the most affected by errors on reciprocal pose estimates, thus demonstrating that the occurrence of moderate antenna gain losses is a key aspect for the assessment of the V2V performance. 

%Questa va rivista completamente con il nuovo problema di ottimizzazione

\section{Sensor-Assisted Dynamic BPC Algorithm}
\label{sec:BAT}

In this section, we propose the sensor-assisted \ac{BPC} strategy in which both Tx power and beamwidths are dynamically controlled at Tx \textit{and} Rx to optimize the performance of V2V systems. The optimization is performed on top of a sensor-aided \ac{BAT}, that has been proved to provide significant benefits in terms of communication overhead and energy consumption with respect to conventional schemes \cite{Mavromatis2017,Perfecto2017,Brambilla_2020}. The timing plays a key role as it determines how frequently vehicles can update their pointing. Signaling latency, if excessive, makes the exchanged information outdated and unable to capture the rapidly varying vehicle dynamics. Furthermore, the optimal beamwidth depends on the position/orientation sensing accuracy and the mobility statistics: wide beams imply relaxing \ac{BAT} requirements, at the price of a higher Tx power, but alignment and tracking of very narrow beams might be too demanding for the available on-board equipment (mostly for insufficient sensor sampling frequency and/or latency). In the following, we first formulate the beamwidth and power optimization problem, where exact instantaneous V2V parameter statistics are needed, and then propose an heuristic BPC mechanism exploiting the outcomes of the Bayesian tracking filter, fitting into a practical V2V system.

\subsection{Beamwidth and Power Optimization}\label{subsec:optimization}
The problem for the optimization of the Tx power and the Tx/Rx beamwidths is formulated as follows:
\begin{equation}\label{eq:optimproblem}
\begin{aligned}
& \underset{\mathbf{\Omega}_1,\mathbf{\Omega}_2}{\mathrm{minimize}}
& & P_{\mathrm{tx}} \\
& \mathrm{subject \,\,\, to}
& & P_{\mathrm{out}} \leq \bar{P}_{\mathrm{out}} & (a)\\
& & & P_{\mathrm{tx}} \, G^{\mathrm{max}}\left(\mathbf{\Omega}_{1}\right) \leq \mathrm{EIRP}_{\mathrm{max}}\,. & (b)
\end{aligned}
\end{equation}
The aim is to minimize the Tx power $P_{\mathrm{tx}}$ over the Tx and Rx beamwidths $\mathbf{\Omega}_1$, $\mathbf{\Omega}_2$ under the constraints of a maximum outage probability $\bar{P}_{\mathrm{out}}$ $(a)$ and a maximum EIRP $(b)$. The relation between $P_{\mathrm{tx}}$ and beamwidths $\mathbf{\Omega}_1$, $\mathbf{\Omega}_2$ is: 
\begin{equation}\label{eq:SNRstatement}
	\mathrm{SNR} = \dfrac{P_{\mathrm{tx}} \,  G_1\left(\delta \alpha_1, \delta \beta_1; \mathbf{\Omega}_1\right) \, G_2\left(\delta \alpha_2, \delta \beta_2; \mathbf{\Omega}_2\right)}{\eta\left(d\right) \, P_{\mathrm{noise}}} \, ,
\end{equation}
where the Tx and Rx angular pointing errors are $(\delta \alpha_1,\,\delta \beta_1)$ and $(\delta \alpha_2,\,\delta \beta_2)$ and $\eta\left(d\right)$ is the path loss for distance $d$. 
While constraint $(b)$ in \eqref{eq:optimproblem} comes from regulatory limits \cite{FCC}, the first constraint $(a)$ explicitly determines the outage probability of the system $P_{\mathrm{out}}$, i.e., the probability of the SNR to be lower than a threshold $\mathrm{SNR}_{\mathrm{min}}$. In this perspective, constraint $(a)$ can be evaluated as:
\begin{equation}\label{eq:outage}
\begin{split}
P_{\mathrm{out}}  =  \, \mathrm{Pr}\left(\mathrm{SNR} < \mathrm{SNR}_{\mathrm{min}}\right)   
=   = \hspace{-4mm} \int\limits_{0}^{\mathrm{SNR}_{\mathrm{min}}} \hspace{-4mm} p_{\mathrm{SNR}}\left(\mathrm{SNR}; \delta \alpha_1,\delta \beta_1,\delta \alpha_2,\delta \beta_2, P_{\mathrm{tx}}, \eta(d), \mathbf{\Omega}_1, \mathbf{\Omega}_2\right) \mathrm{d}\mathrm{SNR} \,,
\end{split}
\end{equation}
i.e., by the integration of the tail of the SNR \ac{PDF} $p_{\mathrm{SNR}}$. Notice that the latter is a multi-dimensional function of the instantaneous joint distribution of the pointing errors $(\delta \alpha_1,\,\delta \beta_1)$ and $(\delta \alpha_2,\,\delta \beta_2)$, to be derived from relations \eqref{eq:estimdist}-\eqref{eq:AzimuthandElevation}, Tx power $P_{\mathrm{tx}}$, distance $d$ and beamwidths $\mathbf{\Omega}_1, \mathbf{\Omega}_2$. Therefore, it requires the knowledge of exact and instantaneous V2V parameters/statistics and its analytical derivation in closed form is made with significant approximations only (e.g., a parabolic approximation of the gain functions $G_1$ and $G_2$ around the maximum). Moreover, the required V2V statistics to evaluate $P_{\mathrm{out}}$ must be known \textit{in advance} to both Tx/Rx terminals, preventing the usage of \eqref{eq:outage} in practical scenarios. %deve essere ottenuta necessariamente con approssimazioni

\subsection*{Misalignment Probability}\label{subsect:Pmis}

In this context, $P_{\mathrm{out}}$ can be approximated from the beam misalignment probability $P_{\mathrm{mis}}$: 
\begin{align}\label{eq:misalignment}
	P_{\mathrm{mis}} = \,& P_{\mathrm{mis,Tx}} \left(1 - P_{\mathrm{mis,Rx}}\right) +  P_{\mathrm{mis,Rx}} \left(1 - P_{\mathrm{mis,Tx}}\right) + P_{\mathrm{mis,Tx}} P_{\mathrm{mis,Rx}}\\
	\approx &  \,P_{\mathrm{mis,Tx}} + P_{\mathrm{mis,Rx}}\,,
\end{align}
where we indicate with $ P_{\mathrm{mis,Tx}} $ the probability that the Tx beam is not aligned with the Rx and with $P_{\mathrm{mis,Rx}} $ the vice-versa. The approximation is valid for small $ P_{\mathrm{mis,Tx}} $ and $ P_{\mathrm{mis,Rx}} $. We proceed by first evaluating the Tx misalignment probability $P_{\mathrm{mis,Tx}}$, as the derivation of $P_{\mathrm{mis,Rx}}$ is similar, and then we approximate $P_{\mathrm{out}}$ by enforcing a minimum SNR constraint, $\mathrm{SNR}_{\mathrm{min}}$. 

Let us define the errors on Tx and Rx positions, expressed in the navigation reference system ($n$-system), as:
\begin{align}\label{eq:position_error_v1_v2}
	\delta\mathbf{p}^n_1 \sim \mathcal{N}(\mathbf{0}, \mathbf{C}^n_{p_1}) \,\,\, , \,\,\,
	\delta\mathbf{p}^n_2 \sim \mathcal{N}(\mathbf{0}, \mathbf{C}^n_{p_2})\,,
\end{align}  
and the errors on Tx and Rx orientations as:
\begin{align}\label{eq:orientation_error_v1_v2_Euler}
\quad \quad
\delta \boldsymbol{\gamma}^{n v_1}_1 \sim \mathcal{N}(\mathbf{0}, \mathbf{C}^{n v_1}_{\gamma_1}) \,\,\, , \,\,\,
\delta \boldsymbol{\gamma}^{n v_2}_2 \sim \mathcal{N}(\mathbf{0}, \mathbf{C}^{n v_2}_{\gamma_2})\,.
\end{align}  
%\begin{align}\label{eq:orientation_error_v1_v2_Quat}
%\delta \mathbf{q}_1 \sim \mathcal{N}(\mathbf{0}, \mathbf{C}_{q_1}), \,\,\,
%\delta \mathbf{q}_2  \sim \mathcal{N}(\mathbf{0}, \mathbf{C}_{q_2})
%\end{align} 
%Notice that an equivalent formulation of \eqref{eq:orientation_error_v1_v2_Euler} can be obtained with quaternions, but Euler angles are more intuitive and preferred for the present derivation.

\begin{figure}[]
	\centering
	\subfloat[$\mathrm{LOS}_1$ and  $\mathrm{LOS}_2$ reference systems \label{subfig:alignment1}]{	\includegraphics[width=0.6\columnwidth]{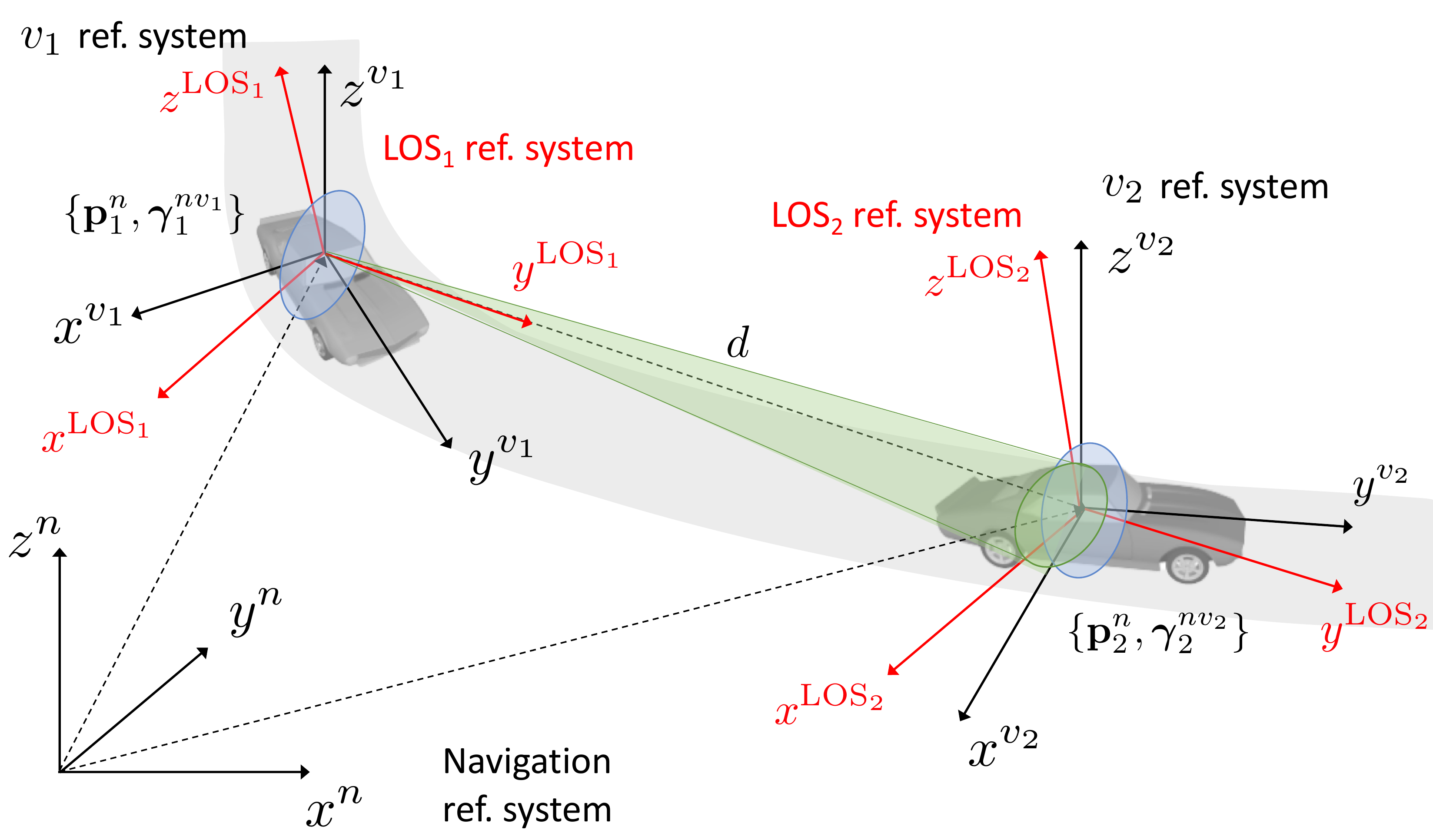}  } \\
	\subfloat[Tx beam misalignment probability \label{subfig:alignment2}]{	\includegraphics[width=0.6\columnwidth]{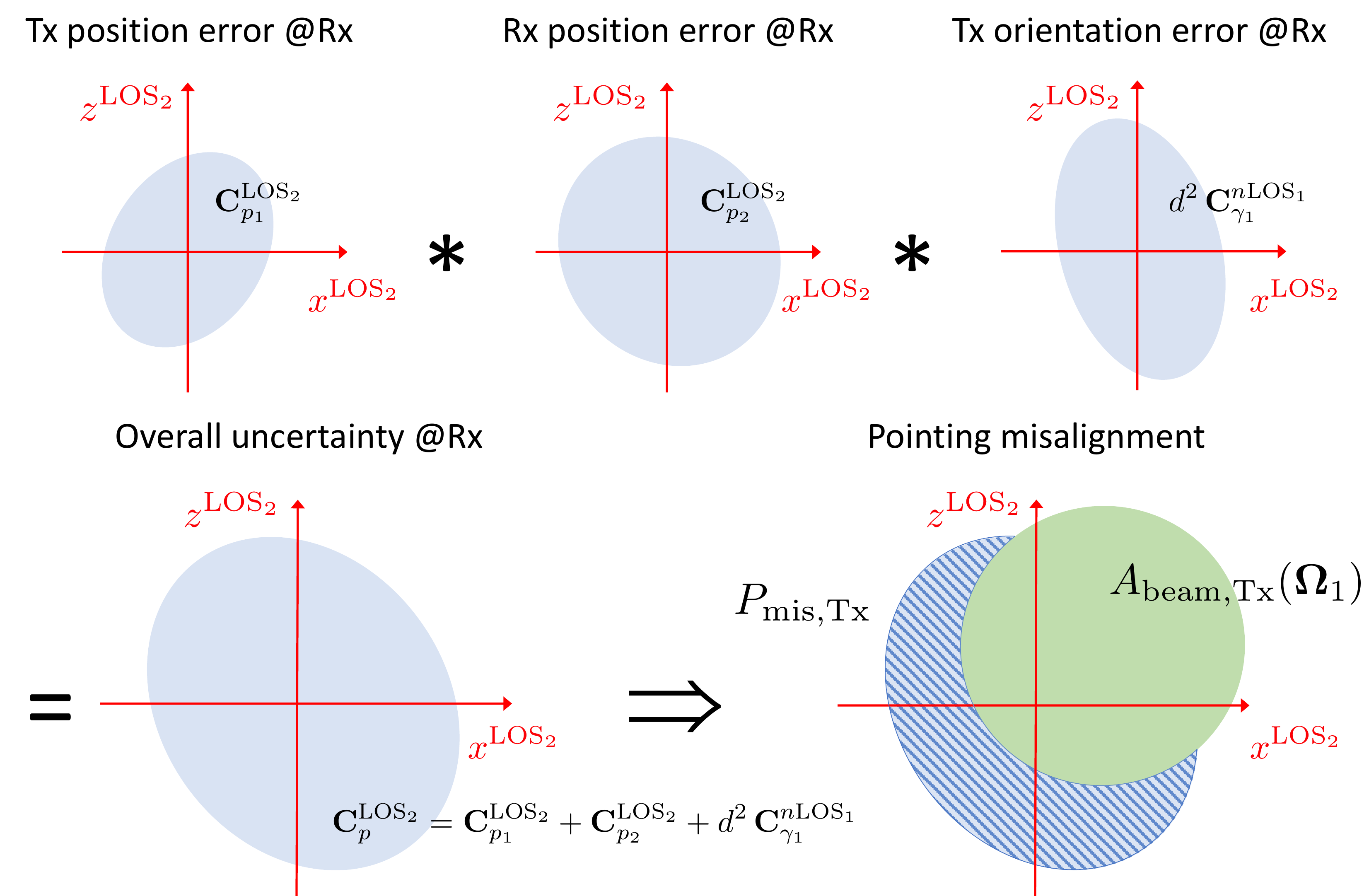}  } \\
	\caption{Representation of the LOS-aligned reference systems used for the evaluation of the Tx beam misalignment probability (\ref{subfig:alignment1}) and composition of the Tx/Rx position and Tx orientation uncertainties defining the Tx beam and the Tx misalignment probability (\ref{subfig:alignment2}). Blue ellipses represent position/orientation uncertainty regions at both Tx and Rx; the Tx beam is the green cone.}
	\label{fig:Alignment_and_Outage}
\end{figure} 

By aligning the Tx and Rx local reference systems ($v_1$- and $v_2$-system, respectively)  with the true LOS direction, as depicted in Fig. \ref{subfig:alignment1}, we obtain the $\mathrm{LOS}_1$ and $\mathrm{LOS}_2$ reference systems, where $y^{\mathrm{LOS}_1}$ and $y^{\mathrm{LOS}_2}$ are LOS-aligned. We are interested in evaluating the impact of both the Tx and Rx position error $\delta\mathbf{p}^n_1$, $\delta\mathbf{p}^n_2$ as well as the Tx orientation error $\delta \boldsymbol{\gamma}^{n v_1}_1$ at the Rx side, in particular on the $(x^{\mathrm{LOS}_2}, z^{\mathrm{LOS}_2})$ plane.

The position error $\delta\mathbf{p}^n_1$ translates from the $n$-system to the $(x^{\mathrm{LOS}_2}, z^{\mathrm{LOS}_2})$ plane as a combination of rotation matrices (details in Appendix \ref{app:appendix2}) as: 
\begin{equation}\label{eq:position_error_1_LOS2}
	\begin{bmatrix}
	\delta x^{\mathrm{LOS}_2}_1 \,\,\,
	\delta z^{\mathrm{LOS}_2}_1
	\end{bmatrix}^{\mathrm{T}}\sim \mathcal{N}(\mathbf{0}, \mathbf{C}^{\mathrm{LOS}_2}_{p_1}) \,,
\end{equation}
where $\mathbf{C}^{\mathrm{LOS}_2}_{p_1}\in\mathbb{R}^{2 \times 2}$ denotes the covariance of the 2D Tx position error on the Rx plane.
Similarly, the Rx position error $\delta\mathbf{p}^n_2$ is mapped onto the $(x^{\mathrm{LOS}_2}, z^{\mathrm{LOS}_2})$ plane as:
\begin{equation}\label{eq:position_error_2_LOS2}
\begin{bmatrix}
\delta x^{\mathrm{LOS}_2}_2 \,\,\,
\delta z^{\mathrm{LOS}_2}_2
\end{bmatrix}^{\mathrm{T}}\sim \mathcal{N}(\mathbf{0}, \mathbf{C}^{\mathrm{LOS}_2}_{p_2}) \,,
\end{equation}
with $\mathbf{C}^{\mathrm{LOS}_2}_{p_2}\in\mathbb{R}^{2 \times 2}$ playing the same role of $\mathbf{C}^{\mathrm{LOS}_2}_{p_1}$ in \eqref{eq:position_error_1_LOS2}.

The effect of an Tx orientation error $\delta \boldsymbol{\gamma}^{n v_1}_1$ at the Rx side can be obtained by first expressing it in the $\mathrm{LOS}_1$-system, i.e., $\delta \boldsymbol{\gamma}^{n \mathrm{LOS}_1}_1$, and then projecting it onto the $(x^{\mathrm{LOS}_2}, z^{\mathrm{LOS}_2})$ plane, at distance $d$. Without detailing the analytical derivation, we can focus only on two components of $\delta \boldsymbol{\gamma}^{n \mathrm{LOS}_1}_1$, i.e., the angular errors corresponding to rotations around the $x^{\mathrm{LOS}_1}$ and $ z^{\mathrm{LOS}_1}$ axes\footnote{An angular error corresponding to a rotation around the $y^{\mathrm{LOS}_1}$ axis does not impact on the $(x^{\mathrm{LOS}_2}, z^{\mathrm{LOS}_2})$ plane, as can be observed from Fig. \ref{subfig:alignment1}.}, namely $\delta \phi^{n \mathrm{LOS}_1}_1$ and $\delta \psi^{n \mathrm{LOS}_1}_1$. These are approximated by a Gaussian distribution as:
\begin{equation}\label{eq:orientation_error_1_LOS1}
\begin{bmatrix}
\delta \phi^{n \mathrm{LOS}_1}_1 \,\,\, \delta \psi^{n \mathrm{LOS}_1}_1
\end{bmatrix}^{\mathrm{T}}\sim \mathcal{N}(\mathbf{0}, \mathbf{C}^{n \mathrm{LOS}_1}_{\gamma_1}) \,.
\end{equation}
The combination of the Tx/Rx position errors and Tx orientation error at the Rx side can finally be evaluated by the convolution of the Gaussian PDFs for the uncorrelated contributions \eqref{eq:position_error_1_LOS2}, \eqref{eq:position_error_2_LOS2} and \eqref{eq:orientation_error_1_LOS1} (the latter multiplied by $d$). The result is:
\begin{align}\label{eq:position_and_orientation_error_LOS2}
\delta \mathbf{p}^{\mathrm{LOS}_2}  = \begin{bmatrix}
\delta x^{\mathrm{LOS}_2} \,\,\,
\delta z^{\mathrm{LOS}_2}
\end{bmatrix}^{\mathrm{T}} \sim \mathcal{N}(\mathbf{0}, \underbrace{\mathbf{C}^{\mathrm{LOS}_2}_{p_1}+ \mathbf{C}^{\mathrm{LOS}_2}_{p_2} + d^2 \, \mathbf{C}^{n \mathrm{LOS}_1}_{\gamma_1}}_{\mathbf{C}^{\mathrm{LOS}_2}_{p}}) \,,
\end{align}
where the covariance of the Tx orientation error $\mathbf{C}^{n \mathrm{LOS}_1}_{\gamma_1}\in\mathbb{R}^{2 \times 2}$ is projected onto the Rx side by multiplying for $d^2$. By choosing a given confidence interval, e.g., $99.7$ percentile, \eqref{eq:position_and_orientation_error_LOS2} defines an ellipse representing the overall uncertainty at the Rx, as depicted in Fig. \ref{subfig:alignment2}. 

\begin{figure}[]
	\centering
	\subfloat[$\bar{\mathbf{\Omega}}$ = 5 deg \label{subfig:misProb_A}]{\includegraphics[width=0.49\columnwidth]{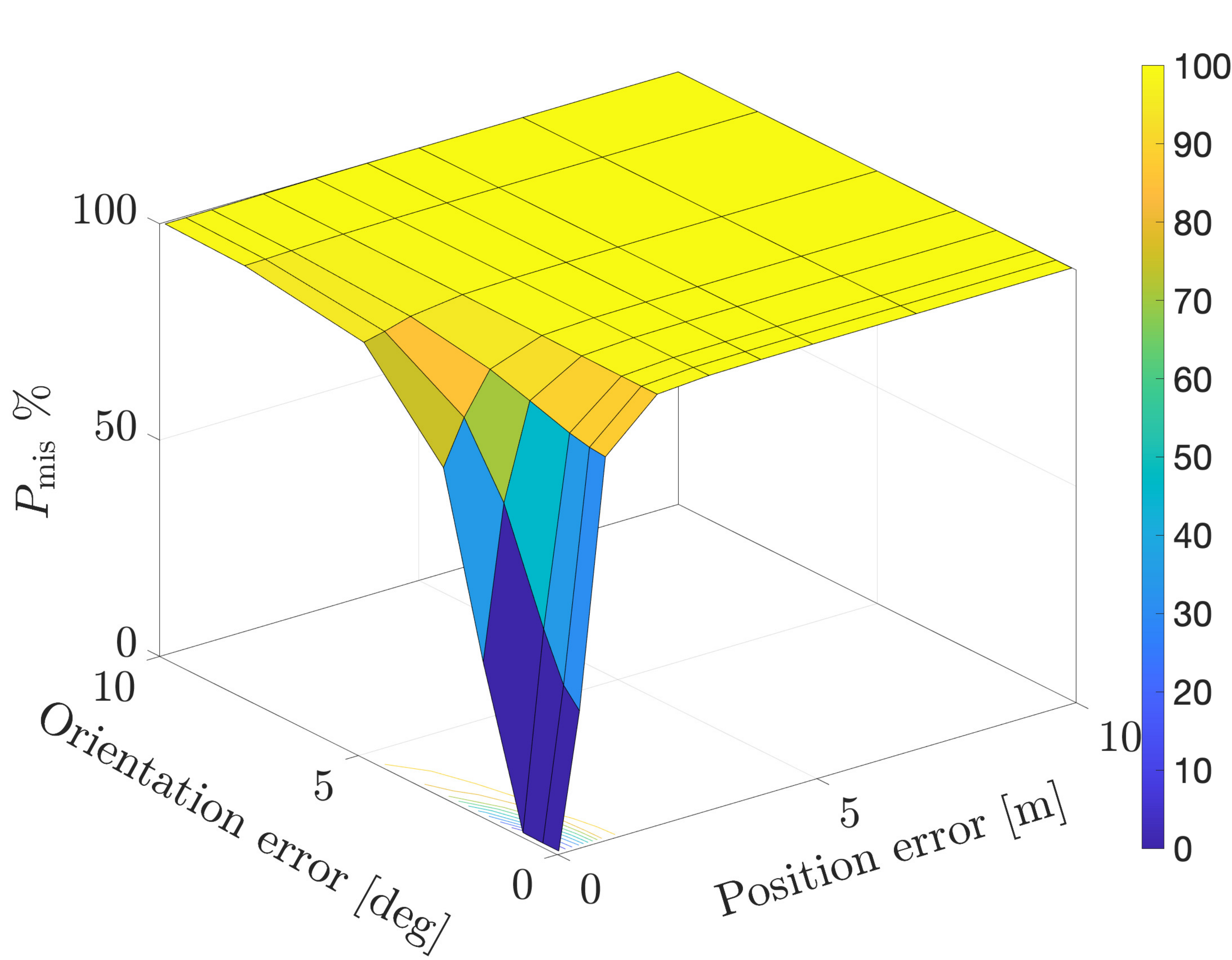}}
	\subfloat[$\bar{\mathbf{\Omega}}$ = 10 deg \label{subfig:misProb_B}]{\includegraphics[width=0.49\columnwidth]{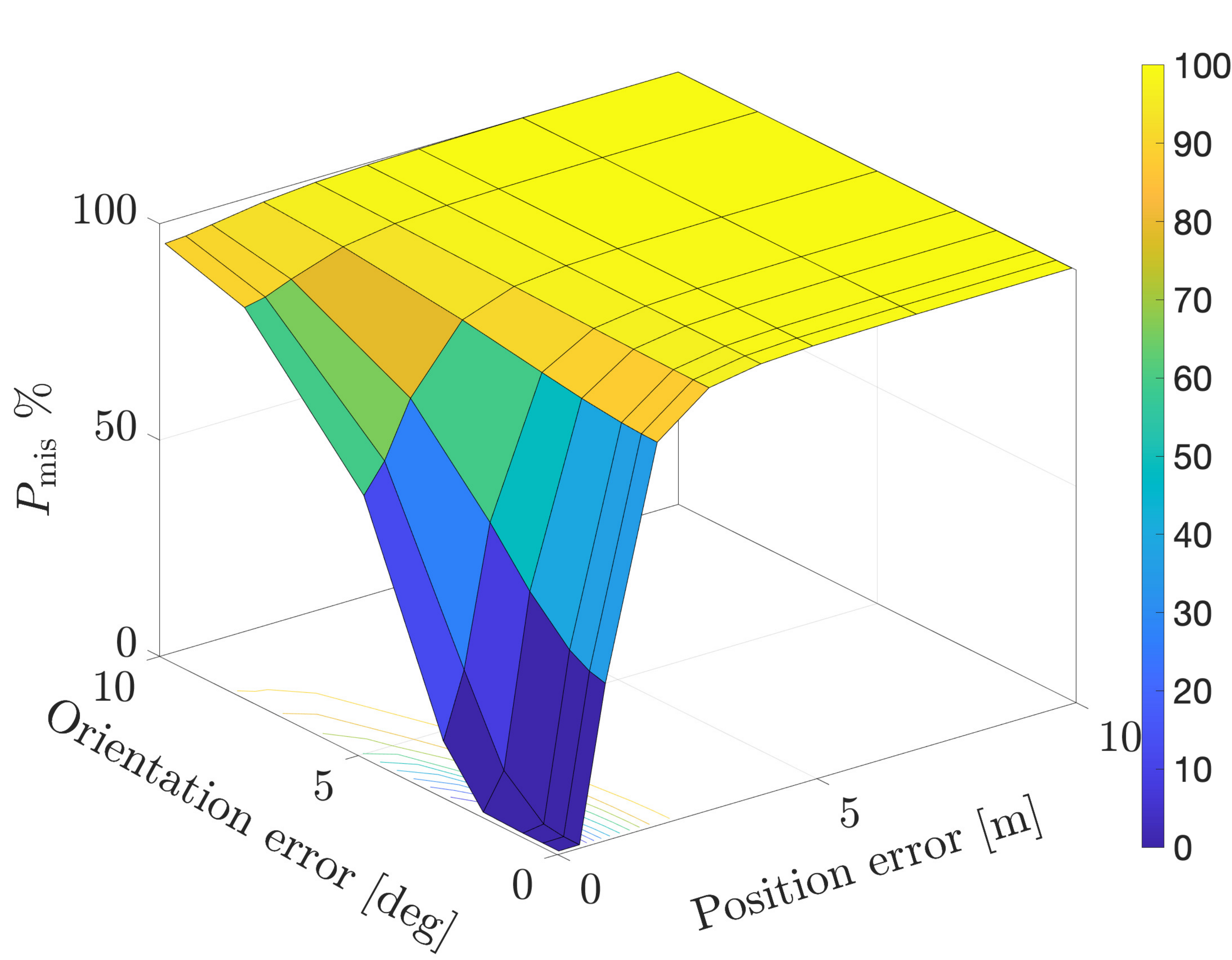}}\\
	\subfloat[$\bar{\mathbf{\Omega}}$ = 20 deg \label{subfig:misProb_C}]{\includegraphics[width=0.49\columnwidth]{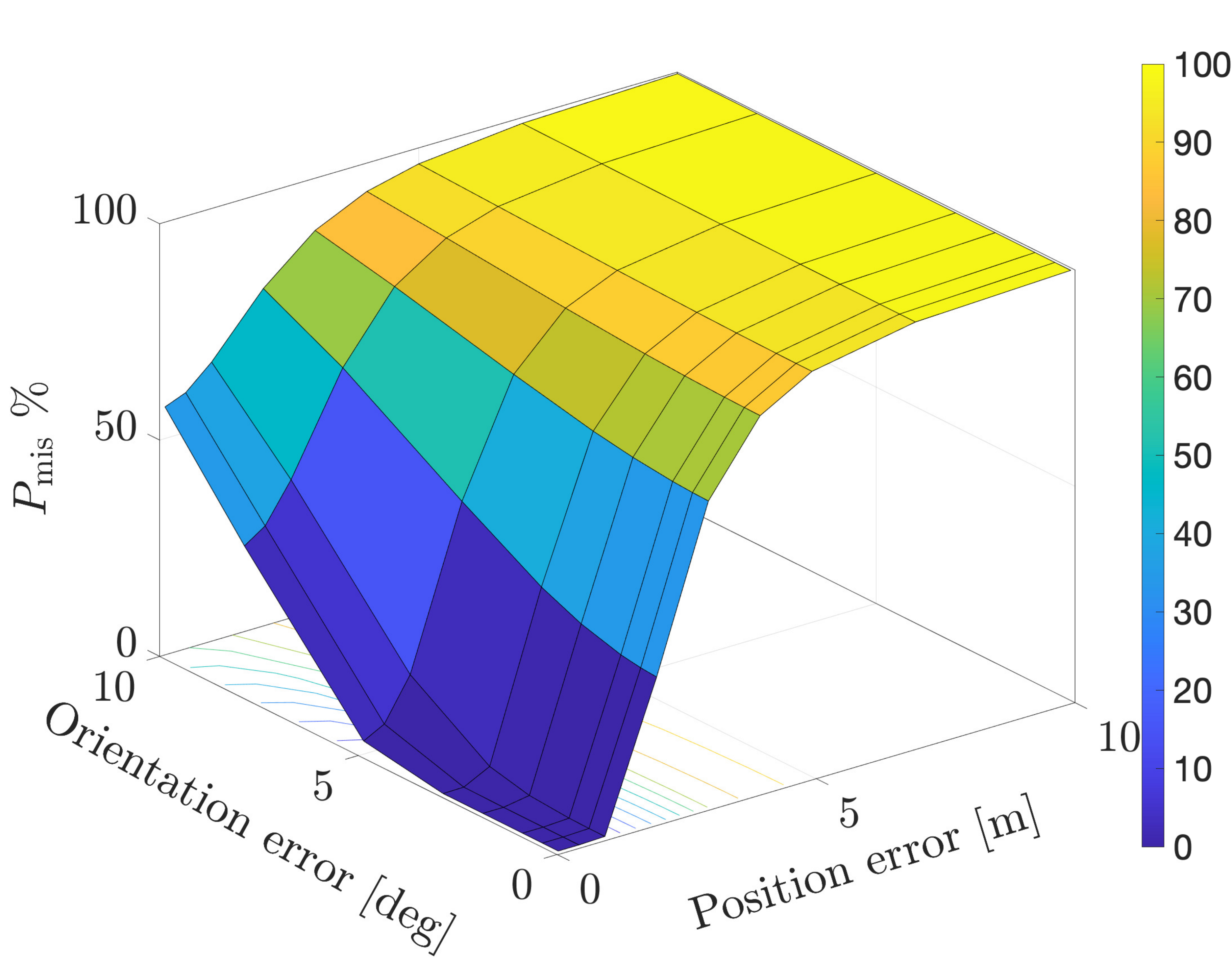}}
	\subfloat[$\bar{\mathbf{\Omega}}$ = 30 deg \label{subfig:misProb_D}]{\includegraphics[width=0.49\columnwidth]{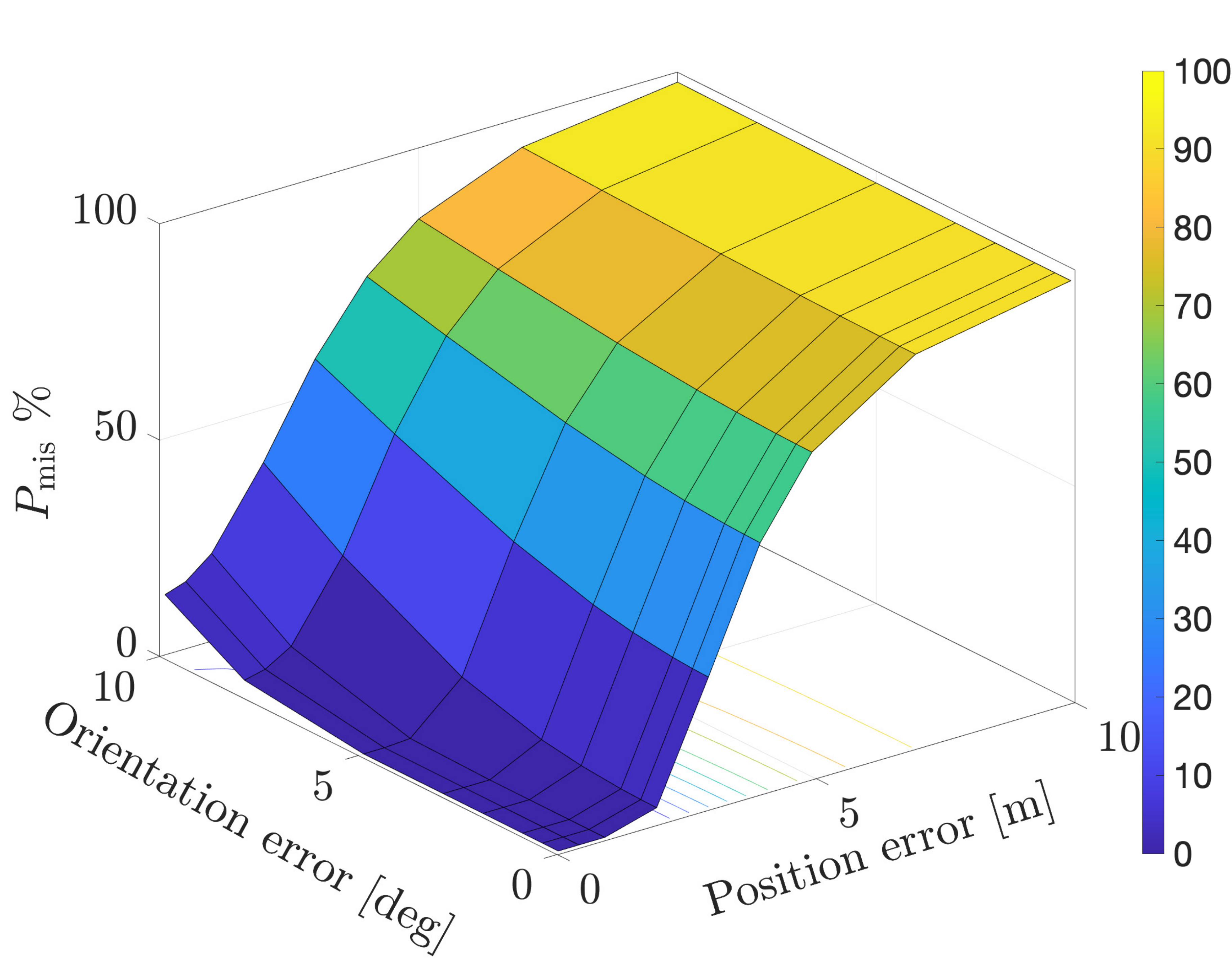}}\\
	\subfloat[ \label{subfig:misProb_E}]{\includegraphics[width=0.49\columnwidth]{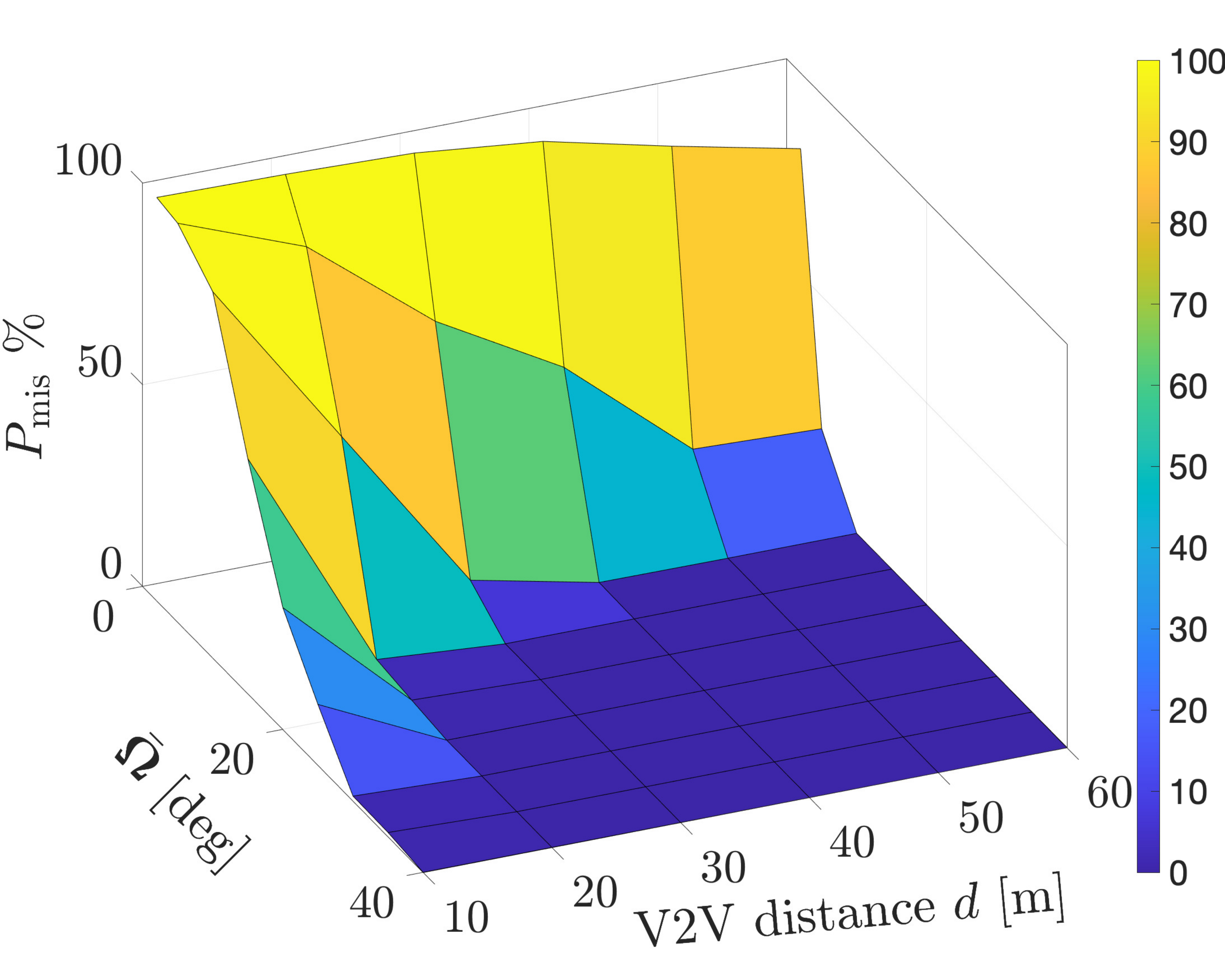}}\\
	\caption{Misalignment probability $P_{\text{mis}}$ vs position and orientation errors for four different beamwidths (\ref{subfig:misProb_A},\ref{subfig:misProb_B},\ref{subfig:misProb_C},\ref{subfig:misProb_D}) fixing the V2V distance $d$ = 20 m; Misalignment probability $P_{\text{mis}}$ vs beamwidth and V2V distance (\ref{subfig:misProb_E}) fixing the position and orientation error to 1 m and 1 deg (std. dev.).}
	\label{fig:Pmis}
\end{figure}

The Tx misalignment probability is therefore $P_{\mathrm{mis,Tx}}=1 - P_{\mathrm{beam,Tx}}(\mathbf{\Omega}_1) $, where:
\begin{equation}\label{eq:TXmisalignment}
\begin{split}
 P_{\mathrm{beam,Tx}}(\mathbf{\Omega}_1) =   \iint\limits_{A_{\mathrm{beam,Tx}}(\mathbf{\Omega}_1)} \!
\frac{\left | \mathbf{C}_{p}^{\mathrm{LOS}_2} \right |^{-\frac{1}{2}}}{2 \pi)} \text{e}^{\left(-\frac{1}{2} \delta {\mathbf{p}^{\mathrm{LOS}_2}}^{\mathrm{T}} \left(\mathbf{C}_{p}^{\mathrm{LOS}_2}\right)^{-1}\delta\mathbf{p}^{\mathrm{LOS}_2} \right)},
\end{split}
\end{equation}
is the integral of the Gaussian PDF  in \eqref{eq:position_and_orientation_error_LOS2} over the projected Tx beam area $A_{\mathrm{beam,Tx}}(\mathbf{\Omega}_1)$ at the Rx, function of the Tx beamwidth $\mathbf{\Omega}_1$. $P_{\mathrm{mis,Tx}}$ is graphically illustrated by the hatched area in Fig. \ref{subfig:alignment2} while Fig. \ref{fig:Pmis} shows an example of $P_{\mathrm{mis}}$ as function of the Tx/Rx position and orientation errors  (\ref{subfig:misProb_A}, \ref{subfig:misProb_B}, \ref{subfig:misProb_C}, \ref{subfig:misProb_D}) for fixed distance $d = 20$ m, and of the Tx/Rx beamwidths $\bar{\mathbf{\Omega}}=\mathbf{\Omega}_1=\mathbf{\Omega}_2$ and V2V distance (\ref{subfig:misProb_E}). 

%SPEIGAZIONE DEL CONCETTO DI ADATTAMENTO DI POTENZA
After the calculation of $P_{\mathrm{mis}}$ from $P_{\mathrm{mis,Tx}}$ and $P_{\mathrm{mis,Rx}}$ \eqref{eq:misalignment}, we consider the required power $P_{\mathrm{tx}}$ to achieve a minimum SNR ($\mathrm{SNR}_{\mathrm{min}}$), 
in worst pointing conditions, i.e., 
\begin{equation}\label{eq:Ptxscaling}
\begin{split}
P_{\mathrm{tx}} & = \dfrac{\mathrm{SNR}_{\mathrm{min}}\,  \eta(d) \, P_{\mathrm{noise}} }{ \dfrac{1}{16} G^{\mathrm{max}}_{1}\left(\mathbf{\Omega}_1\right) G^{\mathrm{max}}_{2}\left(\mathbf{\Omega}_2\right) } \,,
\end{split}
\end{equation}
where the attenuation factor of $1/16$ accounts for the mispointing at $- 3$ dB beamwidth from $G^{\mathrm{max}}$ in both elevation and azimuth directions, for Tx and Rx.
In this setting, $P_{\mathrm{mis}}\approx P_{\mathrm{out}}$ and problem \eqref{eq:optimproblem} can be iteratively solved.

It is important to notice that problem \eqref{eq:optimproblem} admits a single, optimal power/beamwidths configuration when the gain functions $G_1$ and $G_2$ are monotonically decreasing with the pointing errors (e.g., for single-lobe patterns as here). In this situation, the optimal beamwidths and Tx power are those for which constraint $(a)$ in \eqref{eq:optimproblem} is an equality, i.e., when both the Tx and Rx beams perfectly cover the uncertainty regions on the $(x^{\mathrm{LOS}_2}, z^{\mathrm{LOS}_2})$ and $(x^{\mathrm{LOS}_1}, z^{\mathrm{LOS}_1})$ planes, respectively, for a given overall $\bar{P}_{\mathrm{out}}$. 

%remark finale sul multi-veicolare
We remark that, in our work, the power/beamwidths optimization considers a single pair of vehicles. The extension to a generic vehicular network with multiple beam-pairs is beyond the scope of the paper, but in that case the optimization problem \eqref{eq:optimproblem} shall be reformulated to include the effect of mutual V2V interference. The optimal power/beamwidth solution for each vehicle would therefore be a trade-off between the optimality of the solution of \eqref{eq:optimproblem} and the inter-vehicle interference. %Alternatively, the problem could be approached as with machine learning \cite{Chen2020FederatedLearning}.

\subsection{Beamwidth and Power Control (BPC)}\label{subsec:beamadaptation}

To obtain the optimal beamwidth and power configuration from \eqref{eq:optimproblem}, the two involved vehicles must have the \textit{instantaneous} knowledge of the \textit{true} mutual position and related covariance, to be then transferred onto the 2D plane transversal to the true LOS direction, which are not known in BAT. In this regard, we propose a heuristic solution to \eqref{eq:optimproblem}, in which the beamwidths are dynamically controlled according to the position and orientation estimates and corresponding covariance matrices, both obtained as output of the EKF and mutually exchanged between vehicles, without explicitly setting constraints on the outage probability $P_{\mathrm{out}}$. In the proposed \ac{BPC}, the Tx power is continuously adjusted on the basis of the selected Tx/Rx beamwidths, given a threshold performance level in terms of SNR.
The key idea is to exploit the EKF-derived a-posteriori covariance of position and orientation to analytically retrieve the variance of azimuth and elevation pointing angles \eqref{eq:AzimuthandElevation} at both Tx (e.g., $v_1$) and Rx (e.g., $v_2$) sides. These are used to estimate beamwidths $\hat{\boldsymbol{\Omega}}_1$, $\hat{\boldsymbol{\Omega}}_2$ and $\hat{P}_{\mathrm{tx}}$ to operate the V2V link without (or with controlled) outage.

The derivation of the pointing angles' covariance is here reported only for the Tx side, namely for the couple $\left(\alpha_1,\beta_1\right)$, whereas the extension to the Rx's ones is similar.
The first step is the re-formulation of \eqref{eq:estimdist} (relative Tx-Rx position in the $v_1$-system) for EKF-estimated quantities with the rotation matrix:
\begin{equation}\label{eq:vector_distance_rot}
\Delta \hat{\mathbf{p}}_{12}^{v_{1}}  = \mathbf{R}\left \{ \hat{\mathbf{q}}^{v_1 n}_{1}\right\} \left(\hat{\mathbf{p}}^{n}_{2,t}-\hat{\mathbf{p}}^{n}_{1,t}\right),
\end{equation}
explicitly showing dependence of the rotation matrix $\mathbf{R}\left \{ \mathbf{q} \right\}$ on the Tx estimated orientation (quaternion $\hat{\mathbf{q}}^{v_1 n}_{1}$). The relative Tx-Rx position, expressed in the $v_1$-system, is therefore a non-linear combination of multivariate Gaussian \ac{PDF}s, whose covariance can be approximated by linearizing \eqref{eq:vector_distance_rot} as:
\begin{align}\label{eq:vector_distance_rot_linearized}
\delta \Delta \hat{\mathbf{p}}_{12}^{v_{1}} \approx \mathbf{B}_{q,1}\delta \hat{\mathbf{q}}^{v_1 n}_1 + \mathbf{B}_{\Delta p}\delta \Delta \hat{\mathbf{p}}_{12}^{n}\,,
\end{align}
where $\delta \Delta \hat{\mathbf{p}}_{12}^{v_{1}} $ denotes the error on the relative Tx-Rx position, the first term is the linearization with respect to an error on the orientation (by means of the gradient $\mathbf{B}_{q_1}\in\mathbb{R}^{3 \times 4}$) and the second the linearization with respect to an error on the Tx-Rx relative position in the $n$-system (gradient $\mathbf{B}_{\Delta p}\in\mathbb{R}^{3 \times 3}$). The covariance of $\Delta \hat{\mathbf{p}}_{12}^{v_{1}}$ can therefore be approximated as:
\begin{equation}\label{eq:vector_distance_cov}
\hat{\mathbf{C}}^{v_1}_{\Delta p} \approx \mathbf{B}_{q_1}\hat{\mathbf{C}}^{v_1 n}_{q_1} \mathbf{B}^{\mathrm{T}}_{q_1} + \mathbf{B}_{\Delta p} \left(\hat{\mathbf{C}}^{n}_{p_1} + \hat{\mathbf{C}}^{n}_{p_2}\right) \mathbf{B}^{\mathrm{T}}_{\Delta p} \,,
\end{equation}
where $\hat{\mathbf{C}}^{n}_{p_1}\in\mathbb{R}^{3 \times 3}$ and $\hat{\mathbf{C}}^{n}_{p_2}\in\mathbb{R}^{3 \times 3}$ are the EKF-estimated covariance matrices of Tx and Rx position, respectively, and $\hat{\mathbf{C}}^{v_1 n}_{q_1}\in\mathbb{R}^{4 \times 4}$ the estimated covariance of the Tx orientation. The implicit assumption in \eqref{eq:vector_distance_cov} is the independence between position and orientation estimation errors, often verified in practice.

The desired pointing angles' variance is once again derived by linearizing \eqref{eq:AzimuthandElevation} as:
\begin{equation}\label{eq:AzimuthandElevation_linearized}
\delta \alpha_1 \approx \mathbf{b}^{\mathrm{T}}_{\alpha} \delta \Delta \hat{\mathbf{p}}_{12}^{v_{1}} \,, \quad
\delta \beta_1 \approx \mathbf{b}^{\mathrm{T}}_{\beta} \delta \Delta \hat{\mathbf{p}}_{12}^{v_{1}} \,,
\end{equation}
while the two gradients $\mathbf{b}^{\mathrm{T}}_{\alpha} \in \mathbb{R}^{1 \times 3}$ and $\mathbf{b}^{\mathrm{T}}_{\beta} \in \mathbb{R}^{1 \times 3}$ are finally used to obtain:
\begin{equation}
\begin{split}
\hat{\sigma}^2_{\alpha_1}  \approx \mathbf{b}_{\alpha} \hat{\mathbf{C}}^{v_1}_{\Delta p}\mathbf{b}^{\mathrm{T}}_{\alpha} \,,\quad
\hat{\sigma}^2_{\beta_1}  \approx \mathbf{b}_{\beta} \hat{\mathbf{C}}^{v_1}_{\Delta p}\mathbf{b}^{\mathrm{T}}_{\beta} \,.
\end{split}
\end{equation}

\begin{figure}[]
	\centering
	\subfloat[Conceptual block scheme\label{subfig:adaptation1}]{\includegraphics[width=0.8\columnwidth]{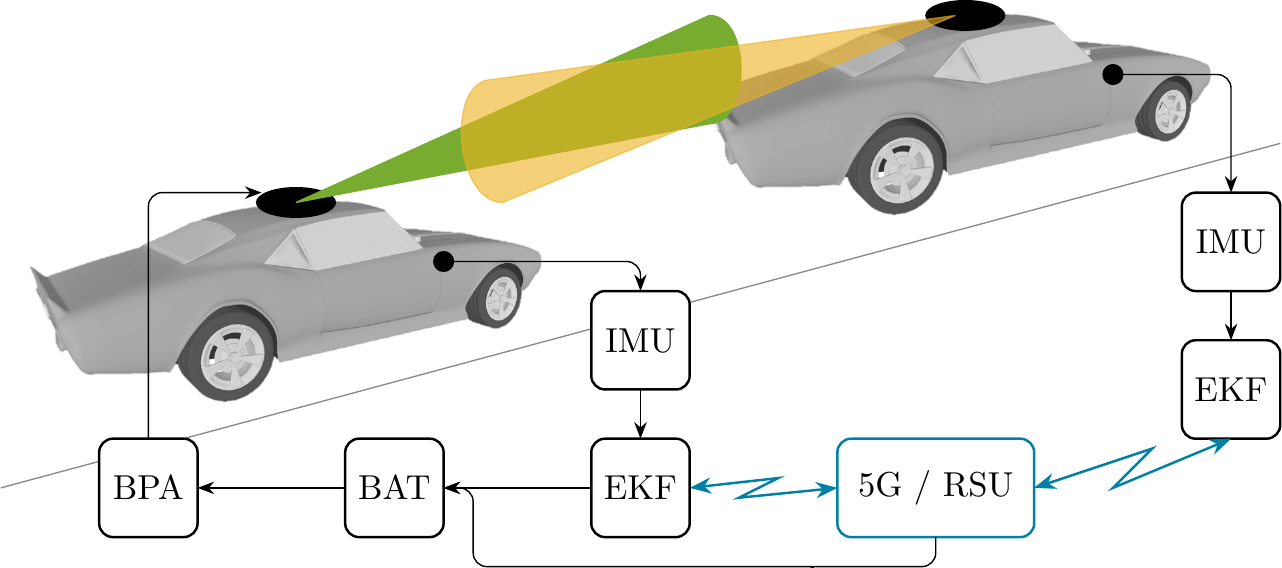}} 
	\hspace{3mm}
	\subfloat[Effect of beam narrowing \label{subfig:adaptation2}]{\includegraphics[width=0.8\columnwidth]{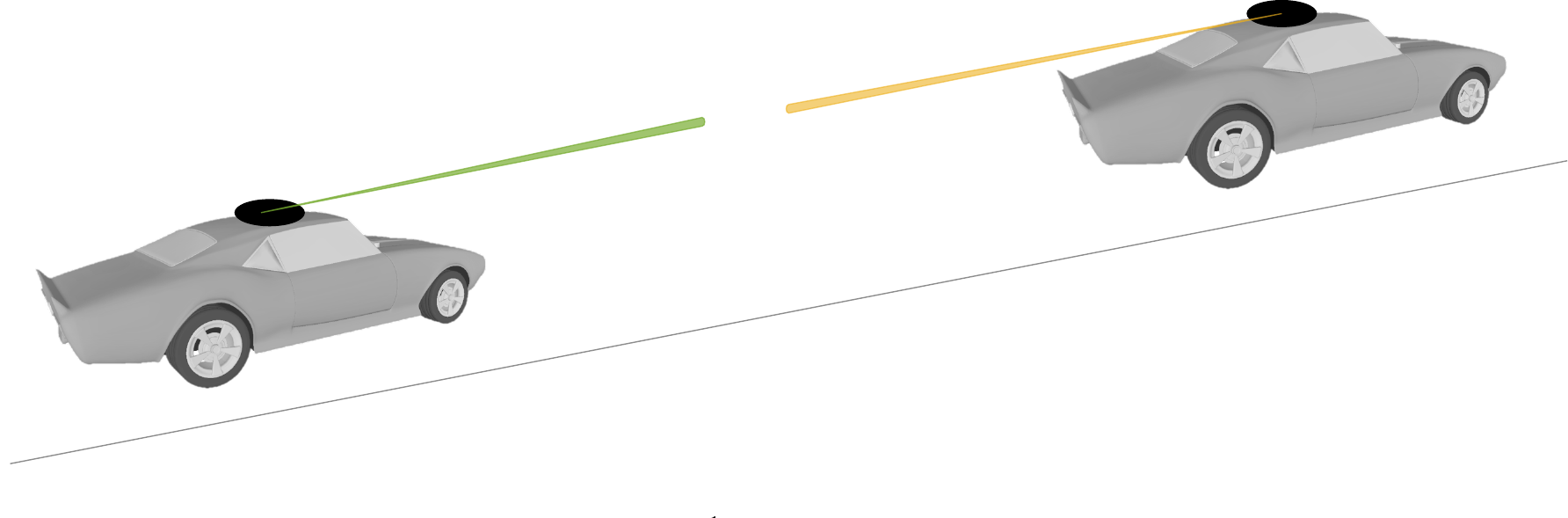}}
	\caption{Proposed BPC: (\ref{subfig:adaptation1}) conceptual block scheme; (\ref{subfig:adaptation2}) effect of beam narrowing with increasing distance.}
	\label{fig:beamadaptation}
\end{figure}

The azimuth and elevation full beamwidths at the Tx are set from $\hat{\sigma}^2_{\alpha_1} $ and $\hat{\sigma}^2_{\beta_1} $ to cover the $\pm k\sigma$ confidence interval (99.7\% in this work, i.e., $k=3$) for the pointing errors:
\begin{equation}\label{eq:beamwidth_selection}
\begin{split}
\hat{\Omega}_{1}^{\mathrm{az}}  =  2 k\hat{\sigma}_{\alpha_1} \,,\hspace{0.5cm} 
\hat{\Omega}_{1}^{\mathrm{el}}  = 2 k\hat{\sigma}_{\beta_1} \,,\\
\end{split}
\end{equation}
where, in the most general case, the beamwidth is chosen in a given codebook (finite resolution of the system). In this case, the outage probability is indirectly ruled by the choice of the confidence interval $k$ on the pointing angles,  which can be set according to the specific application. The proposed beamwidth adaptation is graphically illustrated in Fig. \ref{fig:beamadaptation}. For a given accuracy on the position/orientation estimates, when the mutual distance increases, the Tx vehicle can reduce the beamwidth, and so does the Rx one, allowing a Tx power saving while leaving the communication performance substantially unaffected. Gradients $\mathbf{B}_{q_1}$, $\mathbf{B}_{\Delta p}$ $\mathbf{b}_{\alpha}$, $\mathbf{b}_{\beta}$ are reported in Appendix \ref{app:appendix2}.

On top of beamwidth adaptation, the minimum required Tx power is estimated to guarantee a desired level of performance, expressed in terms of SNR at the Rx side, namely $\mathrm{SNR}_{\mathrm{min}}$. Given the estimated beamwidths $\hat{\mathbf{\Omega}}_1$ and $\hat{\mathbf{\Omega}}_2$, the Tx can estimate the corresponding minimum Tx/Rx gain (in worst pointing conditions) and the corresponding minimum required Tx power as:
\begin{equation}\label{eq:Ptxcontrol}
\begin{split}
\hat{P}_{\mathrm{tx}} & = \dfrac{\mathrm{SNR}_{\mathrm{min}}\,  \eta(\hat{d}) \, P_{\mathrm{noise}} }{ \dfrac{1}{16} G^{\mathrm{max}}_{1}\left(\hat{\mathbf{\Omega}}_1\right) G^{\mathrm{max}}_{2}\left(\hat{\mathbf{\Omega}}_2\right)} \,,
\end{split}
\end{equation}
where $\eta(\hat{d})$ denotes the path loss from the estimated distance $\hat{d}$. In compliance with EIRP regulations, if the requested Tx power exceeds the limit, it is clipped to $\hat{P}_{\mathrm{tx}}  = \mathrm{EIRP}_{\mathrm{max}}/G^{\mathrm{max}}_{1}(\hat{\mathbf{\Omega}}_1)$. Whenever necessary, a due margin on Tx power can be added to reduce the outage probability. 
Each step of the sensor-assisted dynamic \ac{BPC} is summarized in Alg. \ref{alg:adaptive-algorithm}. 

As will be shown in Section \ref{sec:results}, the proposed \ac{BPC} strategy allows to reach a near-optimal system performance (in terms beamwidth and Tx power), approaching the optimization problem in \eqref{eq:optimproblem} with a single-step computation by exploiting only the exchange of information data retrieved from the EKF. 

\begin{algorithm}[!tb]
	\begin{itemize}
		\item $v_1$ and $v_2$ initialize the beamwidth $\mathbf{\Omega}_1$ = $\mathbf{\Omega}_{\mathrm{max}}$;
		\item $v_1$ initializes the Tx power as $P_{\mathrm{tx}}$ = $\mathrm{EIRP}_{\mathrm{max}}/G_{1}^{\mathrm{max}}(\mathbf{\Omega}_1)$;
		\item \textbf{for} each time instant $t$:
		\begin{enumerate}
			\item $v_1$ estimates $\hat{\mathbf{p}}_1$, $\hat{\mathbf{q}}_1$ and the related covariance matrices $\hat{\mathbf{C}}_{p_1}$, $\hat{\mathbf{C}}_{q_1}$ using a sensor fusion from GPS and IMU data, as in Section \ref{sect:EKF}; in the same way, $v_2$ estimates $\hat{\mathbf{p}}_2$, $\hat{\mathbf{q}}_2$ and $\hat{\mathbf{C}}_{p_2}$, $\hat{\mathbf{C}}_{q_2}$;
			\item $v_1$ and $v_2$ mutually exchange $\hat{\mathbf{p}}_1$, $\hat{\mathbf{C}}_{p_1}$ and $\hat{\mathbf{p}}_2$, $\hat{\mathbf{C}}_{p_2}$, respectively, over the control link; 
			\item $v_1$ and $v_2$ estimate the beamforming directions $[\hat{\alpha_1}\,\,\hat{\beta_1}]^\mathrm{T}$ and $[\hat{\alpha_2}\,\,\hat{\beta_2}]^\mathrm{T}$ using  \eqref{eq:AzimuthandElevation};
			\item $v_1$ and $v_2$ estimate the covariance of the beamforming directions and set the beamwidths $\hat{\mathbf{\Omega}}_1$ and $\hat{\mathbf{\Omega}}_2$ as described in subsection \ref{subsec:beamadaptation};
			\item $v_1$ estimates the requested Tx power as in \eqref{eq:Ptxcontrol}, where, if the EIRP limit is exceeded, the Tx power is consequently clipped and the link can experience an uncontrolled outage; 
		\end{enumerate}
		\textbf{end}
	\end{itemize}
	\caption{Sensor-assisted dynamic \ac{BPC} algorithm\label{alg:adaptive-algorithm}}
\end{algorithm}

\section{Data Acquisition Campaign}\label{sect:campaign}

\begin{figure}[]
	\centering
	\includegraphics[width=1\columnwidth]{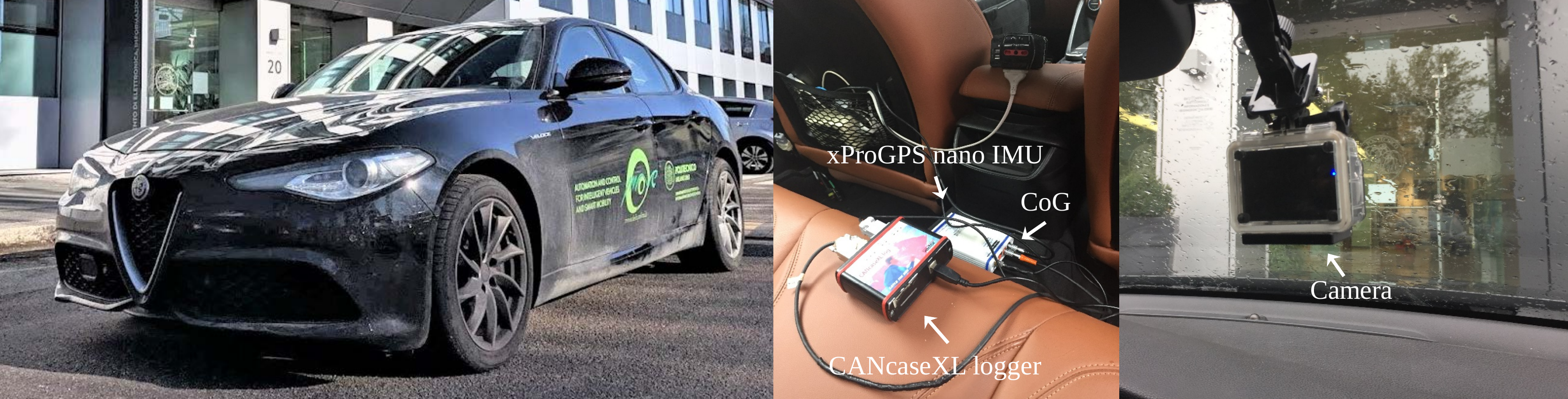}
	\caption{Experimental setup for the data acquisition campaign.}
	\label{fig:Setup}
\end{figure}

For the assessment of the proposed V2V system performance, we carried out an experimental campaign to acquire GPS (position and speed) and IMU (acceleration and angular velocity) data, to be used to simulate a realistic V2V communication on a real vehicle trajectory. The experimental setup is portrayed in Fig. \ref{fig:Setup}. For the experiment, we equipped an Alfa Romeo Giulia car with a xProGPS nano hardware platform from Suchy\textup{\textregistered} data systems. The xProGPS nano integrates a GPS sensor, with 10 Hz sampling frequency, providing 3D position, scalar velocity and heading of the vehicle, plus a 6 \ac{DoF} IMU, with 100 Hz sampling frequency, comprising a 3D gyroscope and a 3D accelerometer. While the IMU is located in the \ac{COG} of the car, the GPS module has been placed on the vehicle roof, the intended position of a mmW transceivers and an high-precision camera. GPS and IMU data  have been recorded and synchronized by a Vector CANcaseXL platform with CANalyzer\textup{\textregistered} software. In addition, we extract data from  other proprietary sensors built-in the car, that are made accessible by the manufacturer on the on-board data bus for research purposes only: four 2D accelerometers on the frontal part of the vehicle (two rigid with the chassis and two with the wheels), one additional 3 \ac{DoF} IMU, four wheel speed sensors and one steering column angle sensor. 

\begin{figure}[]
	\centering
	\subfloat[][Trajectory]{\includegraphics[width=0.4\columnwidth]{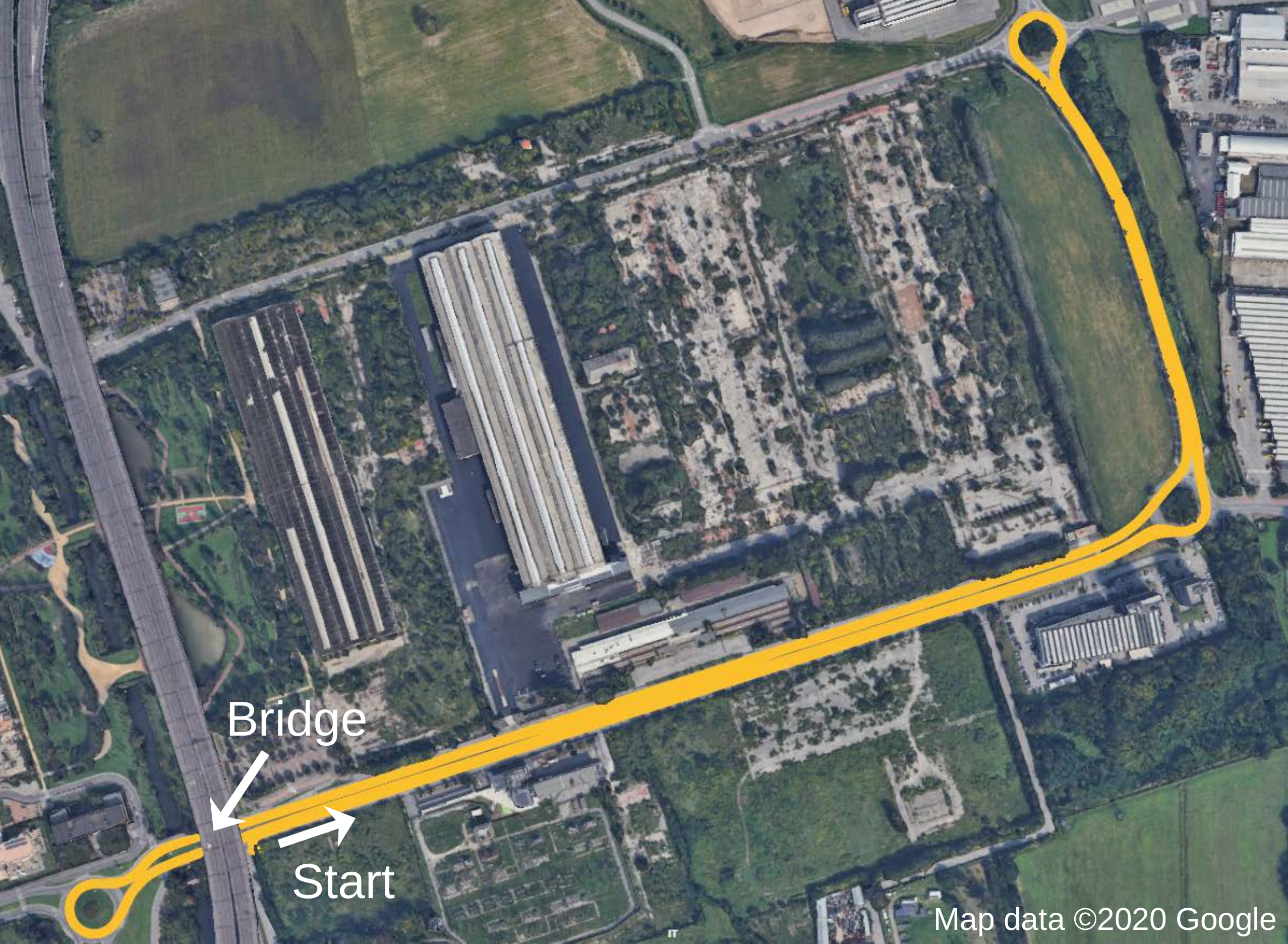}	\label{fig:Trajectory}}
		\subfloat[][Speed profile]{\includegraphics[width=0.4\columnwidth]{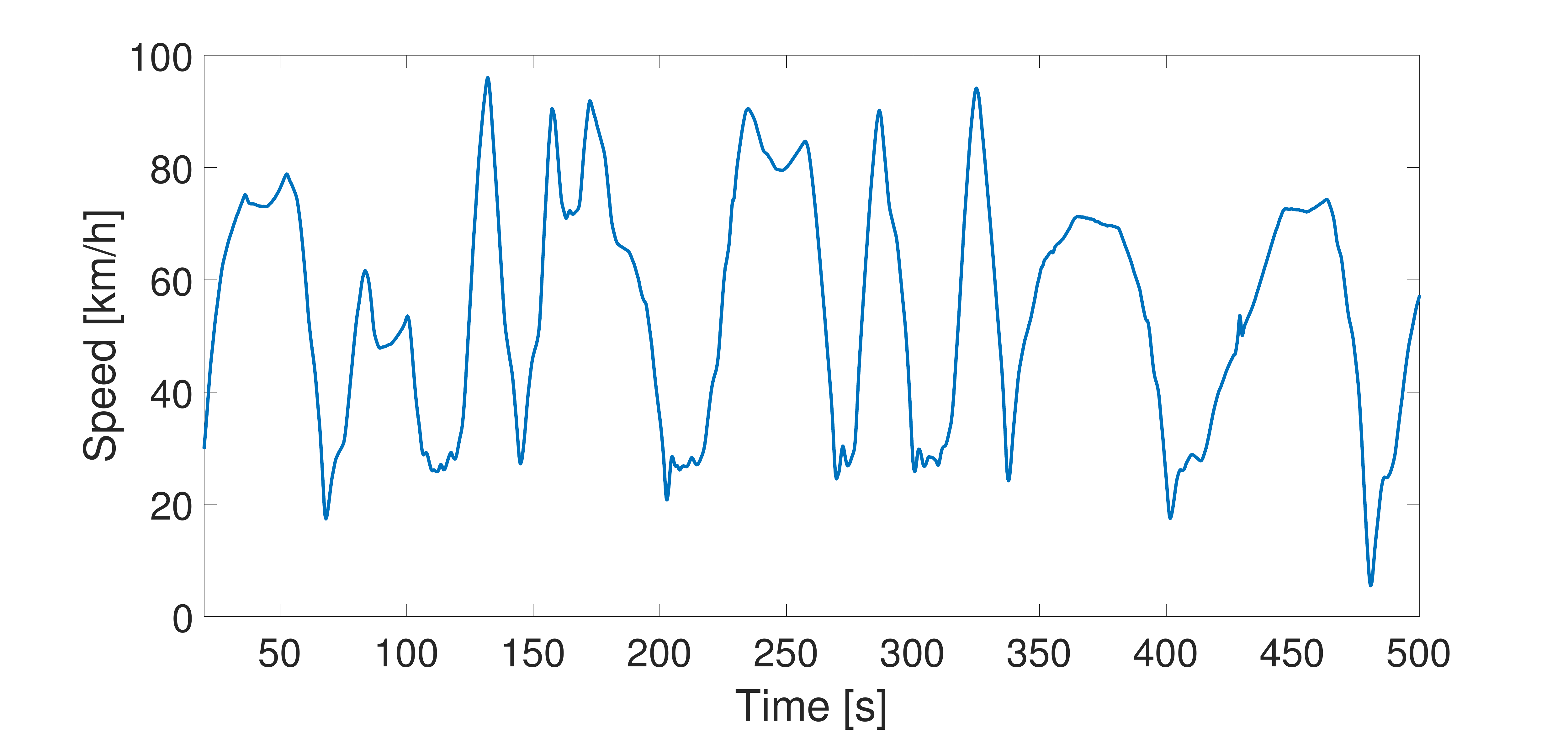}
		\label{fig:Speed}}
	\caption{Selected trajectory (yellow line) for the experimental campaign in the Milan area, Italy (Fig. \ref{fig:Trajectory}) and measured vehicle speed (Fig. \ref{fig:Speed}) over two laps.}
	\label{fig:TrajectoryandSpeed}
\end{figure}
%\begin{figure}[tb!]
%	\centering
%	\subfloat[][Roll]{\includegraphics[width=.35\columnwidth]{Figures/Results_new/roll.eps}
%		\label{subfig:roll}}
%	\subfloat[][Pitch]{\includegraphics[width=.35\columnwidth]{Figures/Results_new/pitch.eps}
%		\label{subfig:pitch}}
%	\subfloat[][Yaw/Heading]{\includegraphics[width=.35\columnwidth]{Figures/Results_new/yaw.eps}
%		\label{subfig:yaw}}
%	\caption{Estimated roll (Fig. \ref{subfig:roll}), pitch (Fig. \ref{subfig:pitch}) and yaw/heading (Fig. \ref{subfig:yaw}) angles for the considered trajectory.}
%	\label{fig:angles}
%\end{figure}

The experimental campaign is carried out along the suburban road route depicted in Fig. \ref{fig:Trajectory}, located in the North-East side of Milan, Italy. This is chosen purposely for being a challenging path in terms of vehicles' motion, alternating relatively fast straight road segments to sharp curves from roundabouts. The absence of buildings allows to simulate a \ac{LOS} V2V communication employing the measured data, focusing on the factors affecting the beam misalignment. Moreover, GPS data are not affected by \ac{NLOS} or urban canyon effects, except for the bridge in the lower left part of Fig. \ref{fig:Trajectory}, where the noisy GPS data are handled by a IMU-assisted extrapolation. The whole trajectory comprises two laps of the selected route in Fig. \ref{fig:Trajectory}, each one of $2.9$ km length. The measured vehicle speed over the whole trajectory is reported in Fig. \ref{fig:Speed}.
GPS and IMU data are fused with the Bayesian approach described in Section \ref{sect:EKF} and Appendix \ref{app:appendix}, to estimate vehicle kinematic, with the corresponding covariance matrix.  
As stated in Section \ref{sect:EKF}, we estimate the instantaneous roll and pitch angles from lateral and longitudinal accelerations with the so called roll and pitch stiffness of the vehicle, following the method in \cite{Corno2019}, while the yaw/heading angle is obtained from the GPS. The employed pitch and roll estimation method leverages on specific calibration that depends on the mechanical characterization of the vehicle and apply only for an IMU placed in the vehicle's \ac{COG}. In case these data were unavailable, a multi-IMU setup can provide an estimation of roll and pitch angles \cite{Bancroft2011}, at the price of an increased cost and computational complexity for the sensor fusion algorithm (not covered here).

\section{Simulation Settings and Numerical Results}
\label{sec:results}

We consider a mmW MIMO V2V link operating at 28 GHz carrier frequency, with a communication bandwidth $B=$ 400 MHz, as specified by 3GPP \cite{3GPP_Rel16_2}. From the measured GPS position and calibrated orientation (Section \ref{sect:campaign}), we simulate a V2V \ac{BAT} between vehicles $v_1$ and  $v_2$, assumed to travel on the same trajectory, separated by an initial time gap $\Delta T$, yielding
\begin{align}
	\mathbf{p}_{1,t} & = \mathbf{p}_{t}\,, \,\, \mathbf{p}_{2,t} = \mathbf{p}_{t-\Delta T}\,,\\
	\mathbf{q}_{1,t} & = \mathbf{q}_{t}\,, \,\, \mathbf{q}_{2,t} = \mathbf{q}_{t-\Delta T}\,.
\end{align}
We set a time gap $\Delta T=3$ s representing a time-varying distance between the two vehicles ranging from a minimum of 6 m to a maximum of 80 m, resulting from the velocity pattern in Fig. \ref{fig:Speed}. We use the \ac{EKF}-derived covariance matrices $\hat{\mathbf{C}}_{p,t}$ and $\hat{\mathbf{C}}_{q,t}$ to generate the experiment setup-specific noisy position/orientation estimates, exchanged between $v_1$ and $v_2$ over the control channel. In all the following, we will use as performance metrics the error on position and orientation as: 
\begin{align}
  \sigma_p =& \sqrt{ \dfrac{1}{N} \sum_{t=1}^{N} \mathrm{tr}\left(\hat{\mathbf{C}}_{p,t}\right)}\, , \,\,\,\,\,
 \sigma_{\gamma} = \sqrt{ \dfrac{1}{N} \sum_{t=1}^{N} \mathrm{tr}\left(\hat{\mathbf{C}}_{\gamma,t}\right)}\, , 
\end{align}
in which $N$ represents the total number of time instants and $\mathbf{C}_{\gamma,t}$ is obtained from $\mathbf{C}_{q,t}$ as in Appendix \ref{app:appendix}. We choose to express the orientation error with the standard deviation on Euler angles to be consistent with available literature and commercial navigation products \cite{NovatelSPAN,MTi-G-710,InertialLabs}. We assume to have free-space propagation and neglect the interaction of the propagating wave with the vehicles' roof. 

We numerically evaluate the performance of the proposed V2V solution in terms of CDF of the SNR and requested Tx power for an extensive number of system parameters, in particular: (\textit{i}) position and orientation accuracies, respectively $\sigma_p$ and $\sigma_{\gamma}$; 
(\textit{ii}) end-to-end latency $\tau$ on the control channel and (\textit{iii}) sensors' sampling frequency $f_\mathrm{data} = 1/T$. 
We assume that the position information is known with an average error that ranges from 0.15 m (as requested to meet the 5G service requirements for eV2X scenarios \cite{3GPP_Rel16_2}) to 1.5 m (as common in nowadays positioning systems), while the orientation information is subject to an average error of 0.15 deg to 1.5 deg. While the latter can reasonably achieved by fusing data from off-the-shelf inertial sensors \cite{kok2017using}, the former represents a lower bound on orientation estimation which can be thought to be provided either with expensive setups \cite{NovatelSPAN,MTi-G-710,InertialLabs} or by the fusion of multiple on-board sensors. The sampling rate varies from $f_\mathrm{data}=100$ Hz, as for \ac{EKF} output data, to $f_\mathrm{data}=1$ kHz, for high-performance sensors. Finally, the end-to-end latency $\tau$ for exchanging pointing data for BPC is chosen to range from $1$ to $100$ ms, with 10 ms being the upper latency requirement for eV2X according to \cite{3GPP_Rel16_2}. Table \ref{table:simulationparameters} reports the complete set of simulation parameters. 
The threshold used to determine the outage condition has been selected as the minimum SNR for an error-free BPSK transmission with a 25\%-overhead Forward Error Correction (FEC) code ($1.3 \times 10^{-2}$ BER on the channel) \cite{ITUG9751-2004}. 
\begin{table*}[]
	\begin{center}
		\caption{Simulation parameters.}
		\label{table:simulationparameters}
		\begin{tabular}{l | c  | c | l | c  | c}
			\hline% <-- Toprule here
			\textbf{Parameter} & \textbf{Symbol} & \textbf{Value} &\textbf{Parameter} & \textbf{Symbol} & \textbf{Value}\\
			\hline
			Carrier frequency & $f_0$ & 28 GHz &
			Signal bandwidth & $B$ & 400 MHz\\
			Noise power & $P_{\mathrm{noise}}$ & $-81$ dBm&
			EIRP limit & $\mathrm{EIRP}_{\mathrm{max}}$ & 43 dBm\\
			V2V gap time & $\Delta T$ & 3 s&
			V2V gap distance & $d$ & 6-80 m \\
			V2V update delay (latency) & $\tau$ & 1, 10, 50, 100 ms &
			Sensors' sampling frequency & $f_{\mathrm{data}}$ & 100, 1000 Hz\\
			Position error std. dev. & $\sigma_p$ & 0.15, 1.5 m &
			Orientation error std. dev. & $\sigma_{\gamma}$ & 0.15, 1.5 deg \\
			Beamwidth (fixed) & $\Omega^{\mathrm{az}}$ & 2.5, 13 deg&
			Beamwidth (fixed) & $\Omega^{\mathrm{el}}$ & 2.5, 13 deg\\
			%			SNR threshold & $\mathrm{SNR}_{\mathrm{thr}}$ & 2.2 dB & & & \\
			\hline
		\end{tabular}
	\end{center}
\end{table*}

 \begin{figure}[]
	\centering
	\subfloat[][Beamwidth]{\includegraphics[width=0.45\columnwidth]{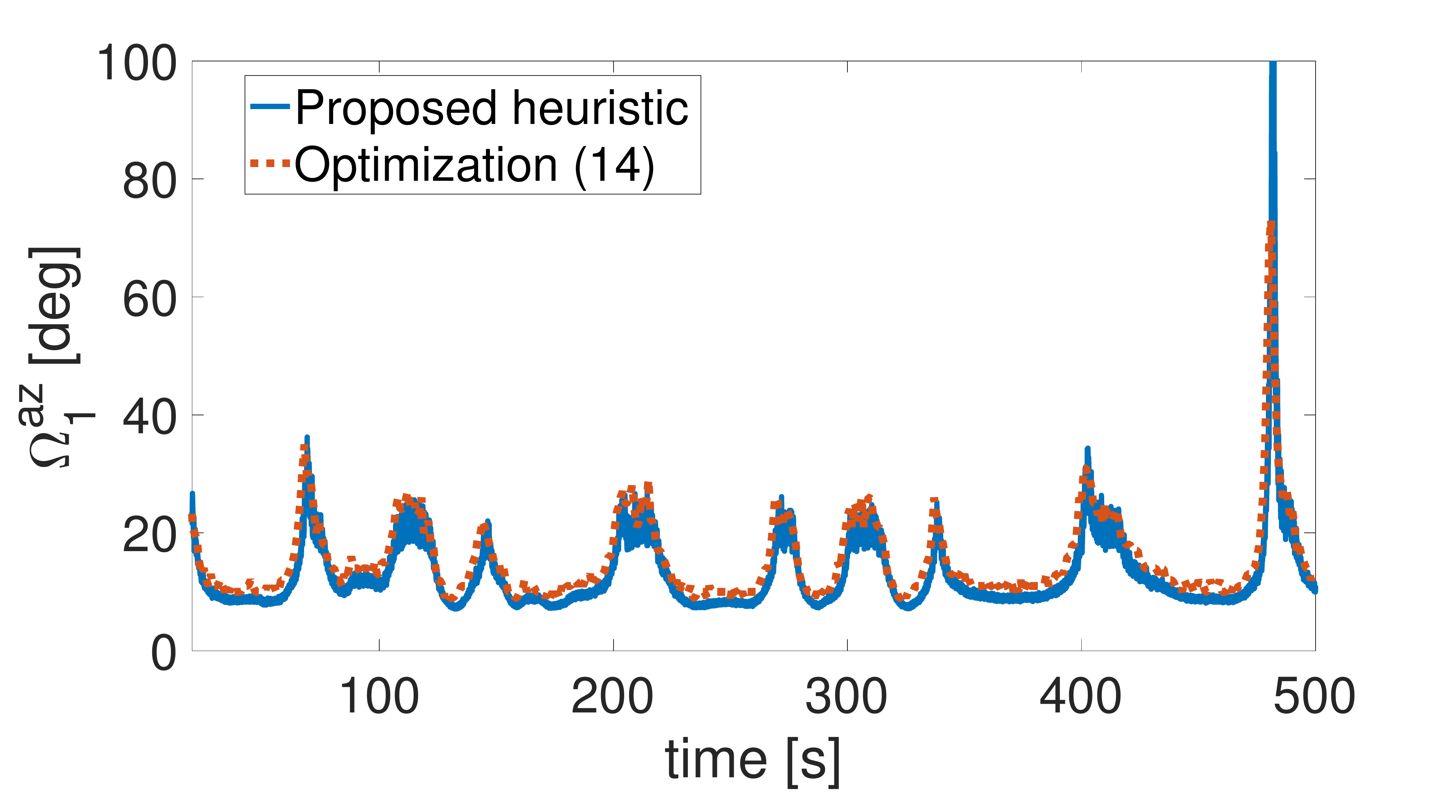}\label{subfig:heuristic_vs_optim_bw}}
	\subfloat[][Tx power]{\includegraphics[width=0.45\columnwidth]{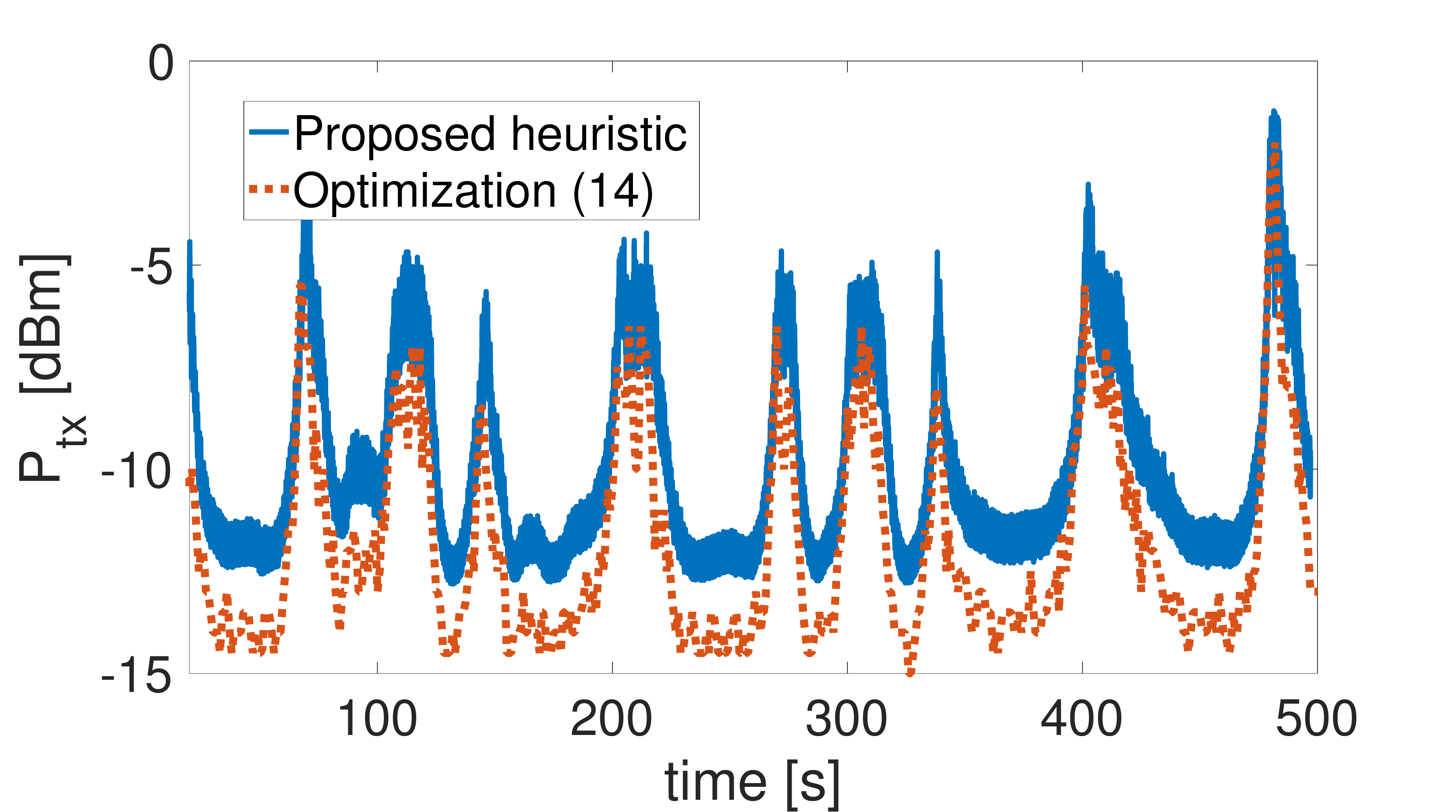}\label{subfig:heuristic_vs_optim_Ptx}}\\
%	\subfloat[][SNR]{\includegraphics[width=\columnwidth]{Figures/Results_new/Heuristic_vs_Optimization_SNR.eps}\label{subfig:heuristic_vs_optim_SNR}}
	\caption{Proposed heuristic (Subsection \ref{subsec:beamadaptation}) vs. optimization \eqref{eq:optimproblem} azimuth beamwidth (\ref{subfig:heuristic_vs_optim_bw}) and Tx power (\ref{subfig:heuristic_vs_optim_Ptx}), for $\sigma_p = 1.5$ m,  $\sigma_{\gamma} = 1.5$ deg and $\tau=10$ ms, $f_{\mathrm{data}}=100$ Hz.}
	\label{fig:heuristic_vs_optim}
\end{figure}

%CONFRONTO OTTIMIZZAZIONE
As a first assessment, we compare the heuristic \ac{BPC} (Subsection \ref{subsec:beamadaptation}) to the ideal optimization solution, where the problem \eqref{eq:optimproblem} is solved for the best beamwidth/Tx power for a maximum outage probability $\bar{P}_{\mathrm{outage}} = 6 \times 10^{-4}$ (corresponding to a $\pm 3 \sigma$ confidence on the Tx and Rx pointing errors for the heuristic solution). The optimization \eqref{eq:optimproblem} is used as benchmark, though it would be unpractical in real V2V systems. The comparison is provided in Fig. \ref{fig:heuristic_vs_optim}, for the beamwidth (Fig. \ref{subfig:heuristic_vs_optim_bw}) and the transmitted power (Fig. \ref{subfig:heuristic_vs_optim_Ptx}). The results are obtained for position/orientation errors of $\sigma_p = 1.5$ m and $\sigma_{\gamma} = 1.5$ deg, latency $\tau=10$ ms and sensors' sampling frequency $f_{data}=100$ Hz (the latter two used in the heuristic approach only). For the same outage probability $\bar{P}_{\mathrm{outage}}$, the proposed heuristic \ac{BPC} uses slightly narrower beams and higher power ($\approx \! 1-2$ dB) with respect to the optimization case, which makes use of the true and instantaneous V2V parameters. Overall, the performance of the heuristic approach practically matches the optimal one, with negligible penalty.
%However, fluctuations in the SNR may cause occasional outage events (SNR below the threshold of 2.2 dB). On the other hand, as expected, the optimum solution of \eqref{eq:optimproblem} provides better stability over time and it does not lead to outage events on the tested trajectory.

\begin{figure}[]
	\centering
	\subfloat[][Beamwidth]{\includegraphics[width=0.45\columnwidth]{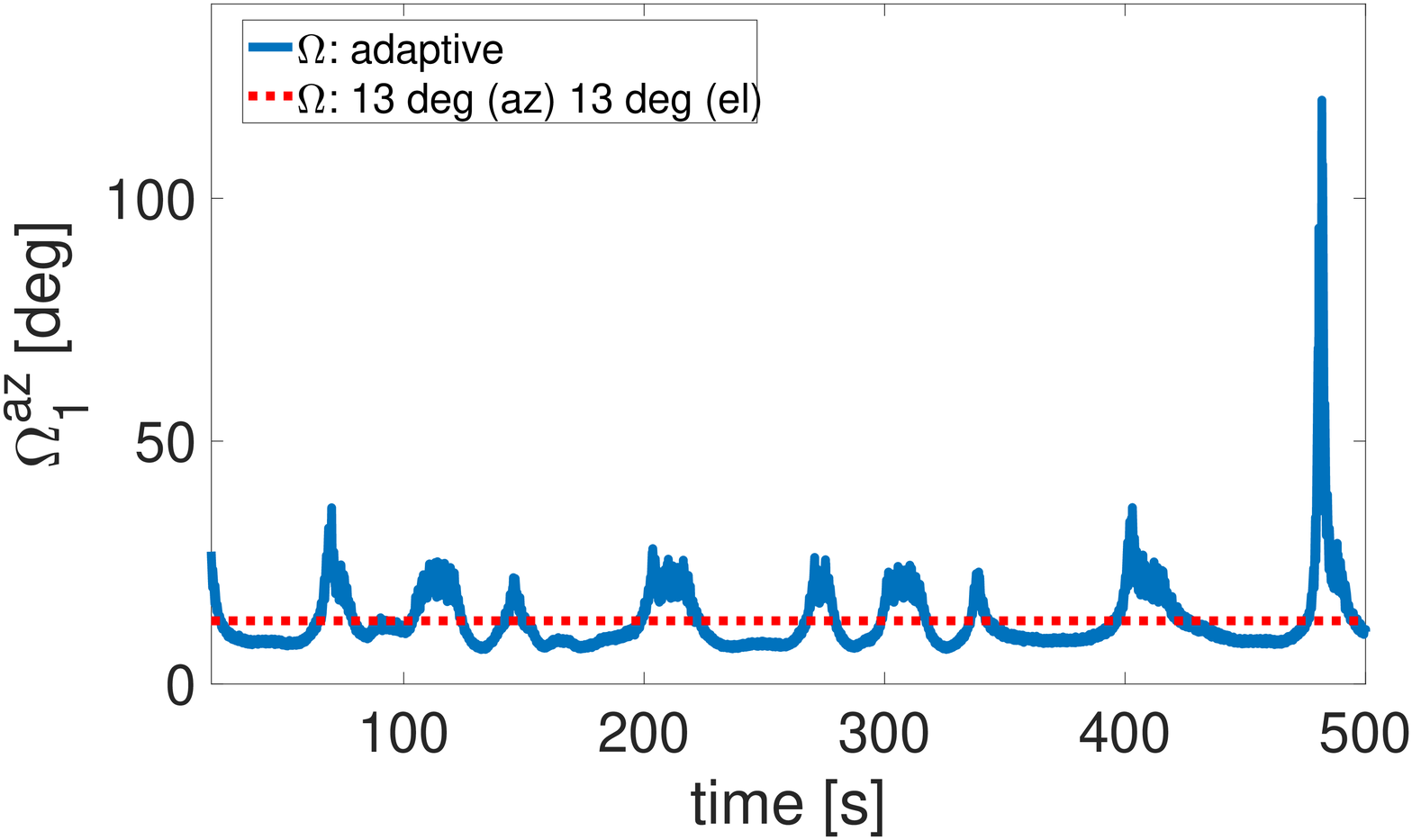}\label{subfig:heuristic_vs_fixedbw_bw_1}} 
	\subfloat[][Beamwidth]{\includegraphics[width=0.45\columnwidth]{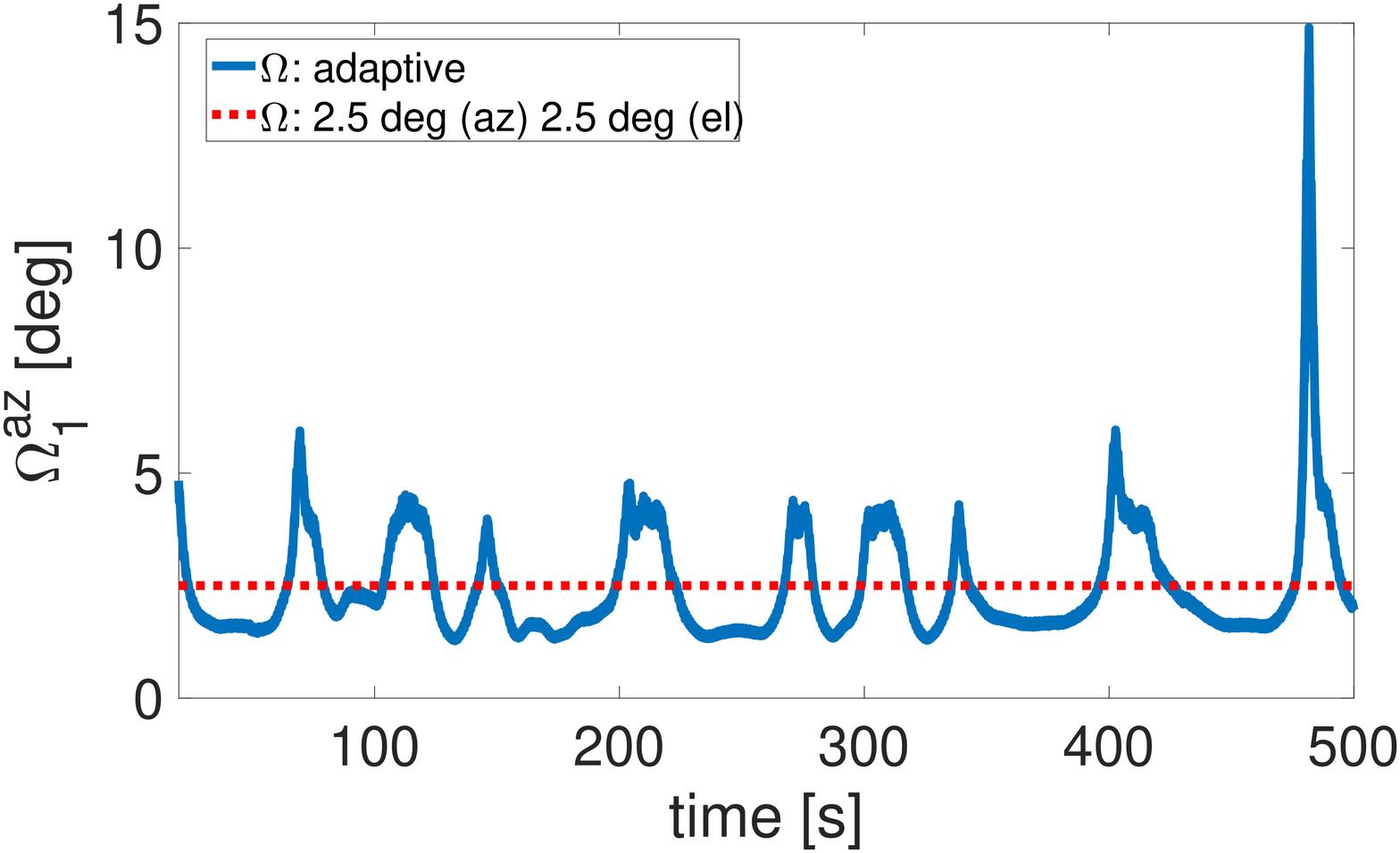}\label{subfig:heuristic_vs_fixedbw_bw_2}} \\ \vspace{-0.38cm}
	\subfloat[][Tx power]{\includegraphics[width=0.45\columnwidth]{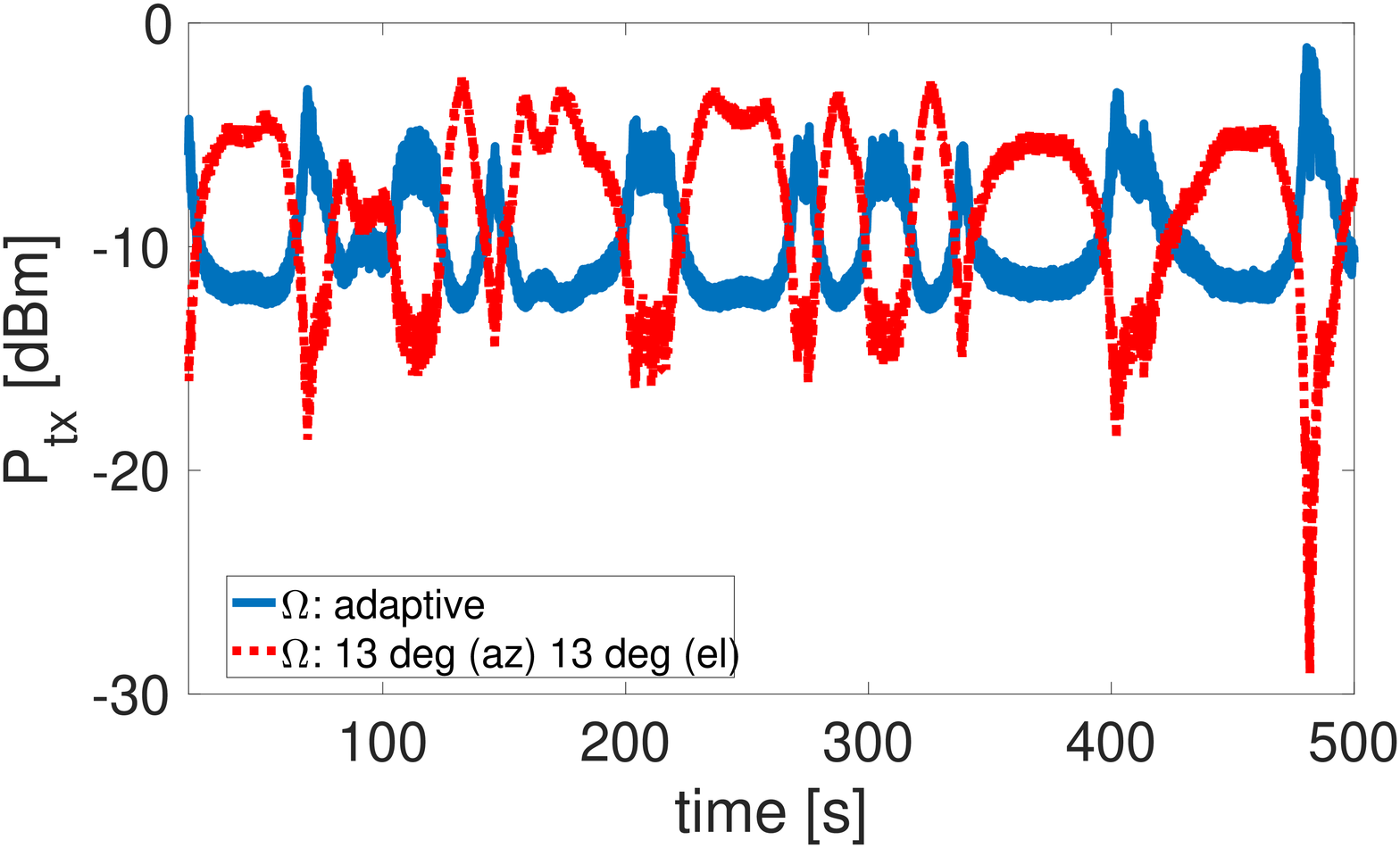}\label{subfig:heuristic_vs_fixedbw_Ptx_1}} 
	\subfloat[][Tx power]{\includegraphics[width=0.45\columnwidth]{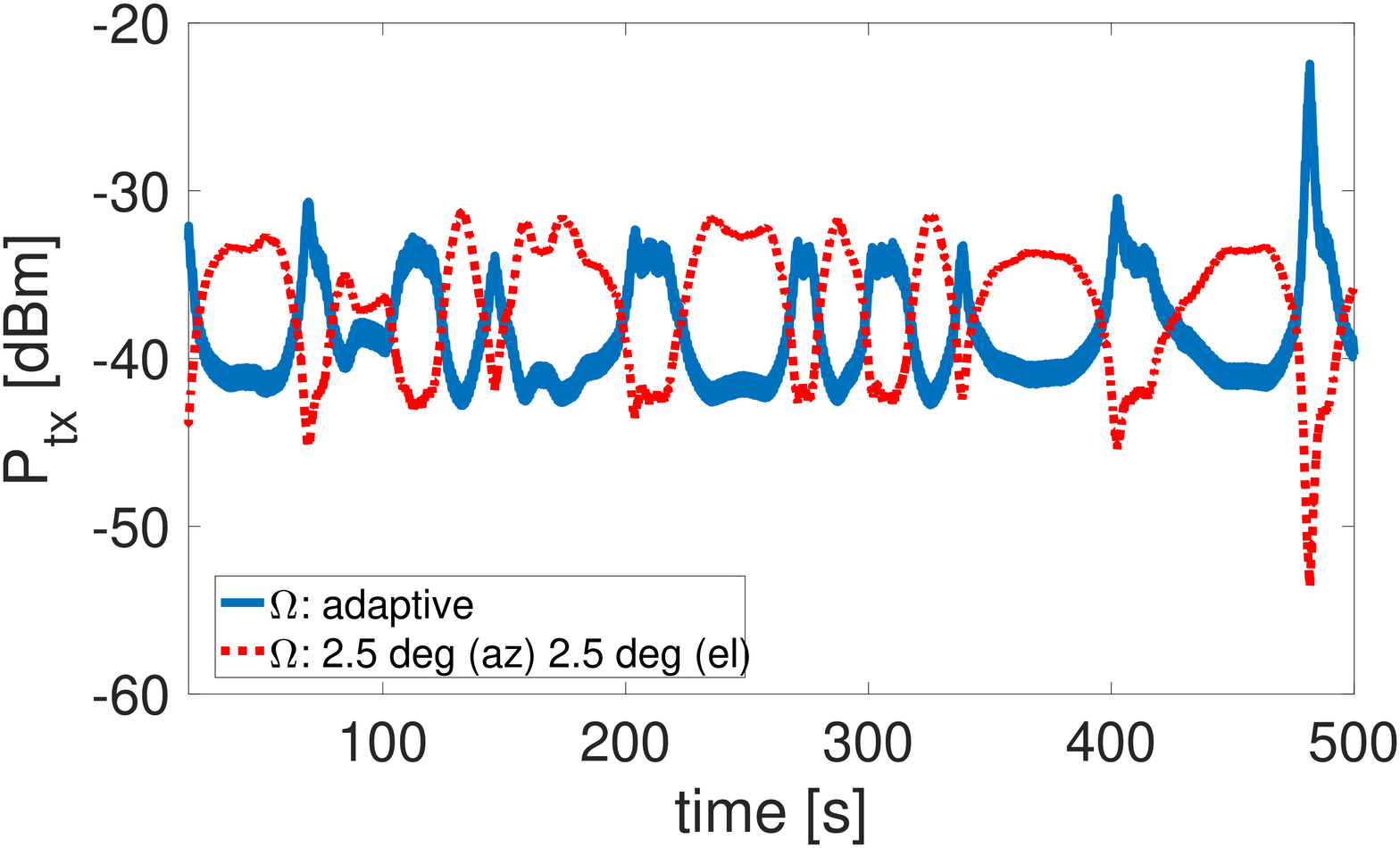}\label{subfig:heuristic_vs_fixedbw_Ptx_2}} \\ \vspace{-0.38cm}
	\subfloat[][SNR]{\includegraphics[width=0.45\columnwidth]{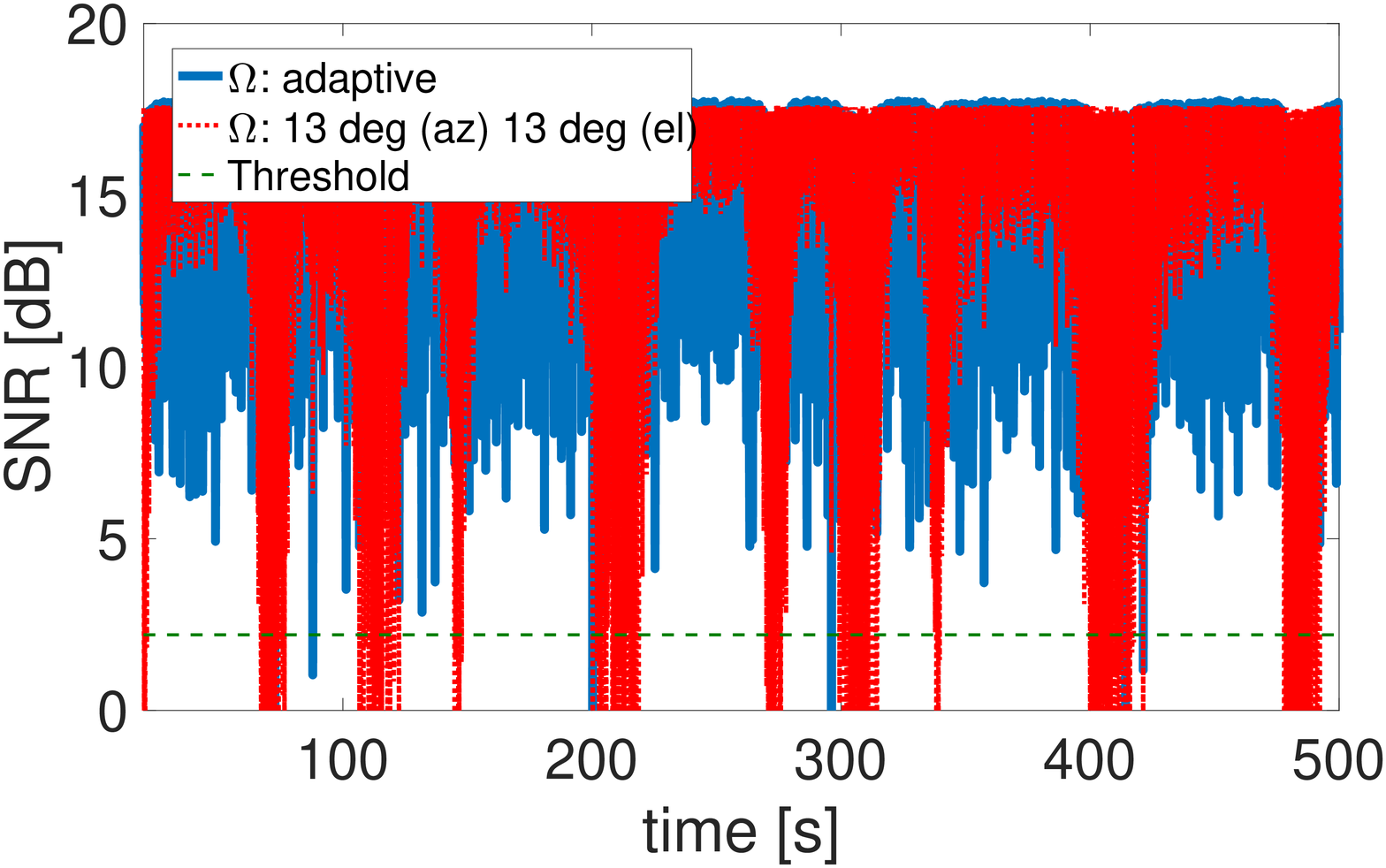}\label{subfig:heuristic_vs_fixedbw_SNR_1}} 
	\subfloat[][SNR]{\includegraphics[width=0.45\columnwidth]{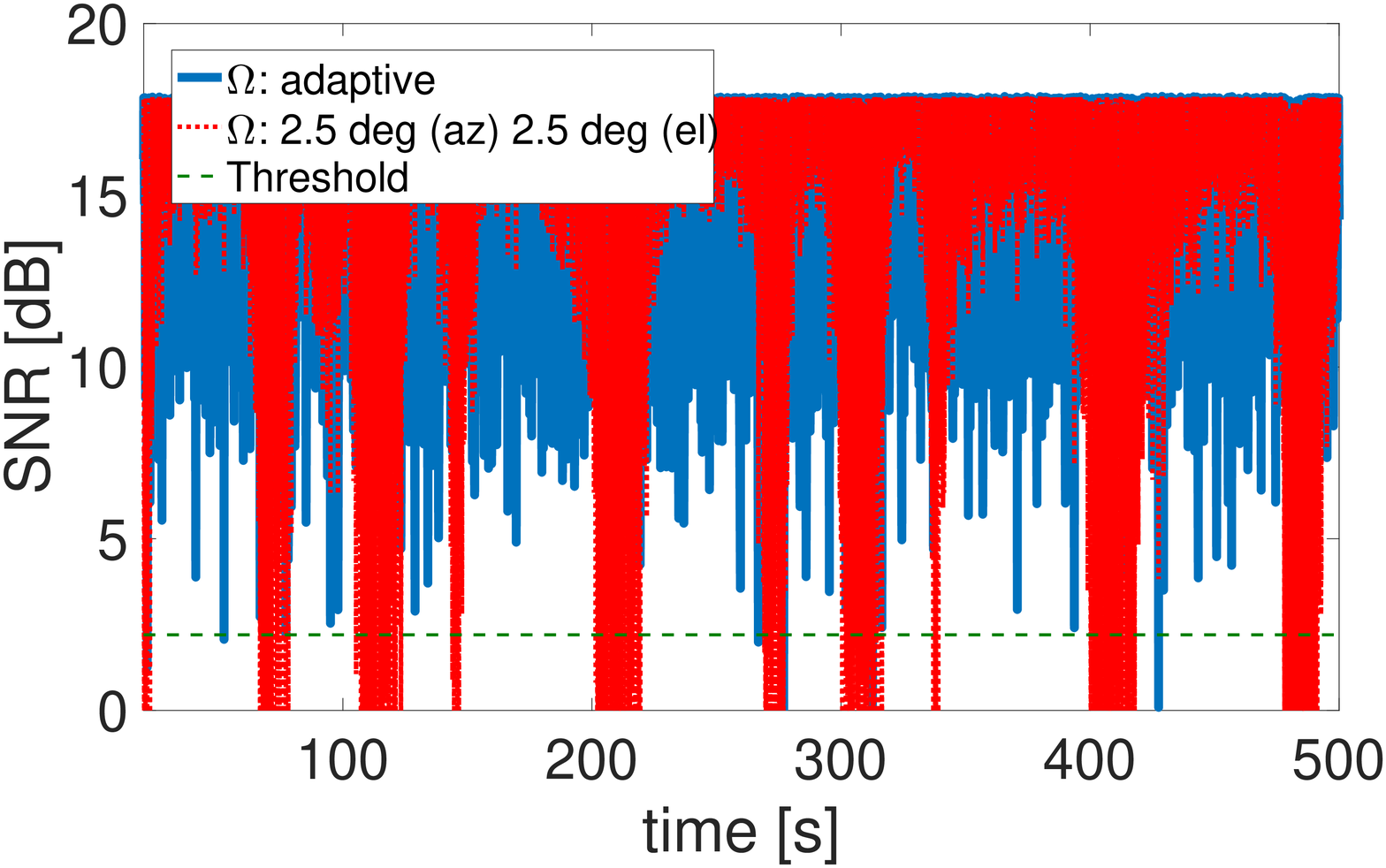}\label{subfig:heuristic_vs_fixedbw_SNR_2}} \\ \vspace{-0.38cm}
	\subfloat[][SNR CDF]{\includegraphics[width=0.45\columnwidth]{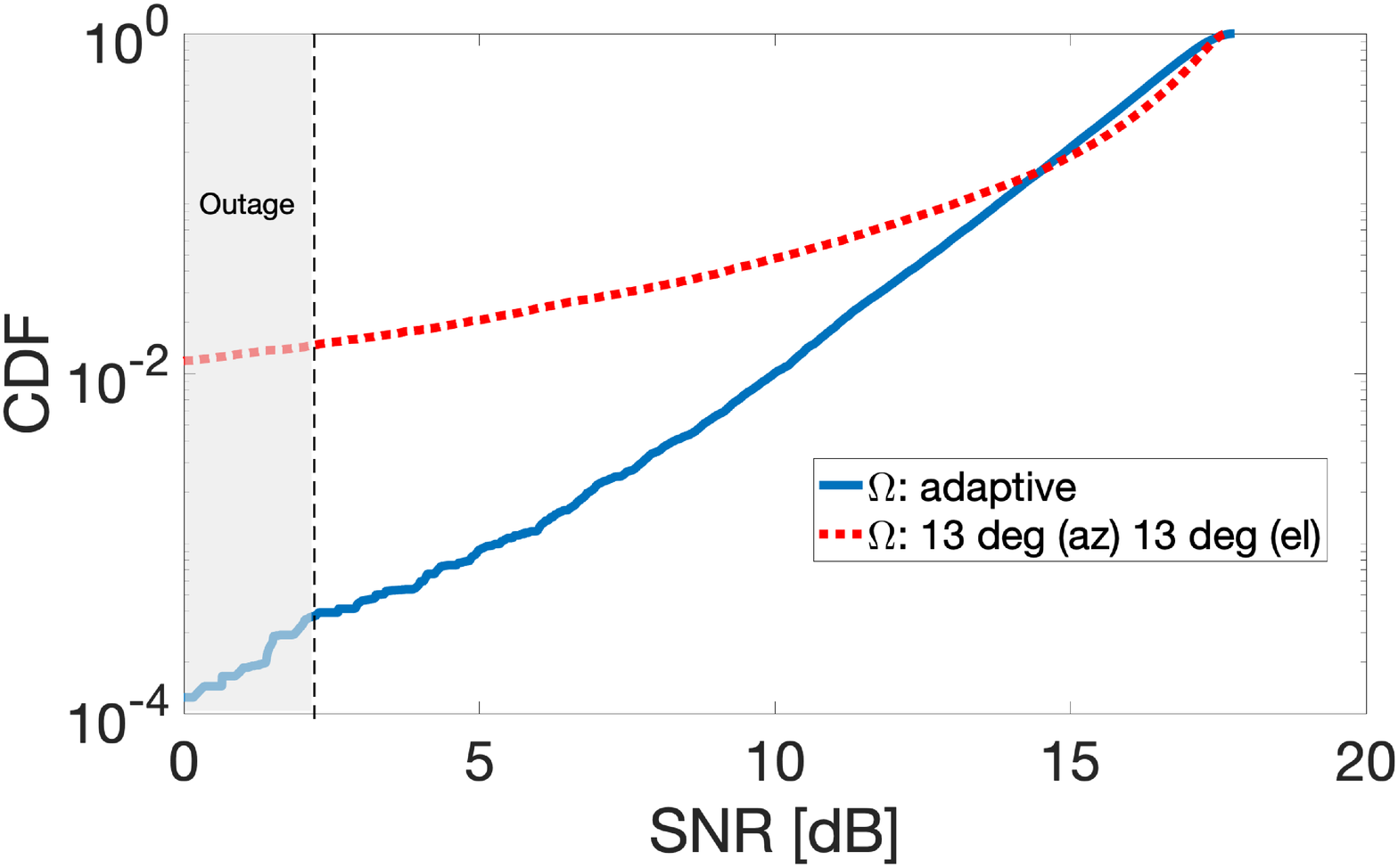}\label{subfig:heuristic_vs_fixedbw_CDFSNR_1}} 
	\subfloat[][SNR CDF]{\includegraphics[width=0.45\columnwidth]{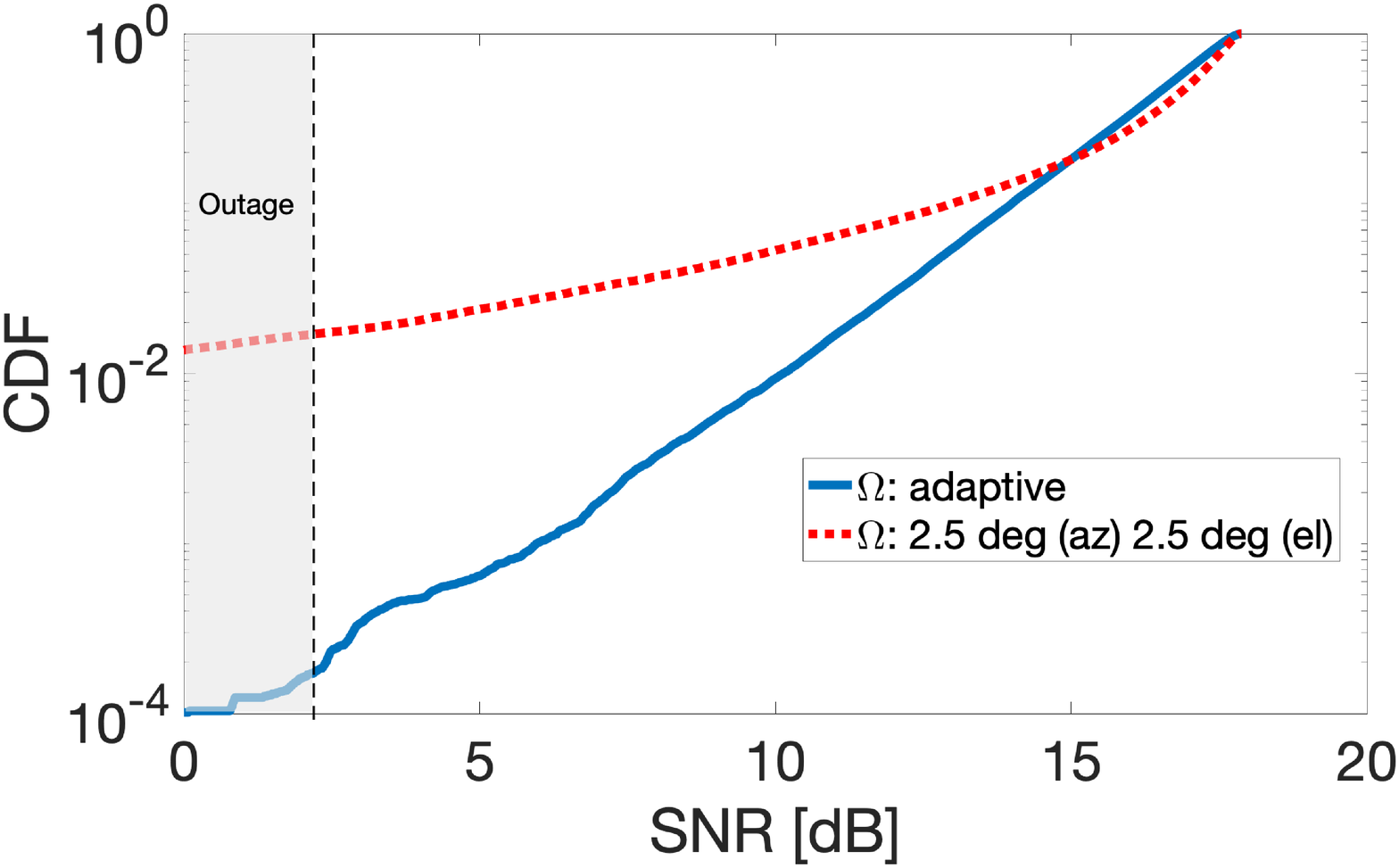}\label{subfig:heuristic_vs_fixedbw_CDFSNR_2}} \\ 
	\caption{Proposed heuristic BPC (blue lines) vs. fixed-beamwidth BAT (red lines) performance, for $\sigma_p = 1.5$ m,  $\sigma_{\gamma} = 1.5$ deg (left column), $\sigma_p = 0.15$ m,  $\sigma_{\gamma} = 0.15$ deg (right column), and $\tau=10$ ms, $f_{\mathrm{data}}=100$ Hz.}
	\label{fig:heuristic_vs_fixedbw}
\end{figure} 

To show the advantages of the proposed sensor-assisted dynamic \ac{BPC}, we use as comparison a fixed beamwidth \ac{BAT} system where only the Tx power is adjusted according to the estimated reciprocal distance and the SNR threshold value. In other words, $v_1$ and $v_2$ do not exploit the covariance of the position estimates. Therefore, $v_1$ adapts the Tx power as in Subsection \ref{subsec:beamadaptation}, but the beamwidth $\mathbf{\Omega}$ is kept fixed and equal to the $v_2$'s one:
\begin{equation}\label{eq:Ptxcontrol_fixedbw}
\begin{split}
\hat{P}^{\mathrm{fixed}}_{\mathrm{tx}} & = \dfrac{\mathrm{SNR}_{\mathrm{min}}\,  \hat{\eta}(\hat{d}) \, P_{\mathrm{noise}} }{ \dfrac{\left(G^{\mathrm{max}}\left(\mathbf{\Omega}\right)\right)^2}{16}}\, .
\end{split}
\end{equation}
To this aim, we select the fixed beamwidth to be equal to the average value of the heuristic ones, to enable the fairest comparison between the two algorithms. Results are reported in Fig. \ref{fig:heuristic_vs_fixedbw} where the left column refers to currently available sensors' accuracies ($\sigma_p = 1.5$ m and $\sigma_{\gamma} = 1.5$ deg)  while the right column is intended for next-generation mobility, where precise position and orientation information is required to fulfill automated driving capabilities ($\sigma_p = 0.15$ m and $\sigma_{\gamma} = 0.15$ deg). To facilitate the reader in analyzing the results, we will refer to this two driving scenarios as:
 	\begin{enumerate}
 		\item S1: current mobility,
 		\item S2: next-generation mobility,
 	\end{enumerate}
where the distinction is merely provided by quality of position/orientation estimation.
The latter directly impacts on the beamwidth dimension: precise position/orientation information allows to use narrower beams. This relation is shown in Fig. \ref{subfig:heuristic_vs_fixedbw_bw_1} and Fig. \ref{subfig:heuristic_vs_fixedbw_bw_2}, where the beam assumes values in the interval $\left[10,120\right]$ deg for S1, while for S2 the range of values is $\left[1.8,15\right]$ deg. On the other hand, a fixed beamwidth algorithm selects circular beams of $13$ deg in S1 and of $2.5$ deg in S2. The use of narrow beams allows a reduction in power consumption at the Tx side, as the emitted power is concentrated in a narrower spatial region, at the expenses of being more susceptible to fast variations in the link geometry. The evolution of the transmit power $P_{\mathrm{tx}}$ over time is  shown in Fig. \ref{subfig:heuristic_vs_fixedbw_Ptx_1} and Fig. \ref{subfig:heuristic_vs_fixedbw_Ptx_2}, for S1 and S2, respectively. We recall that the adaptation of $P_{\text{tx}}$ for the fixed beamwidth BAT method is based only on the estimate V2V distance, while the proposed \ac{BPC}  also considers the available information on position/orientation accuracy. Results shows a fluctuating $P_{\mathrm{tx}}$ around $-9$ dBm for S1, a mean value that decreases to $-37$ dBm for S2. It is important to notice that the mean power for the proposed \ac{BPC} is lower with respect to the fixed beamwidth case ($6.5$ dBm less in S1 and $7$ dBm less in S2).
The joint combination of beamwidth selection at each vehicle and of the Tx power directly impacts on the V2V link quality \eqref{eq:Prx}. Figs. \ref{subfig:heuristic_vs_fixedbw_SNR_1} and \ref{subfig:heuristic_vs_fixedbw_SNR_2} illustrate the SNR over time, where the threshold on the SNR is also indicated as to discriminate if the V2V link is in outage or not. It can be appreciated that a fixed beamwidth BAT presents many outage events in correspondence of a reduction in $P_{\mathrm{tx}}$, indicating that relying only on a distance information is not enough to guarantee a reliable communication. On the other hand, an improved stability over time of the SNR is experienced with the proposed \ac{BPC}, reducing at minimum the outage events. A better insight on the outage is provided in terms of CDF of the SNR in Figs. \ref{subfig:heuristic_vs_fixedbw_CDFSNR_1} and \ref{subfig:heuristic_vs_fixedbw_CDFSNR_2}, for S1 and S2, respectively. These two plots present a similar behavior as the use of a narrower beam is compensated by a significant reduction in the emitted power. Once more, it is shown how the proposed \ac{BPC} provides an improved robustness to the nearly-instantaneous variations in the V2V link geometry caused by the reciprocal variation of dynamics, significantly reducing the drops in SNR. 

\begin{figure}[]
	\centering
	\includegraphics[width=0.8\columnwidth]{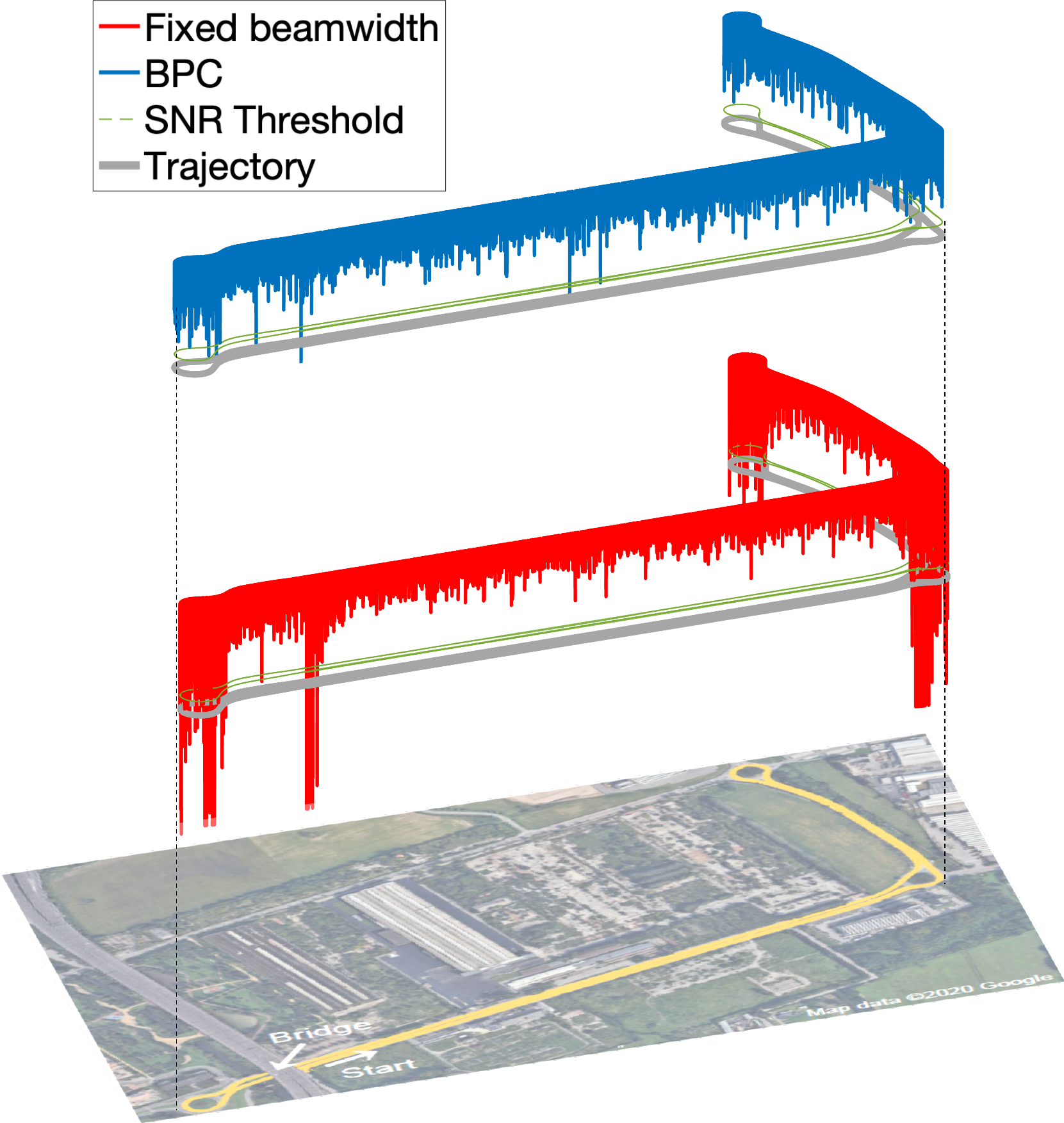}
	\caption{Spatial representation along the road trajectory of the SNR obtained with the proposed heuristic \ac{BPC} (blue line) vs. fixed-beamwidth BAT system SNR (red line), for $\sigma_p=0.15$ m and $\sigma_{\gamma}=0.15$ deg and $\tau=10$ ms, $f_{\mathrm{data}}=100$ Hz.}
	\label{fig:heuristic_vs_fixedbw_SNR3D}
\end{figure}

\begin{figure}[]
	\centering
	\subfloat[][$f_{\mathrm{data}}=100$ Hz]{\includegraphics[width=0.45\columnwidth]{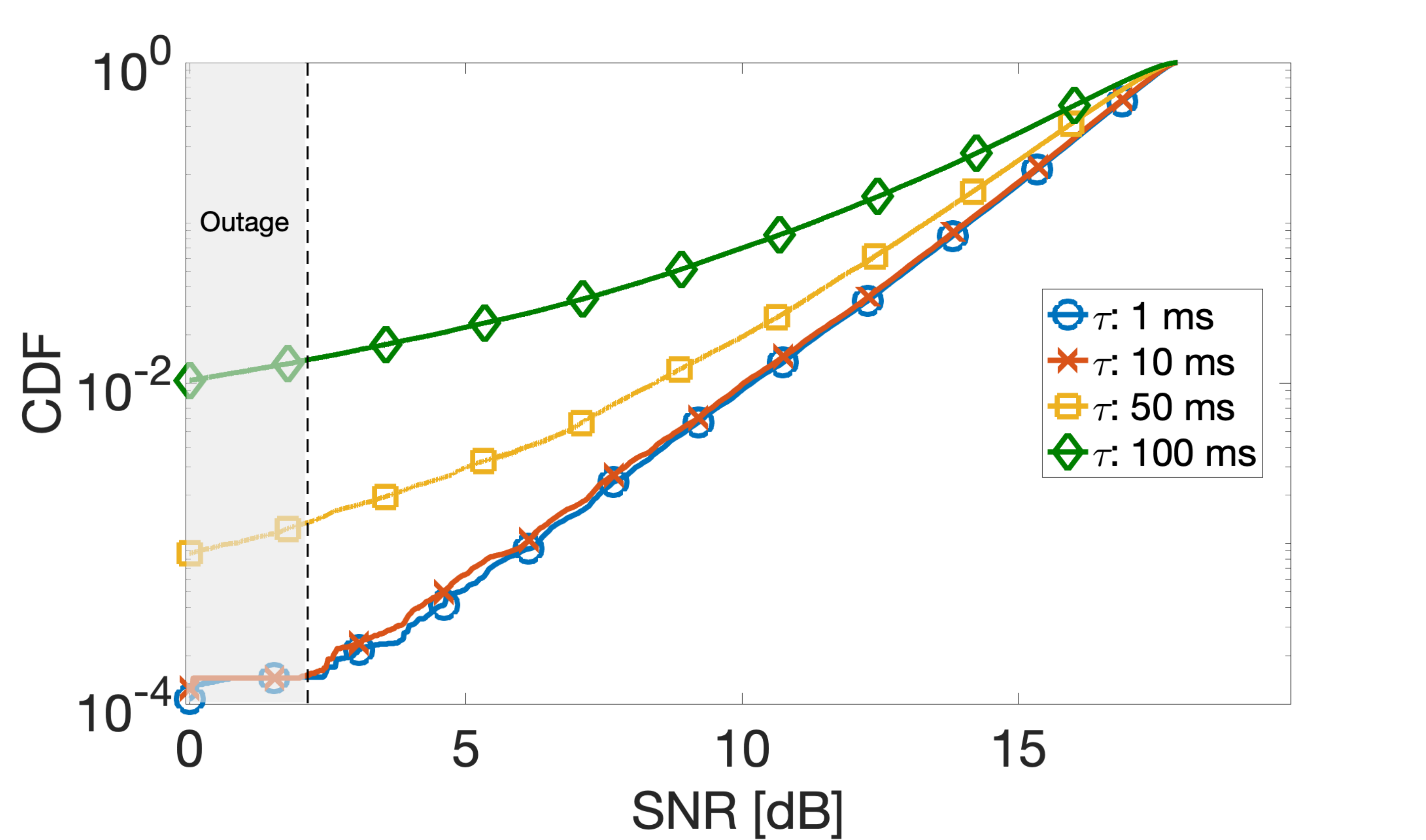}\label{subfig:heuristic_vs_parameters_100Hz}} 	
	\subfloat[][$f_{\mathrm{data}}=1$ kHz]{\includegraphics[width=0.458\columnwidth]{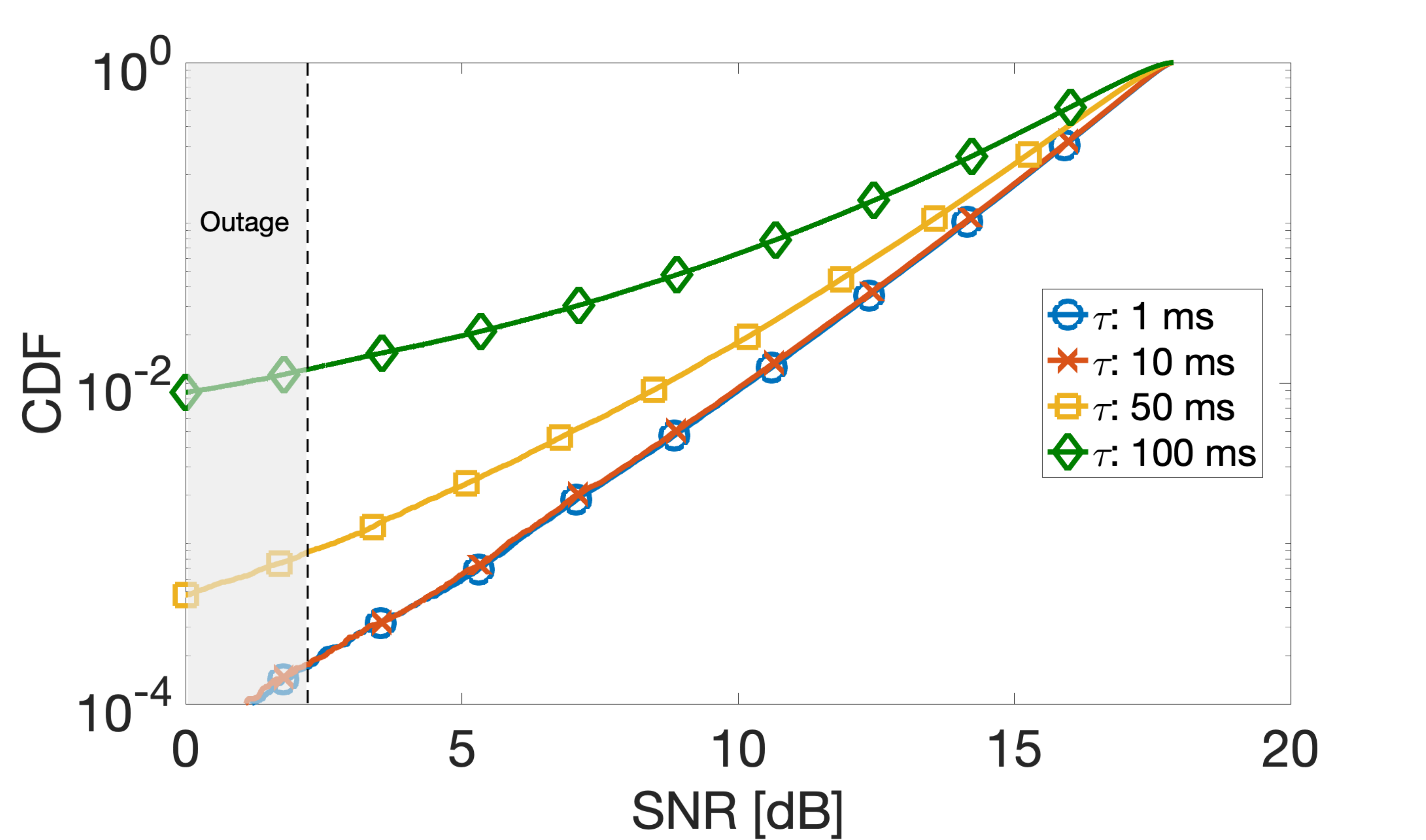}\label{subfig:heuristic_vs_parameters_1000Hz}}\\ 
	\caption{Heuristic (Alg. 1) SNR CDF, for $f_{\mathrm{data}}=100$ Hz (\ref{subfig:heuristic_vs_parameters_100Hz}) and $f_{\mathrm{data}}=1$ kHz (\ref{subfig:heuristic_vs_parameters_1000Hz}), for $\sigma_p = 0.15$ m,  $\sigma_{\gamma} = 0.15$ deg, varying the latency $\tau$.}
	\label{fig:heuristic_vs_paramters}
\end{figure}

In order to look at the results in Figs. \ref{subfig:heuristic_vs_fixedbw_SNR_1} and \ref{subfig:heuristic_vs_fixedbw_SNR_2} with more detail, we report in Fig. \ref{fig:heuristic_vs_fixedbw_SNR3D} the SNR over the trajectory for $\sigma_p=0.15$ m and $\sigma_{\gamma}=0.15$ deg. The adoption of a sensor-assisted dynamic BPC allows to avoid the SNR drops corresponding to fast curves, where the alignment of fixed, narrow beams still BAT is nearly impossible, even with an accurate position/orientation information and 5G-compliant latency ($10$ ms).

As last remark on Fig. \ref{fig:heuristic_vs_fixedbw} and Fig. \ref{fig:heuristic_vs_fixedbw_SNR3D}, we want to underline that the outage performance of the proposed BPC are practically dependent on the selected confidence interval employed to determine the beamwidth \eqref{eq:beamwidth_selection}: by increasing $k$, the outage probability can be reduced to suit the stringent requirements of high LoA applications, at the price of an higher Tx power.

The last set of results we present are aimed to show the impact of latency $\tau$ and sensor's sampling frequency $f_{\mathrm{data}}$ on the proposed \ac{BPC} algorithm, analyzed for $\sigma_p=0.15$ m and $\sigma_{\gamma}=0.15$ deg, where their impact is expected to be the greatest. The results are reported in Fig. \ref{fig:heuristic_vs_paramters}.  We can notice that, increasing the sampling frequency leads to better performance, but, in practice, the difference between a 100 Hz and a 1 kHz system is negligible, as 100 Hz is basically enough to capture the vehicle dynamics. The latency over the control link $\tau$ shows its effect only for values larger than 10 ms, which however corresponds to the upper limit envisioned for 5G. For the sake of comparison, in case of a less accurate position/orientation estimation ($\sigma_p=1$ m, $\sigma_{\gamma}=1$ deg, $f_{\mathrm{data}} = 10$ Hz) the performance of the sensor-assisted dynamic \ac{BPC} system would be practically independent on $\tau$. For this reason, we can conclude that both the sensors' sampling frequency and the latency on the control link are not critical parameters for the proposed system, even in case of a very accurate position/orientation estimation.

\section{Conclusion}
\label{sec:conclusion}

This paper proposes a sensor-assisted dynamic BPC method
for beam-based mmW or sub-THz Vehicle-to-Anything
(V2X) communications. Data from on-board sensors (GPS
and IMU) are used to estimate vehicle dynamics (position
and orientation), which is exchanged over a low-frequency
control link with other vehicles to acquire the instantaneous
knowledge of the nearby system geometry. The proposed
BPC strategy leverages the position/orientation and related
uncertainty (covariance) to control the Tx/Rx beamwidths
and Tx power, explicitly avoiding constrained optimization
approaches, which require huge amount of information to
be solved, incompatible with rapidly time-varying V2V
scenarios. The proposed approach is validated in simulation
for a LOS V2V communication system from data of real
trajectory (with measured GPS and IMU data) from a
dedicated an experimental campaign. The results show that
the proposed BPC allows to practically match the optimal
achievable performance (up to 1-2 dB of excess power).
Meaningful improvements are experienced by adopting the
proposed BPC system with respect to a fixed-beamwidth
one. This confirms the need of V2V apparata to continuously
adapt to the fast mobility of vehicles, preventing the design
and use of a ‘deterministic/static’ configuration in place of
a re-configurable one. Future investigations will extend the
experimental campaign including multiple on-board sensors
in the position/orientation estimation and will consider a
more general vehicular network, where the mutual V2V
interference must be considered in the BPC algorithm.

\appendices
\section{} \label{app:appendix}
This appendix contains the details on the \ac{EKF} implementation mentioned in Section \ref{sect:EKF}.

The state evolution in \eqref{eq:state_transition} is expanded as a function of the single state elements (position, velocity, orientation) as:
\begin{equation}\label{eq:state_equation_split}
	\begin{split}
		\mathbf{p}_{t|t-1} &= \mathbf{p}_{t-1} +  T \, \mathbf{v}_{t-1} + \frac{ T^2}{2}\left(\mathbf{R}_{t}\left \{\mathbf{q}_{t-1} \right \} \left(\mathbf{z}^b_{a,t-1}-\mathbf{b}_{a,t-1}\right) + \mathbf{g}_{t-1} + \mathbf{w}_{a,t}\right) \,,\\
		\mathbf{v}_{t|t-1} &= \mathbf{v}_{t-1} +  T \left(\mathbf{R}_{t}\left \{\mathbf{q}_{t-1} \right \} \left(\mathbf{z}^b_{a,t-1}-\mathbf{b}_{a,t-1}\right) + \mathbf{g}_{t-1} + \mathbf{w}_{a,t}\right) \,,\\
		\mathbf{q}_{t|t-1} &= \mathbf{q}_{t-1} \odot \exp_{\mathrm{q}}\left \{ \frac{ T}{2} \left(\mathbf{z}^b_{\omega,t-1} - \mathbf{b}_{\omega,t-1} + \mathbf{w}_{\omega,t}\right)\right \} \,,
	\end{split}
\end{equation}	
where $\mathbf{w}_t = \left[\mathbf{w}_{a,t}\,\,\,\mathbf{w}_{\omega,t}\right]^{\mathrm{T}}\in\mathbb{R}^{6\times 1}$, $T$ is the fundamental sampling of the tracking system (usually dictated by the IMU),  vectors $\mathbf{b}_{a,t}\in\mathbb{R}^{3 \times 1}$ and $\mathbf{b}_{\omega,t}\in\mathbb{R}^{3 \times 1}$ indicate the bias of 3D accelerometer and 3D gyroscope, while $\mathbf{g}_{t}$ comprises the components of gravitational acceleration.

The \ac{EKF} alternates one (or more) prediction steps for mean and covariance 
\begin{align}\label{eq:prediction}
\hat{\boldsymbol{\theta}}_{t|t-1} & = \mathbf{f}\left(\hat{\boldsymbol{\theta}}_{t-1|t-1},\mathbf{u}_{t-1}, \mathbf{w}_{t} \right)  \, ,\\
\mathbf{P}_{t|t-1} & = \mathbf{F}_{t-1}\mathbf{P}_{t-1}\mathbf{F}_{t-1}^{\mathrm{T}} + \mathbf{G}_{t-1}\mathbf{C}_{s,t}\mathbf{G}_{t-1}^{\mathrm{T}} \, , 
\end{align}
with the update steps
\begin{align}\label{eq:update}
\hat{\boldsymbol{\theta}}_{t|t} & = \hat{\boldsymbol{\theta}}_{t|t-1} + \mathbf{K}_{t} \left(\mathbf{z}_{t}- \mathbf{h}(\hat{\boldsymbol{\theta}}_{t|t-1} ) \right)  \, , \\
\mathbf{P}_{t|t} & = \mathbf{P}_{t|t-1} - \mathbf{K}_{t}  \mathbf{H}_{t} \mathbf{P}_{t|t-1}
\, .
\end{align}
Matrices $\mathbf{F}_{t}$ and $\mathbf{G}_{t}$ indicate, respectively, the gradients of the state equation in \eqref{eq:state_equation_split} with respect to $\boldsymbol{\theta}_{t}$ and $\mathbf{w}_{s,t}$, while matrix $\mathbf{K}_{t} = \mathbf{P}_{t|t-1} \mathbf{H}_{t}^{\mathrm{T}} (\mathbf{H}_{t}\mathbf{P}_{t|t-1}\mathbf{H}_{t}^\mathrm{T} + \mathbf{C}_{n,t})^{-1}$ is the Kalman filter gain.

The process noise covariance $\mathbf{C}_{w,t}$ is
\begin{equation}\label{eq:Cs}
\mathbf{C}_{w,t} = 
\begin{bmatrix}
\sigma_{a,t}^2\mathbf{I}_3 & \mathbf{0}_{33}  & \mathbf{0}_{33} \\
 \mathbf{0}_{33} & \sigma_{a,t}^2\mathbf{I}_3 & \mathbf{0}_{33} \\
\mathbf{0}_{33}  &   \mathbf{0}_{33} & \sigma_{\omega,t}^2\mathbf{I}_3
\end{bmatrix},
\end{equation}
while the observation covariance $\mathbf{C}_{n,t}$ is
\begin{equation}\label{eq:Cm}
\mathbf{C}_{n,t} = 
\begin{bmatrix}
\sigma_{\mathrm{GNSS},t}^2\mathbf{I}_3  & \mathbf{0}_{31} & \mathbf{0}_{34} \\
\mathbf{0}_{13} & \sigma_{v}^2 & \mathbf{0}_{14}  \\
\mathbf{0}_{43} &  \mathbf{0}_{41} & \mathbf{C}_{q,t} \\
\end{bmatrix}\, ,
\end{equation}
where the covariance matrix of the quaternion is obtained from the covariance matrix of Euler angles, a-priori known and computed as \cite{kok2017using}:
\begin{equation}\label{eq:euler_quat_mapping}
\mathbf{C}_{q,t}  = \frac{\partial \mathbf{m}}{\partial \boldsymbol{\gamma}_{t}} \, \mathbf{C}_{\gamma,t} \left(\frac{\partial \mathbf{m}}{\partial \boldsymbol{\gamma}_{t}}\right)^{\mathrm{T}}
\end{equation}
by using the mapping from quaternions to Euler angles defined as:
\begin{equation}
\mathbf{q}_{t}  = \mathbf{m}\left(\boldsymbol{\gamma}_{t}\right). 
\end{equation}

The gradients $\mathbf{F}_{t}$, $\mathbf{G}_{t}$ and $\mathbf{H}_{t}$ are computed as:
\begin{equation}\label{eq:Fmatrix}
	\mathbf{F}_{t} = 
	\begin{bmatrix}
	\mathbf{I}_3 &  T\,\mathbf{I}_3 & \dfrac{T^2}{2} \dfrac{\partial \mathbf{R}\left \{\mathbf{q}_{t|t} \right \}}{\partial \mathbf{q}_{t|t}} \mathbf{z}^b_{a,t}\\
	\mathbf{0}_{33} & \mathbf{I}_3 & \ T\,\dfrac{\partial \mathbf{R}\left \{\mathbf{q}_{t|t} \right \}}{\partial \mathbf{q}_{t|t}} \mathbf{z}^b_{a,t} \\
	\mathbf{0}_{43} & \mathbf{0}_{43} & \left(\exp_{\mathrm{q}} \left \{ \dfrac{ T}{2} \mathbf{y}^b_{\omega,t}\right \} \right)^{\mathrm{R}}
	\end{bmatrix}\, ,
\end{equation}
\begin{equation}\label{eq:Gmatrix}
    \mathbf{G}_{t} = 
	\begin{bmatrix}
\mathbf{I}_6 & \mathbf{0}_{63} \\
\mathbf{0}_{46} & -\dfrac{ T}{2}\left(\hat{\mathbf{q}}_{t|t}\right)^{\mathrm{L}}\dfrac{\partial\exp_{\mathrm{q}} \left \{ \mathbf{w}^b_{\omega,t}\right \} } {\partial \mathbf{w}^b_{\omega,t}}
    \end{bmatrix}\, ,
\end{equation}
\begin{equation}\label{eq:Hmatrix}
\mathbf{H}_{t} = 
	\begin{bmatrix}
	\mathbf{I}_3 & \mathbf{0}_{33} & \mathbf{0}_{34} \\
	\mathbf{0}_{13} & \frac{\left(\mathbf{v}^n_{t|t}\right)^{\mathrm{T}}}{\norm{\mathbf{v}^n_{t|t}}_2} & \mathbf{0}_{14} \\
	\mathbf{0}_{43} & \mathbf{0}_{43} &  \mathbf{I}_4
    \end{bmatrix}\, ,
\end{equation}
where we indicate with $(\cdot)^{\mathrm{L}}$ and $(\cdot)^{\mathrm{R}}$ two different matrix representations of a quaternion, as defined in \cite{kok2017using}. When unit quaternions are employed, a renormalization is necessary after each update step \cite{kok2017using}:
\begin{equation}\label{eq:renormalization}
\begin{split}
\hat{\mathbf{q}}_{t|t} & = \dfrac{\widetilde{\mathbf{q}}_{t|t}}{\norm{\widetilde{\mathbf{q}}_{t|t} }_2}\,, \\
\mathbf{P}_{t|t} & = \mathbf{J}_{t} \widetilde{\mathbf{P}}_{t|t} \mathbf{J}_{t}^{\mathrm{T}}\,,
\end{split}
\end{equation}
where we indicate with $\widetilde{\mathbf{q}}_{t|t}$ and $\widetilde{\mathbf{P}}_{t|t}$ the quaternion and its covariance before the renormalization and matrix $\mathbf{J}_{t}$ is:
\begin{equation}\label{eq:J}
\mathbf{J}_{t} = 
\begin{bmatrix}
\mathbf{I}_6 & \mathbf{0}_{64} \\
\mathbf{0}_{46} & \dfrac{1}{\norm{\widetilde{\mathbf{q}}_{t|t}}^3_2} \widetilde{\mathbf{q}}_{t|t} \widetilde{\mathbf{q}}^{\mathrm{T}}_{t|t}
\end{bmatrix} \, .
\end{equation}

\section{} \label{app:appendix2}
This appendix reports the detailed derivation of the effect of a Tx/Rx position error and Tx orientation error on the $(\delta x^{\mathrm{LOS}_2},z^{\mathrm{LOS}_2} )$ plane (Subsection \ref{subsec:optimization}) and the expression of the gradients used in the proposed heuristic BPC method (Section \ref{subsec:beamadaptation}).

The effect of a Tx position error in the $n$-frame in the $\mathrm{LOS_2}$-system can be evaluated by 
considering the scheme in Fig. \ref{fig:Alignment_and_Outage}. The position error $\delta\mathbf{p}^{n}_1$ is firstly rotated by the matrix $\mathbf{R}\left \{ \boldsymbol{\gamma}^{v_1 n }_1\right \}$ to align to the $v_1$-system (the Tx local coordinate system) and then further rotated by the matrix $\bar{\mathbf{R}}_1$ to be aligned with the $\mathrm{LOS_1}$-system, equivalent to the $\mathrm{LOS_2}$-system, as:
\begin{equation}
	\delta\mathbf{p}^{\mathrm{LOS}_2}_1 = \underbrace{\bar{\mathbf{R}}_1 \mathbf{R}\left \{ \boldsymbol{\gamma}^{v_1 n}_1\right \}}_{\mathbf{Q}_{p_1}} \delta\mathbf{p}^{n}_1 \sim \mathcal{N}(\mathbf{0}, \underbrace{\mathbf{Q}_{p_1} \mathbf{C}^n_{p_1}\mathbf{Q}^{\mathrm{T}}_{p_1}}_{\widetilde{\mathbf{C}}^{\mathrm{LOS}_2}_{p_1}})\,,
\end{equation}
from which 
\begin{equation}
\begin{bmatrix}
\delta x^{\mathrm{LOS}_2}_1\\
\delta z^{\mathrm{LOS}_2}_1
\end{bmatrix} \sim \mathcal{N}\left( \begin{bmatrix}
0\\
0
\end{bmatrix} , \mathbf{C}^{\mathrm{LOS}_2}_{p_1} \right) \,,
\end{equation}
where 
\begin{equation}
	\mathbf{C}^{\mathrm{LOS}_2}_{p_1} = \begin{bmatrix}
	\left[\widetilde{\mathbf{C}}^{\mathrm{LOS}_2}_{p_1}\right]_{11} \, \, \left[\widetilde{\mathbf{C}}^{\mathrm{LOS}_2}_{p_1}\right]_{13} \\
	\left[\widetilde{\mathbf{C}}^{\mathrm{LOS}_2}_{p_1}\right]_{31} \, \, \left[\widetilde{\mathbf{C}}^{\mathrm{LOS}_2}_{p_1}\right]_{33} 
	\end{bmatrix} \in \mathbb{R}^{2 \times 2}.
\end{equation}
A similar derivation is made for the Rx position error $\delta\mathbf{p}^{\mathrm{LOS}_2}_2$ and $\begin{bmatrix}
\delta x^{\mathrm{LOS}_2}_2 \,\,\,
\delta z^{\mathrm{LOS}_2}_2
\end{bmatrix}^{\mathrm{T}}$.

%DESCRIZIONE DELLA DERIVAZIONE DELL'ERRORE ANGOLARE NEL SISTEMA LOS-2
The effect of an orientation error $\delta \gamma^{n v_1}_1$ on the $(\delta x^{\mathrm{LOS}_2},z^{\mathrm{LOS}_2} )$ plane is obtained by first computing the distribution of the orientation error expressed in the $\mathrm{LOS_1}$-system, $\delta \gamma^{n \mathrm{LOS_1}}_1$, and then projecting onto the  $(\delta x^{\mathrm{LOS}_2},z^{\mathrm{LOS}_2} )$ plane by multiplying the result by $d$, the V2V true distance. The first step is achieved by considering, as stated in Subsection \ref{subsec:optimization}, two components of $\delta \boldsymbol{\gamma}^{n \mathrm{LOS}_1}_1$, namely $\delta \phi^{n \mathrm{LOS}_1}_1$ and $\delta \psi^{n \mathrm{LOS}_1}_1$, which are defined as:
\begin{align}\label{eq:AzimuthandElevation_LOSframe}
    %elevazione
	\delta \phi^{n \mathrm{LOS}_1}_1 &= \mathrm{asin}\left(\frac{\Delta {p}_{12,z}^{\mathrm{LOS_1}}}{ \norm{\Delta\mathbf{p}_{12}^{\mathrm{LOS_1}}}_2}\right) \,, \\
	%azimuth
	\delta \psi^{n \mathrm{LOS}_1}_1 &= \mathrm{atan}\left(\frac{\Delta {p}_{12,y}^{\mathrm{LOS_1}}}{\Delta {p}_{12,x}^{\mathrm{LOS_1}}}\right) \,,
\end{align}
which are basically the definitions of elevation and azimuth angles in \eqref{eq:AzimuthandElevation}, expressed in the $\mathrm{LOS_1}$-system. The derivation of the distribution of $\delta \phi^{n \mathrm{LOS}_1}_1$ and $\delta \psi^{n \mathrm{LOS}_1}_1 $ follows the linearization steps outlines in subsection \ref{subsec:beamadaptation}, retrieving first the gradient of $\Delta\mathbf{p}_{12}^{\mathrm{LOS_1}}$ with respect to $\delta \gamma^{n v_1}_1$ as:
\begin{align}
	\delta \Delta \mathbf{p}_{12}^{\mathrm{LOS}_1} \approx \underbrace{\dfrac{\partial (\bar{\mathbf{R}}_1 \mathbf{R}\left \{\gamma^{n v_1}_1 \right \})}{\partial \gamma^{n v_1}_1}}_{\mathbf{B}_{\gamma_1}} \delta \gamma^{n v_1}_1 \,,
\end{align}
and then by linearizing the relations \eqref{eq:AzimuthandElevation_LOSframe} with respect to $\delta \Delta \mathbf{p}_{12}^{\mathrm{LOS}_1}$, similarly to \eqref{eq:AzimuthandElevation_linearized}. The composition of gradients provides matrix $\mathbf{C}^{\mathrm{LOS}_1}_{\gamma_1}$ of \eqref{eq:orientation_error_1_LOS1}.
%\begin{align}
%\delta \Delta \mathbf{p}_{12}^{\mathrm{LOS}_1} \approx \underbrace{ = \hat{\mathbf{q}}_{1}}\Delta \hat{\mathbf{p}}_{12}^{n}}_{\mathbf{B}_{\delta\gamma,1}}\delta \hat{\mathbf{q}}_1 +  \underbrace{\mathbf{R}^{v_1 n}\left \{ \hat{\mathbf{q}}_{1}\right\}}_{\mathbf{B}_{\Delta p}} \delta \Delta \hat{\mathbf{p}}_{12}^{n}  \,,
%\end{align}

The gradients in Subsection \ref{subsec:beamadaptation} are computed as:

\begin{align}
\mathbf{B}_{q,1} & = \dfrac{\partial \mathbf{R}\left \{\mathbf{q}^{v_1 n}_1 \right \}}{\partial \mathbf{q}^{v_1 n}_1} \biggr\rvert_{\mathbf{q}^{v_1 n}_1 = \hat{\mathbf{q}}^{v_1 n}_{1}}\Delta \hat{\mathbf{p}}_{12}^{n}\,,\\
\mathbf{B}_{\Delta p} & = \mathbf{R}\left \{ \hat{\mathbf{q}}^{v_1 n}_{1}\right\}\,,\\
\mathbf{b}^{\mathrm{T}}_{\alpha} & = \dfrac{\partial \alpha_1\left( \Delta \mathbf{p}_{12}^{v_{1}}\right)}{\partial \Delta \mathbf{p}_{12}^{v_{1}}} \biggr\rvert_{\Delta \mathbf{p}_{12}^{v_{1}}=\Delta \hat{\mathbf{p}}_{12}^{v_{1}}} \,, \\
\mathbf{b}^{\mathrm{T}}_{\beta} & = \dfrac{\partial \beta_1\left( \Delta \mathbf{p}_{12}^{v_{1}}\right)}{\partial \Delta \mathbf{p}_{12}^{v_{1}}}\biggr\rvert_{\Delta \mathbf{p}_{12}^{v_{1}}=\Delta \hat{\mathbf{p}}_{12}^{v_{1}}} \,.
\end{align}

%%------------------------------------------------------
%%DERIVAZIONE  DELL'INTEGRALE DELLA GAUSSIANA 2D SU AREA ELLITTICA
%
%\section{} \label{app:appendix3}

\section*{Acknowledgment} \label{app:acknowledgment}
The authors want to acknowledge Sergio Savaresi and Luca Franceschetti of Systems and Control group of Politecnico di Milano, for the cooperation in the experimental campaign.

\bibliographystyle{IEEEtran}
\bibliography{Bibliography}

\end{document}